\documentclass{article}

\newcommand{\vir}[1]{``#1"}

\newcommand{\ignore}[1]{}
\usepackage{arxiv}
\usepackage[utf8]{inputenc} % allow utf-8 input
\usepackage[T1]{fontenc}    % use 8-bit T1 fonts
\usepackage{url}            % simple URL typesetting
\usepackage{booktabs}       % professional-quality tables
\usepackage{amsfonts}       % blackboard math symbols
\usepackage{nicefrac}       % compact symbols for 1/2, etc.
\usepackage{microtype}      % microtypography
\usepackage{lipsum}		% Can be removed after putting your text content
\usepackage{graphicx}
\usepackage{doi}

\usepackage{graphics,amssymb,amsmath,epsfig,color}
\usepackage{graphicx}

\title{Pareto Optimal Compression of Genomic Dictionaries, with or without Random Access in Main Memory}

\author{ \href{https://orcid.org/0000-0002-6286-8871}{\includegraphics[scale=0.06]{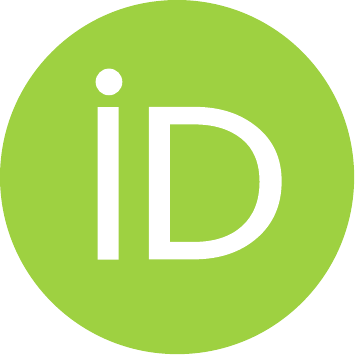}\hspace{1mm}Raffaele ~Giancarlo}\thanks{Corresponding author, \href{email:email-id.com} raffaele.giancarlo@unipa.it} \\
	Department of Mathematics and Computer Science\\
	University of Palermo\\
	Palermo, 90121, Italy \\
	\texttt{raffaele.giancarlo@unipa.it} \\
	\And
	\href{https://orcid.org/0000-0001-6001-5050}{\includegraphics[scale=0.06]{orcid.pdf}\hspace{1mm}Gennaro~Grimaudo} \\
	Department of Engineering\\
	University of Palermo\\
	Palermo, 90121, Italy \\
	\texttt{gennaro.grimaudo@unipa.it} \\
}

\renewcommand{\shorttitle}{\textit{Pareto Optimal Compression of Genomic Dictionaries} }

\hypersetup{
pdftitle={A template for the arxiv style},
pdfsubject={q-bio.NC, q-bio.QM},
pdfauthor={Raffaele~Giancarlo, Gennaro~Grimaudo},
pdfkeywords={genomic data compression, Alignment-Free, compact genomic representations, succinct data structures, Pareto optimal genomic compression},
}

\graphicspath{./images/}

\begin{document}
\maketitle

\begin{abstract}
        {\textbf{Motivation:}} A Genomic Dictionary, i.e., the set of the $k$-mers appearing in a genome, is a fundamental source of genomic information: its collection is the first step in strategic computational methods ranging from assembly to sequence comparison and phylogeny. Unfortunately, it is  costly to store. This motivates some recent studies regarding the compression of those  $k$-mer sets. However, such an area does not have the maturity of genomic compression, lacking an homogeneous and methodologically sound experimental foundation that allows to fairly compare the relative merits of the available solutions, and that takes into account also the rich choices of compression methods that can be used.   \\
        \textbf{Results:}  We provide such a foundation here, supporting it with an extensive set of experiments that use reference datasets and a carefully selected set of representative data compressors. Our results highlight the spectrum of compressor choices one has in terms of Pareto Optimality of compression vs. post-processing, this latter being important when the Dictionary needs to be decompressed many times. In addition to the useful indications,  not available elsewhere, that this study offers to the researchers interested in storing $k$-mer dictionaries in compressed form, a software system that can be readily used  to explore the Pareto Optimal solutions available  r a given Dictionary is also provided. \\
        \textbf{Availability:} The software system is available at {\url{https://github.com/GenGrim76/Pareto-Optimal-GDC}}, together with user manuals and installation instructions.\\
        \textbf{Contact:} \href{email:email-id.com}{raffaele.giancarlo@unipa.it}\\
        \textbf{Supplementary information:} Additional data are available in the Supplementary Material. 
\end{abstract}

% keywords can be removed
\keywords{genomic data compression \and Alignment-Free \and compact genomic representations \and succinct data structures \and Pareto optimal genomic compression}

\section{Introduction} \label{sec:introduction}

    A well established fact in Biological Sequence Analysis, as well as Genome Assembly, is that  $k$-mer Dictionaries, i.e., the set  of length $k$ sub-sequences with their frequencies appearing in a collection of genomic sequences representing a genome, are a fundamental structure. As a consequence, a great deal of effort has been made in order to  devise methods to compute them efficiently,  on architectures ranging from  conventional computers \cite{swati2018} to the Cloud  \cite{FADE,Persson21}. Because of their ubiquitous use in Computational Biology, e.g., Alignment-free Genomic Analysis \cite{Romualdi21, Zielezinski2019}, those Dictionaries  can be seen as a valid alternative to store with respect to the genomic sequences they have been derived from. That is, rather than  computing the Dictionary of a genome every time it is needed, one computes it once and then keeps it on disk for future use, as it may be case Alignment-free Genomic Analysis. 
    Once that such an \emph{alter ego} role  with respect to the genomic sequences is found of interest in a given application domain, it is natural to ask for succinct representations of $k$-mer Dictionaries. Indeed, the size of a Dictionary grows exponentially in $k$, although it cannot exceed the size of the corresponding genome by a multiplicative factor more than  $k$. Yet, in applications such as Genome Assembly or Genome-Scale Phylogeny, the size of the Dictionary may well be at least that of the genomic sequences it represents. Therefore, it is not surprising that some effort has been made in order to devise succinct representations of a $k$-mer Dictionary. This area is well presented in \cite{Chikhi21}, with additional relevant results reported in \cite{Rahman21, Rahman21dataset}. Despite those contributions, the degree of maturity of this area, as indicated in \cite{Rahman21dataset}, is not comparable to the one achieved by the area of genomic sequence compression (see \cite{giancarlo2014compressive, numanagic2016comparison}). The following technical consideration are relevant to place our contribution in a proper light. 
    
    Given a $k$-mer Dictionary compression method, the evaluation of its merits should be measured in terms of compression effectiveness, compression and decompression time.  A standard approach would be to resort to  the classic Data Compression measures such as compression ratio \cite{storer1987data}. In fact, this latter is the measure of choice for genomic compression \cite{numanagic2016comparison}. Unfortunately and rather subtly, when one considers the available solutions for the compression of a $k$-mer dictionary as we do here, care must be exercised in order to obtain a fair and informative comparison.  Another aspect, largely ignored in genomic compression but very relevant here is the 
    \vir{compress once and decompress many times}  paradigm (see \cite{bicriteria2019}), which is now a standard in the Data Compression literature. In particular, the trade-off compression/post-processing time is crucial to evaluate a compression method.  Such an aspect is hardly considered in the area of genomic data compression, although quite relevant here since we expect to compress a Dictionary once and to re-use it many times. Therefore, the time to recover the data from its compressed version is a key parameter for the evaluation of a compression method.  
    Finally, Random Access over $k$-mer Dictionaries is desirable in many applications, as discussed in \cite{Chikhi21}. However,  it is reported to be costly \cite{Rahman21} and it has received very little attention in this area.
    
    Motivated by this State of the Art regarding a crucial data structure for genomic studies, the goal of this research is to carry out a critical and comprehensive analysis of the solutions available for $k$-mer Dictionary compression, in order to have a solid ground on which to continue the development of this area.
    Following \cite{Rahman21dataset}, we consider two application areas, as follows. Application Area 1:  the Dictionary cannot be reconstructed from the input genomic sequences and  Application Area 2:  the Dictionary can be reconstructed from the input genomic sequences. The interested reader can find specific details and examples regarding those two application areas in \cite{Rahman21dataset}.

    Our first contribution is a methodological one, as we extensively characterize the area of $k$-mer Dictionaries compression by identifying  several scenarios  of interest and we compare the various solutions fairly and informatively. In this respect, we focus on compression effectiveness and post-processing time. Those would be the two extreme cases of Pareto optimal solutions with respect to the trade-off compression/post-processing time (see \cite{mornati2013pareto} for an introduction to Pareto Optimality). We go further than that and highlight the benefit of knowing Pareto optimal solutions, on benchmark datasets. 
    Our second contribution is to give a series of useful indications to the researchers interested in working with $k$-mer Dictionaries on the use of various tools in order to save space, time, or both. Our third contribution is to provide a software system, i.e., \textbf{Pareto-Optimal-GDC} (Pareto Optimal Genomic Dictionaries Compression) that can actually be used  in any specific context to establish which are the best choices to compress a given $k$-mer Dictionary, among the paradigmatic ones introduced and discussed here. 

\section{Materials and Methods} \label{Materials_Methods}
    \subsection{\bf Problem Statement and Synopsis of the  Compression/Decompression Scenarios} 
        Consider a $k$-mer Dictionary obtained  from a set of genomic sequences,  for a given $k$.   Each $k$-mer in the Dictionary has an associated frequency, indicating  how many times it appears in the set of sequences.  The Dictionary can be represented as a pair $(D_k, F_k)$, where $D_k$ is the set of $k$-mers  and $F_k$ is the set of associated frequencies. It is to be noted that the order in which $k$-mers and frequencies appear in the respective sets provides a bijection. The goal is to compress $D_k$ and $F_k$. For each of those two tasks, which can be handled separately, we consider two scenarios for compression/decompression that apply to both of the Application Areas mentioned in the Introduction. As for Application Area 2, we need to introduce an additional scenario, as we motivate in what follows.  Within each scenario, we consider different cases. We anticipate that, when a $k$-mer counter is needed for the extraction of the Dictionary, among the many possible choices \cite{swati2018}, we use  {\bf DSK} \cite{rizk2013dsk} as the base method, to be consistent with research in \cite{Rahman21, Rahman21dataset}. As a consequence of the {\bf{DSK}} output, we consider only canonical $k$-mers, according to the order relation defined by {\bf DSK}: $A<C<T<G$. We also account for the choice of $k$-mer counters other than {\bf DSK}, in particular the \emph{de facto} standard \textbf{KMC3} \cite{Kokot17}.
	
	\begin{itemize}
    		\item {\bf Succinct on Disk, Random Access in Main Memory}. For brevity, we refer to this scenario as \textbf{Succinct on Disk} and we use the acronym {\bf SD-RAM} in the captions of Figures and Tables.  $D_k$ and $F_k$ are compressed via succinct data structures that can (a) be stored on disk and then uploaded in Main Memory; (b) once in Main Memory, given a query $k$-mer, one can recover efficiently from the succinct data structures the number of its occurrences in the Dictionary; (c) once in Main Memory, one can recover $D_k$ from the succinct data structure representing it and then $F_k$ by querying the succinct data structure representing it. Among the many available compressed data structures for string collections (see the classic review \cite{Navarro07}), for storing $D_k$, we use the \emph{de facto} standard {\bf FM-index} \cite{FMInd}, as implemented in the {\bf SDSL-lite}  software \cite{gbmp2014sea,GogGithub}. With reference to a recent review on the related topic of representing sets of DNA $k$-mers  \cite{Chikhi21}, we do not consider relevant succinct data structure suggested there. Namely,  the one based on work by Conway and Bromage \cite{ConwayBromage11}, since there are pathological cases in which that data structure would require  $\Omega{(|D_k|k)}$ space, i.e., when   $D_k$ is very sparse with respect to the universe of $k$-mers. Moreover,  we have also excluded   the {\bf Bloom Filter Trie} \cite{Holley2016} since, to the best  of our knowledge, the corresponding software \cite{HolleyGithub16} is not able to handle arbitrary values of $k$. As for $F_k$, we use the succinct data structure {\bf BCSF} \cite{shibuya2022space}, implemented in \cite{locomgithub21}, since it is the best one for this task, to date.  
    		
    		\item {\bf Compressed on Disk, No Random Access in Main Memory}. For brevity, we refer to this scenario as \textbf{Compressed on Disk} and we use the acronym {\bf CD-NRAM} in the captions of Figures and Tables. In this case, $D_k$ and $F_k$ are simply compressed and stored on disk. To this end, and as far as $D_k$ is concerned,  we follow a recent study on high performance compression for FASTA/Q files \cite{FerraroPetrillo2021}, chhosing the ones able to deal very well with  that data format. They are a mix of standard textual compressors, as well as specialized ones: {\bf bzip2} \cite{Bzip2}, {\bf lz4} \cite{Lz4}, {\bf Zstandard} ({\bf zstd}, for short) \cite{Zstd}, {\bf MFCompress} ({\bf{MFC}}, for short) \cite{pinho2013mfcompress} and {\bf SPRING} \cite{spring}.  As far as $F_k$ is concerned, its compression can be seen as an instance of inverted index compression problem, which has received quite a bit of attention, resulting in the proposal of many compressors specialized to integer sequences compression, either sorted or not.  A recent benchmarking study  \cite{Pibiri20} summarizes very well the State of the Art and provides useful indications, derived via rigorous experimental standards, of which methods perform best, considering compression rate as well as compression/decompression speed. Based on that study, in addition to the standard text compression methods mentioned earlier, we use the following specialized integer compressors: {\bf BIC} \cite{Moffat2000} and {\bf Opt-PFOR} \cite{Yan09}. Indeed,  the former compresses quite well, being  slow in compression and decompression time, while the second is a representative of a class of methods that perform well in compression and offer fast decompression time. As for their implementations, we use the {\bf Inverted Index Benchmark} software  \cite{iibench20},  associated to the benchmarking study. In turn, that software makes extensive use of the {\bf JavaFastPFOR} library \cite{JavaFastPFOR}.  
  
	\end{itemize}        

	 \begin{itemize}
	 	
            \item {\bf The Base Case Scenario}. It applies only to Application Area 2, where the Dictionary can be recovered from the genomic sequences. Its aim to evaluate how convenient is to compress/decompress the Dictionary with respect to recovering it from the input genomic sequences, that are stored and compressed on disk. In this case, the genomic sequences are decompressed and the Dictionary is recovered via {\bf{DSK}}. We distinguish two sub-case. One in which Random Access in Main Memory is required. For brevity, we refer to this scenario as \textbf{B-Succinct on Disk} and we use the acronym \textbf{B-SD-RAM}. The second is when  Random Access in Main Memory is not required. For brevity, we refer to this scenario as \textbf{B-Compressed on Disk} and we use the acronym \textbf{B-CD-NRAM}.
	 
	 \end{itemize}
	
	\subsection{\bf Compression/Decompression Scenarios: Details}
	   
            \subsubsection{\bf The Succinct on Disk Scenario} We have the following two cases. 
	
	\begin{itemize}
		\item {\bf  Case  $D_k$ and $F_k$ are given Explicitly.}  We refer to this case simply as Explicit, since the context will indicate that we are considering the \textbf{Succinct on Disk Scenario}. We use {\bf DSK} to obtain $D_k$ and $F_k$. The input is $D_k$ and $F_k$ and the output is the compressed versions of each set, via succinct data structures. The compression ratio is given by the number of bytes of the output divided by the number of bytes of the input.
		
		\begin{itemize} 
			\item {\bf  Compression.}  Each set is given \emph{verbatim} to the software that builds the {\bf FM-index}  and the {\bf BCSF}, respectively. This latter data structure is further compressed with one of the standard compressors used in this study, e.g., {\bf zstd} to fix ideas.  We remark that a further compression of the file containing the {\bf FM-index} brings no significant space reduction. This step is denoted as \emph{pre-processing} and its time performance is given by summing the time taken by all of its sub-steps, including the use of {\bf DSK}. It is to be noted that we also consider the case in which $F_k$ is represented as an offset and a file of gaps $G_k$. For conciseness, we do not explicitly account for this case here and in the other Scenario, but it will be a part of our experiments. 
			
			\item {\bf  Decompression.} The compressed version of {\bf BCSF} is decompressed, e.g. via {\bf zstd}, and loaded in Main Memory, together with the {\bf  FM-index}. If the intent is simply to process membership queries efficiently, with possibly the retrieval of the corresponding frequency, no further action is required. If $D_k$ and $F_k$ are needed, then $D_k$ is extracted via the {\bf FM-index} and each $k$-mer  frequency is extracted from {\bf BCSF} to reconstruct $F_k$. We refer to those stages as \emph{post-processing}. We account only for the case in which it is needed to recover $D_k$ and $F_k$, and we do not consider the processing of sequences of membership queries. 
	    \end{itemize}

        \item {\bf Case $D_k$ is given Implicitly and $F_k$ is given Explicitly.} We refer to this case simply as Implicit, since the context will indicate that we are considering the \textbf{Succinct on Disk Scenario}. $D_k$ is implicitly provided via a Spectrum Preserving String Set $S$. Such a set of strings, introduced in  \cite{Brinda2021, Rahman21}, is a set of sequences such that (a) the set of $k$-mers contained in strings in $S$ is $D_k$; (b) each string in $S$ has length at least $k$. We use {\bf ESSCompress} \cite{Rahman21}  to compute $S$. We also use  {\bf DSK} to obtain $F_k$. The input is $S$ and $F_k$ and the output is the compressed versions of each part of the input, via succinct data structures. The compression ratio is given by the number of bytes of the output divided by the number of bytes of the input. 

        \begin{itemize}
	       \item  {\bf  Compression.}  The {\bf FM-index}  for $S$ is built and stored on disk. Likewise, {\bf BCSF} is built for $F_k$, then compressed via one of the standard methods mentioned earlier, e.g. via {\bf zstd} for fix ideas, and finally stored on disk. The time for this step is taken in analogy with what specified in the previous case. The time accounts also for the use of {\bf ESSCompress} and {\bf DSK}. 
	
	       \item {\bf Decompression.}  The compressed version of {\bf BCSF} is decompressed, e.g. via {\bf zstd}, and loaded in Main Memory, together with the {\bf FM-index}. Now, care must be exercised both for single queries and for the recovery of $D_k$ and $F_k$. Indeed, $S$ is composed of both canonical and non-canonical $k$-mers. Given a query $k$-mer, which we assume to be in canonical form (recall that $D_k$ stores only canonical $k$-mers),  we must search for its occurrence in $S$, via the {\bf FM-index}, both in its canonical and non-canonical form, while the frequency of the query $k$-mer can be readily recovered from {\bf BCSF}. In order to recover $D_k$ in full, we extract $S$ via the {\bf  FM-index}. Then, we scan $S$ with a window of size $k$. If the $k$-mer in the window is canonical, we store it in $D_k$. Else, we transform it into its canonical form and store it in $D_k$. Finally, for each $k$-mer in $D_k$, we query {\bf BCSF} for its frequency. The final result is $F_k$. We refer to those stages as \emph{post-processing}. As in the previous case, we account only for the time to recover $D_k$ and $F_k$. 
	
        \end{itemize}

    \end{itemize}

	\subsubsection{\bf The Compressed on Disk Scenario. }\label{sec:STC}
	Again, we have two cases, described next. 
	\begin{itemize}
	
\item 	{\bf  Case $D_k$ and $F_k$ are given Explicitly.} Again, we refer to this case simply as Explicit, since the context will indicate that we are considering the \textbf{Compressed on Disk Scenario}. We use {\bf DSK} to obtain $D_k$ and $F_k$. Then, we have the following three sub-cases. 

\begin{itemize}
	\item {\bf Verbatim}. The input is $D_k$ and $F_k$ and each is processed \emph{as is}. The output is the compressed version of each set and the corresponding compression factor is given by the ratio of output and input sizes in bytes. For later reference,  we denote such a sub-case as {\bf DP0}.
	
	\begin{itemize}
	
	\item {\bf Compression.} $D_k$ is compressed via one of the standard and specialized textual compression methods used in this research, e.g., via {\bf MFC} to fix ideas. Likewise, $F_k$ is compressed via one of the standard textual compressors used in this study, e.g., via {\bf zstd} to fix ideas, in addition to the ones specialized to integers. Both compressed files are stored on disk. We refer to this stage as  \emph{pre-processing} and its total time is given by the sum of the times of each of its steps, including the use of {\bf DSK}. 
	
	\item {\bf  Decompression. } The compressed files are loaded from disk and decompressed, via the methods used for compression, e.g. via {\bf MFC} to recover $D_k$, and via {\bf zstd} to recover $F_k$. We refer to this stage as  \emph{post-processing} and its total time is given by the sum of the times of each of its steps.
	
	\end{itemize}
	
	\item{ {\bf  Case $D_k$ is sorted prior to processing. }} For later reference,  we denote such a sub-case as {\bf DP1}.  Rearranging the content of a multi-sequence textual file may yield better compression \cite{Bzip2}. We explore this idea here for $D_k$, using the most natural of the rearrangements, i.e., sorted.  
	In particular, $D_k$   is lexicographically sorted via the Linux Ubuntu sort routine \cite{sortUBUNTU}, from now on referred to simply as  {\bf sort}.  In order to preserve the implicit  bijection  between $D_k$ and $F_k$, this latter is permuted according to the sorting permutation used for $D_k$. Compression and Decompression are as in {\bf DP0}, except that for the \emph{pre-processing} time, we also  charge for  the sorting of  $D_k$ and rearrangement of $F_k$.

	\item {{\bf   Case $F_k$ is sorted prior to processing. }} For later reference,  we denote such a sub-case as {\bf DP2}.  It is well known that sorted sequences of integers can be compressed more effectively than their  unsorted versions, e.g., \cite{iibench20}.  We explore this idea here for $F_k$. Indeed, we use {\bf sort} to obtain a new file. Then $D_k$ is permuted accordingly to preserve the implicit bijection between the two sets. The remaining details regarding {\bf DP2} are as for {\bf DP0}. 
	
\end{itemize}

\item 	{\bf  Case  $D_k$ is given Implicitly and $F_k$ is given Explicitly.} Again, we refer to this case simply as Implicit, since the context will indicate that we are considering the \textbf{Compressed on Disk Scenario}. For later reference, we denote such a case as {\bf DP3}. This is the same case we have already considered for the {\bf Succinct on Disk Scenario}, except that here we are not allowing random access to the files. This complicates the process of maintaining the implicit bijection between $D_k$ and $F_k$, since the former now is given implicitly via $S$. The input is $S$ and $F_k$ and the output is the compressed versions of each of them. The compression ratio is given by the number of bytes of the output divided by the number of bytes of the input. We proceed as follows.

\begin{itemize}
	
	\item  {\bf Compression}. We scan $S$ and, for each $k$-mer in $S$, we convert it into its canonical representation, and associate to it its position in $S$. Let $D'_k$ be the set of such pairs. We sort those pairs lexicographically according to the first field, via {\bf sort}. We sort lexicographically $D_k$, again via {\bf sort} and permute $F_k$  accordingly to preserve the bijection. In order to assign a frequency to each $k$-mer in $S$, we take the lexicographically sorted $D'_k$ and $D_k$ and scan them, in analogy with the well known {\bf merge} routine in \textbf{Mergesort}. For each $k$-mer match, we have the pair $(i,j)$, where $i$ is the position of the matched $k$-mer  in $S$ and $j$ is its frequency. We sort those pairs according to the first component, to obtain a new version of $F_k$ where the implicit bijection is now between positions in $S$ and frequencies.  
	Then, $S$ and $F_k$ are compressed via one of the appropriate methods, e.g., {\bf MFC} for $S$ and {\bf zstd} for $F_k$ to fix ideas, and stored on disk.   This is the \emph{pre-processing} step. Its time is given by the sum of the times of each of its sub-steps, including the use of  {\bf ESSCompress} and {\bf DSK}.

	\item {\bf Decompression}. We decompress both files on disk, via the methods used in the \emph{pre-processing} step, e.g. {\bf MFC} and {\bf zstd}, to obtain $S$ and $F_k$. We then scan $S$ in order to obtain $D_k$, as already outlined in the \textbf{Succinct on Disk Scenario}. The bijection between $D_k$ and $F_k$ is ensured by the way we have arranged $F_k$	in the \emph{pre-processing} step. This is the \emph{post-processing} step and its time is computed as in the previous analogous cases.  
\end{itemize}

	\end{itemize}

\subsubsection{\bf The Base Case Scenario}
	 
    \begin{itemize}
	
		\item {\bf B-Succinct on Disk}. The genomic sequences are represented on disk succinctly via the {\bf FM-index}. This would be the  \emph{pre-processing } step. Once that the Dictionary is needed, the {\bf FM-index} is uploaded in Main Memory. Then, the genomic sequences are recovered from it and we use {\bf DSK}  to obtain $D_k$ and $F_k$. This would be the \emph{post-processing} step. 
		
		\item {\bf B-Compressed on Disk}. As in the previous case, except that now we use one of the generic or specialized textual compressors, e.g. {\bf MFC} for the sake of  discussion.
		
    \end{itemize}
	
    We also introduce measures whose intent is to indicate whether it is more convenient to keep the Dictionary compressed according to a chosen case in our scenarios, with respect to keeping the compressed genomic sequences on disk, decompress them, and reconstruct the Dictionary from scratch.  They are described in detail in Section {\bf Comparison with the Base Case Scenario} in the Supplementary File.

    \subsection{\bf Experimental Set-up} \label{Experimental_Setup}
        %It is reported in the Section with the same name 

        The hardware is described in the Section with the same name in the Supplementary File. We experiment with a mix of datasets, in FASTA format, of different sizes and that have been used in previous related studies. Namely, \emph{Staphylococcus Aureus} \cite{Staphylococcus}, representing small datasets, i.e, in the MBs; \emph{Human Chromosome 14} \cite{ Rahman21dataset}, representing medium datasets, i.e., in the hundreds of MBs; \emph{Assembled Plants Genomes} \cite{Zielezinski2019}, representing large datasets, i.e., in the few GBs. As for values of $k$,  following the Literature, e.g., \cite{Brinda2021, Rahman21, rizk2013dsk}, 	we experiment with $k=4, 8, 16, 32, 48, 64$. However,  on the largest of the selected datasets, {\bf BCSF}, {\bf ESSCompress} and the  {\bf FM-index} run out of memory  or they are too slow in processing (at least 4 days) for $k=48,64$. Therefore, the corresponding experiments have been excluded from this research. 

	\section{Results}

        For all the Scenarios considered in this research, we have performed experiments 
        in agreement with the datasets, values of $k$ and software described in the previous section. In presenting the results of those experiments, we consider both application areas mentioned in the Introduction. It is useful to recall that, in Application Area 1, we are given the Dictionary, but it cannot be rebuilt from the corresponding genomic sequences. Therefore, in this case, we can only compress and then decompress the Dictionary. As for Application Area 2, it is possible to rebuild the Dictionary from the genomic sequences. Therefore, the {\bf Base Case Scenario} described in the previous section must be considered.

        For both of the mentioned application areas, as motivated in the Introduction, we consider the circumstance in which compression is more important than post-processing time and its complement,  referring to the first as favouring compression and to the second as favouring post-processing time. However, the following remarks apply to both Scenarios. 
        
        First, it is to be noted that, for each of the scenarios and cases considered in this research, the specific performance, both in terms of compression and post-processing time is determined by the compressors that are used, e.g.,  the post-processing time of  \textbf{DP0} with the configuration $D_k$ compressed with \textbf{bzip2} and $F_k$ compressed with \textbf{zstd} is may be different than the configuration with other choices. Therefore, there is a variety of setting to be considered. Among all those settings, for each Scenario, Case and Sub-Case, we report only the best in terms of compression and the best in terms of post-processing time. That is,  the configuration that outputs the least bytes on disk and the one that takes the least time in terms of post-processing.    Those two points are the \vir{extreme points} of the Pareto optimal configurations  ruling the trade-off  compression ability vs post-processing time. Therefore, they are a natural choice for the presentation of our results. However, in the {\bf Discussion} section, we present Pareto optimal results to outline the set of choices one has available.

        Second, being the input of the various cases different, e.g., {\bf DP0,  DP3}, the classic measure of compression ratio cannot be used to select the best case, for each Scenario. Therefore, this part is provided for completeness in the Supplementary File (sections 
       {\bf Succinct on Disk Scenario - Ability to Compress the Input} and {\bf Compressed on Disk Scenario - Ability to Compress the Input}).

    \subsection{\bf Succinct on Disk Scenario}
     
        \subsubsection{\bf Application Area 1}
        Here we are interested in establishing which of the Explicit and Implicit Case to pick, with respect to the dataset size and value of $k$. We consider the configurations yielding the minimum  number of  bytes to be stored on disk. We then compute the ratio  Explicit/Implicit of those two quantities. Those latter are reported in Table 1 in the Supplementary File for the three datasets used in this research. Since the \emph{Human Chromosome 14} dataset offers a good summary of our results and for conciseness, we report the corresponding data in Table \ref{tab:HC14_Tab1sup_Tab2sup} (panel {\bf Area 1}), and we anticipate that we will adhere to such a procedure for all of our experiments. In this Scenario, the possible configurations in terms of compressor choices are limited only to their use to compress {\bf BCSF}. As a consequence of this limited choice,  
        Table 1 in the Supplementary File and Table  \ref{tab:HC14_Tab1sup_Tab2sup} (panel {\bf Area 1})
        also account for the circumstance in which post-processing time is more important than compression. 
        
        As evident from those Tables, the Explicit Case is only marginally of interest and use for practice. Therefore, our experiments suggest that when the Dictionary cannot be reconstructed from the genomic sequences and must be stored in the succinct data structures granting random access, the Implicit Case, i.e., Spectrum Preserving String Set approaches to $k$-mer Dictionary compression, must be used.

        \subsubsection{\bf Application Area 2}       
        We have also performed experiments in order to assess how convenient is the {\bf{Succinct on Disk Scenario}} with respect to the corresponding {\bf{B-Succinct on Disk Scenario}}. 
        In view of the disappointing performance of the Explicit Case and for brevity, we concentrate only on the Implicit Case. The compression and post-processing time ratios are as defined in the section \textbf{Comparison with the Base Case Scenario} (Succinct vs. B-Succinct on Disk Scenarios) in the Supplementary File. Those results are reported in Table 2 in the Supplementary File.  Table \ref{tab:HC14_Tab1sup_Tab2sup} (panel {\bf Area 2}) reports the corresponding results for the \emph{Human Chromosome 14} dataset and in logic terms. That is, Y indicates that the Implicit Case is more convenient than the {\bf{Base Case} Scenario} and N indicates the complement. 
     
     From the mentioned Tables, it turns out that the Implicit Case is more convenient than the  {\bf Base Case Scenario}, either for small or large values of $k$,  as the size of the genomic dataset grows. The reason for this is rather subtle, as we now explain. Letting $S$ be the sequence produced by {\bf ESSCompress},  what seems to influence the convenience or not of the Implicit Case is the ratio between the lengths of $S$ and that of the genomic dataset $G$. Such a ratio, as a function of $k$, seems to be concave, with values larger than one that split its domain into two parts (not necessarily equal). This is exemplified in Figure S7 in the Supplementary File. Therefore, depending on $k$ and the ratio $|S|/|G|$, the Implicit Case grants random access, good compression, and fast post-processing time with respect to the {\bf{Base Case Scenario}}. It is worth pointed out that Spectrum Preserving String Sets have not been considered in regard to succinct data structures and random access in main memory. Therefore, it is even more relevant that our experiments bring to light a very clean picture of when to use such an approach.

        %TAB.1 HC14: Succinct on Disk Vs. Base Case
        \begin{table*}[!ht]
            \centering
            \resizebox{1.00\textwidth}{!}{%        
                \begin{tabular}{|c|ccc|cc|}
                \hline
                \multicolumn{1}{|l|}{} & \multicolumn{3}{c|}{\textbf{Area 1}} & \multicolumn{2}{c|}{\textbf{Area 2}} 
                \\ 
                \hline
                 
                \textbf{k} & \multicolumn{1}{c|}{\textbf{B}} & \multicolumn{1}{c|}{\textbf{C}} & \textbf{T} & \multicolumn{1}{c|}{\textbf{C}} & \textbf{T} 
                \\ 
                \hline
             
                \textbf{4} & \multicolumn{1}{c|}{ FM(Dk)-zstd(BCSF(Fk)) / FM(S)-zstd(BCSF(Fk))} & \multicolumn{1}{c|}{1,019E+00} & 1,887E-01 & \multicolumn{1}{c|}{Y} & Y 
                \\ 
                \hline
            
                \textbf{8} & \multicolumn{1}{c|}{FM(Dk)-zstd(BCSF(Fk))  /    FM(S)-zstd(BCSF(Fk))} & \multicolumn{1}{c|}{1,471E+00} & 1,097E+00 & \multicolumn{1}{c|}{Y} & Y 
                \\ 
                \hline
                
                \textbf{16} & \multicolumn{1}{c|}{FM(Dk)-zstd(BCSF(Fk))  /    FM(S)-zstd(BCSF(Fk))} & \multicolumn{1}{c|}{6,015E+00} & 5,935E+00 & \multicolumn{1}{c|}{N} & N 
                \\ 
                \hline 
                
                \textbf{32} & \multicolumn{1}{c|}{FM(Dk)-zstd(BCSF(Fk))  /    FM(S)-zstd(BCSF(Fk))} & \multicolumn{1}{c|}{2,784E+01} & 2,200E+01 & \multicolumn{1}{c|}{N} & Y 
                \\ 
                \hline
                 
                \textbf{48} & \multicolumn{1}{c|}{FM(Dk)-zstd(BCSF(Fk))  /    FM(S)-zstd(BCSF(Fk))} & \multicolumn{1}{c|}{4,214E+01} & 2,961E+01 & \multicolumn{1}{c|}{Y} & Y 
                \\ 
                \hline
                  
                \textbf{64} & \multicolumn{1}{c|}{FM(Dk)-zstd(BCSF(Fk))  /    FM(S)-zstd(BCSF(Fk))} & \multicolumn{1}{c|}{5,497E+01} & 3,683E+01 & \multicolumn{1}{c|}{Y} & Y 
                \\ 
                \hline
                
                \end{tabular}
            }
            \vspace{5pt}
            \caption{\textbf{\textit{Human Chromosome 14} Dataset - SD-RAM Scenario.} The Table is divided into two panels, corresponding to  \textbf{Application Area 1}, and \textbf{Application Area 2}, respectively. For the first panel, the three columns indicate the following. Column \textbf{B}: the ratio to be taken between the \textbf{Succinct on Disk Scenario} Explicit Case and the \textbf{Succinct on Disk Scenario} Implicit Case, both taken in the best compression configuration with respect to the \textbf{BCSF} succinct data structure. In Column \textbf{C} that ratio accounts for compression and, in Column \textbf{T}, it accounts for post-processing time. As for the second panel, it reports the comparison of the {\bf Succinct vs. B-Succinct on Disk Scenarios}, in logic terms, and as detailed in the Main text.}
            \label{tab:HC14_Tab1sup_Tab2sup}
            \end{table*}

    \subsection{\bf Compressed on Disk Scenario}

        \subsubsection{\bf Application Area 1}
        \begin{itemize}
            \item {{\bf Favouring  Compression}. Here we are interested in establishing which case, and with which choice of compressors,  provides the smallest output, in terms of bytes to be stored on disk. The relevant results, extracted from our experiments, are reported in Table 3 in the Supplementary File, for all of the datasets used in this research, while for the \emph{Human Chromosome 14} dataset the results are reported in Table \ref{tab:HC14_CD-NRAM_Compr_PostProc} (panel {\bf Compression}). Indeed, for each organism and each value of $k$ considered in this research, we report the best performing case in terms of output size in bytes, together with the compressors used to achieve it.  When one of the  Explicit Cases wins, we also report the best (minimum output size in bytes) compression obtained in the Implicit Case, with an indication of the loss in terms of output size (the numeric value in brackets is the ratio of the Implicit Case output in bytes with the same of the winning Explicit Case). When the Implicit Case wins, we proceed symmetrically.
            
            As somewhat expected, there is no absolute winner across datasets and values of $k$.  The Implicit Case, i.e., {\bf DP3}, 
            seems to be the one of choice for the small and medium datasets, depending on the value of $k$. Being that case based on Spectrum Preserving String Sets techniques, that are considered cutting edge,   we place them in the proper light with respect to the many choices available to compress Dictionaries, highlighting also their limitations in terms of computational resources they need. Indeed, the largest dataset, with $k=48,64$,  could not be processed on the hardware used here, as specified in the previous section ({\bf Experimental Set-up}). 
            Another important contribution to the advancement of the State of the Art is given by the finding that a few important common aspects emerge in terms of data compressor choices for the compression of the frequencies, a largely ignored aspect so far. Indeed, {\bf{bzip2}} is the method of choice to compress the frequencies, with  {\bf{zstd}} also worth of consideration. This is surprising, since two cutting edge  specialized integer compressors seem to be of no use in the context of genomic Dictionary Compression. No consistent indication emerges  with respect to the advantage of using Gap Encoding. As for textual compression, {\bf{MFC}} is the method of choice, as somewhat expected, with {\bf{zstd}} also worth of consideration. }

            \item {{\bf Favouring  Post-Processing Time}. Here we are interested in establishing which case and with which choice of compressors,  provides the fastest post-processing time. The relevant results, extracted from our experiments, are reported in Table 4 in the Supplementary File, for all of the datasets used in this research. Such a table is obtained in analogy with Table 3 in that File, except that the role of compression and post-processing are exchanged. We also report in percentage, for each value of $k$ and each organism,  how much the best case in terms of post-processing loses in terms of output size in bytes with respect to the winner in compression reported in Table 3 in the  Supplementary File. This is the first numeric value. As for the second, it indicates the percentage loss in post-processing time,  if we use the best compression Case reported in Table 3 in the Supplementary File. Those two values provide some estimate of the compression/post-processing time trade-off. For the \emph{Human Chromosome 14} dataset,  the results are reported in Table \ref{tab:HC14_CD-NRAM_Compr_PostProc} ({\bf Post-Processing} panel).

            The first relevant result is that the Implicit Case is never a winner when post-processing time is important. This is due to the fact that, with respect to the Explicit Cases, its post-processing step needs a scanning of the sequences $S$, in addition to decompression. Indeed, in terms of post-processing time, such a scanning takes on average over $60\%$ of the overall post-processing time. Moreover, independently of the winning Explicit Case, {\bf zstd} seems to be the compression method of choice both for $D_k$ and $F_k$. Those results further contribute to place Spectrum Preserving String Sets methods in the proper context in the Literature, together with the choice of specialized integer compression methods. Finally, it is evident that, if one favours the best solution in terms of post-processing time, the loss in terms of compression can be of an order of magnitudo. A symmetric situation holds for favouring compression with respect to post-processing time. }

            %TAB.2 HC14: CD-NRAM Favour Compression and Post-Processing Time
            \begin{table*}[!ht]
                \centering
                \resizebox{1.00\textwidth}{!}{%
                    \begin{tabular}{|c|c|c|}
                    \hline
                    \textbf{k} & \textbf{Compression} & \textbf{Post-Processing} \\ \hline
                    \textbf{4} & DP2(zstd(Dk), zstd(Gk)) - DP3(zstd(S), zstd(Fk)) {[}1.01E+00{]} & DP1(zstd(Dk)), Opt-PFOR(Fk)) {[}1.88E+00{]}, {[}2.50E+00{]} \\ \hline
                    \textbf{8} & DP2(MFC(Dk), bzip2(Gk)) - DP3(bzip2(S), bzip2(Fk)) {[}1.25E+00{]} & DP1(zstd(Dk)), lz4(Fk)) {[}1.98E+00{]}, {[}2.87E+02{]} \\ \hline
                    \textbf{16} & DP3(zstd(S), bzip2(Fk)) - DP2(zstd(Dk), bzip2(Gk)) {[}2.05E+00{]} & DP2(zstd(Dk)), zstd(Fk)) {[}2.05E+00{]}, {[}2.28E+01{]} \\ \hline
                    \textbf{32} & DP3(MFC(S), bzip2(Fk)) - DP2(zstd(Dk), bzip2(Gk)) {[}2.05E+01\} & DP1(zstd(Dk)), zstd(Fk)) {[}2.05E+01{]}, {[}2.04E+01{]} \\ \hline
                    \textbf{48} & DP3(MFC(S), bzip2(Fk)) - DP0(MFC(Dk), bzip2(Fk)) {[}3.16E+01{]} & DP0(zstd(Dk), zstd(Fk)) {[}4.01E+01{]}, {[}1.08E+01{]} \\ \hline
                    \textbf{64} & DP3(MFC(S), bzip2(Fk)) - DP0(MFC(Dk), bzip2(Fk)) {[}3.59E+01{]} & DP2(zstd(Dk), zstd(Fk)) {[}5.77E+01{]}, {[}1.06E+01{]} \\ \hline
                    \end{tabular}
                }
                \vspace{5pt}
                \caption{\textbf{\emph{Human Chromosome 14} Dataset - CD-NRAM Scenario: Favour Compression and Post-Processing Time.} The table is divided into two panels: \textbf{Compression}, and \textbf{Post-Processing}, respectively. For each panel, the entries are as specified in the Main text. }
                \label{tab:HC14_CD-NRAM_Compr_PostProc}
            \end{table*}

            %TAB.3 HC14: Synopsis of the CD-NRAM Scenarios Vs. The Base Case Scenario: with MFC (favour compression), and with zstd (favour post-processing time)    
            \begin{table*}[!ht]
                \centering
                \resizebox{.40\textwidth}{!}{%
                    \begin{tabular}{|c|clclcl|clclcl|}
                    \hline
                     & \multicolumn{6}{c|}{\textbf{Compression}} & \multicolumn{6}{c|}{\textbf{Post-Processing}} \\ \hline
                    \textbf{k} & \multicolumn{2}{c|}{\textbf{C}} & \multicolumn{2}{c|}{\textbf{T}} & \multicolumn{2}{c|}{\textbf{B}} & \multicolumn{2}{c|}{\textbf{C}} & \multicolumn{2}{c|}{\textbf{T}} & \multicolumn{2}{c|}{\textbf{B}} \\ \hline
                    \textbf{4} & \multicolumn{2}{c|}{Y} & \multicolumn{2}{c|}{Y} & \multicolumn{2}{c|}{DP2} & \multicolumn{2}{c|}{Y} & \multicolumn{2}{c|}{Y} & \multicolumn{2}{c|}{DP1} \\ \hline
                    \textbf{8} & \multicolumn{2}{c|}{Y} & \multicolumn{2}{c|}{Y} & \multicolumn{2}{c|}{DP2} & \multicolumn{2}{c|}{Y} & \multicolumn{2}{c|}{Y} & \multicolumn{2}{c|}{DP1} \\ \hline
                    \textbf{16} & \multicolumn{2}{c|}{N} & \multicolumn{2}{c|}{Y} & \multicolumn{2}{c|}{DP3} & \multicolumn{2}{c|}{N} & \multicolumn{2}{c|}{Y} & \multicolumn{2}{c|}{DP2} \\ \hline
                    \textbf{32} & \multicolumn{2}{c|}{N} & \multicolumn{2}{c|}{Y} & \multicolumn{2}{c|}{DP3} & \multicolumn{2}{c|}{N} & \multicolumn{2}{c|}{Y} & \multicolumn{2}{c|}{DP1} \\ \hline
                    \textbf{48} & \multicolumn{2}{c|}{N} & \multicolumn{2}{c|}{Y} & \multicolumn{2}{c|}{DP3} & \multicolumn{2}{c|}{N} & \multicolumn{2}{c|}{Y} & \multicolumn{2}{c|}{DP0} \\ \hline
                    \textbf{64} & \multicolumn{2}{c|}{N} & \multicolumn{2}{c|}{Y} & \multicolumn{2}{c|}{DP3} & \multicolumn{2}{c|}{N} & \multicolumn{2}{c|}{Y} & \multicolumn{2}{c|}{DP2} \\ \hline
                    \end{tabular}
                }
                \vspace{5pt}
                \caption{\textbf{\textit{Human Chromosome 14} Dataset -  CD-NRAM Scenarios Vs. The Base Case Scenario - Logic Indication:  with MFC (favour compression), and with zstd (favour post-processing time).}  The table is divided into two panels: Compression and Post-Processing. For the first panel, the three columns indicate the following. Column \textbf{B} indicates the best \textbf{Compressed on Disk Scenario} for the Case favouring compression. Columns \textbf{C} and \textbf{T} indicate, respectively, the compression ratio and the post-processing time ratio between the \textbf{Compressed on Disk Scenario}, indicated in the column \textbf{B}, and the corresponding \textbf{Base Case Scenario} with \textbf{MFC}, as specified in the Main text. As for the second panel, its entries are analogous to the ones of the first one, except that here we favor post-processing time and consider the \textbf{Base Case Scenario} with the \textbf{zstd} compressor. }
                \label{tab:HC14_CD-NRAM_Vs_BaseCase_ComprMFC_PostProcZSTD}
            \end{table*}            
            
        \end{itemize}
         
       \subsubsection{\bf Application Area 2} For this scenario, we have also performed experiments in order to assess how convenient it is with respect to the {\bf{B-Compressed on Disk Scenario}}, both when favouring compression and post-processing time. 
            
            \begin{itemize}
            
                \item {\bf {Favouring Compression}}. {\bf{MFC}} is the best, in terms of compression of the given genomic sequences, for all of the datasets considered in this research (data not shown and available upon request). Therefore, it is the most challenging for a comparison of the {\bf{Compressed on Disk Scenario}} with respect to the {\bf{B-Compressed on Disk Scenario}}, when compression performance is important. For such a comparison, for each organism, we select the best {\bf{Compressed on Disk}} Case in terms of compression (see again Table 3 in the Supplementary File). We then take the ratio of the number of bytes output by that method and the number of bytes output by the {\bf{Base Case Scenario}} with {\bf MFC}, in agreement with the criteria outlined in the section \textbf{Comparison with the Base Case Scenario} (Compressed vs. B-Compressed on Disk Scenarios) in the Supplementary File. For each value of $k$ considered in this research, those results are reported in Table 5 in the Supplementary File, in logic terms, i.e., Y indicates that the given case is more convenient than the {\bf{B-Compressed on Disk Scenario}} with {\bf{MFC}}, and with numeric values. 
                Table \ref{tab:HC14_CD-NRAM_Vs_BaseCase_ComprMFC_PostProcZSTD}, panel {\bf Compression}, reports the same results in logic form for the \emph{Human Chromosome 14} dataset. As evident from those Tables and with reference to the logic indication, the cases of the {\bf{Compressed on Disk Scenario}} are convenient with respect to reconstructing the Dictionary from the compressed genomic sequences only for small values of $k$ (see Column {\bf C} in the mentioned Tables). However, quite remarkably, they are convenient in terms of post-processing time across datasets and values of $k$ we have used for this research (see Column {\bf T} in the mentioned Tables). The numeric values in Table 5 in the Supplementary File provide a quantification. That is, when compression is important, they lose in compression to the approach of reconstructing the Dictionary from the genomic sequences but gain in post-processing. With reference to this finding, it is to be noted that, for the \vir{intermediate} dataset, i.e., \emph{Human Chromosome 14}, little is lost in compression, and little is gained in post-processing time. On the large dataset \emph{AP}, one can lose a lot in compression, with possibly significant gains in post-processing time.

                For completeness, Tables 6-9 in the Supplementary File provide results analogous to the ones presented here for {\bf MFC} in regard to the other compressors used in this research. They show the vast variety of choices one can make regarding the compression of the Dictionary with respect to rebuilding it from the compressed genomic sequences.

                \item {\bf {Favouring Post-Processing Time}}. {\bf{zstd}} is the best, in terms of post-processing time, for all of the datasets considered in this research (data not shown and  available upon request). Therefore, it is the most challenging for the {\bf{Compressed on Disk Scenario}}, when post-processing time performance is important. 
                We proceed in analogy with the case in which we favour compression, discussed earlier. The relevant results are in Table 10 in the Supplementary File and in Table \ref{tab:HC14_CD-NRAM_Vs_BaseCase_ComprMFC_PostProcZSTD}, panel {\bf Post-Processing}. From those tables, it is evident that, again remarkably, the {\bf Compressed on Disk Scenario} are convenient with respect to reconstructing the Dictionary from the compressed genomic sequences, when post-processing time is important.  In terms of compression,  once that post-processing time is again privileged, they are better only for small values of $k$.
                
                For completeness, Tables 11-14 in the Supplementary File provide results analogous to the ones presented here for {\bf zstd} in regard to the other compressors used in this research. They show the vast variety of choices one can make regarding the compression of the Dictionary with respect to rebuilding it from the compressed genomic sequences when the post-processing time is important. 
                
                Finally, since we use {\bf DSK} to recover the Dictionary from the genomic sequences in the {\bf B-Compressed on Disk Scenario}, which is not the fastest $k$-mer counter available, it is natural to ask how robust are our results regarding post-processing time with respect to the choice of a $k$-mer counting program. In order to shed light on this point, we have conducted additional experiments, reported in section {\bf On the Choice on the $k$-mer Counter for Post-Processing Speed} in the Supplementary File, with the use of the  \emph{de facto} standard {\bf KMC3} \cite{Kokot17}. Those experiments show that our findings are robust with respect to the choice of the $k$-mer counting program. 
            \end{itemize}

	\section{Discussion}

        One common shortcoming that emerges from our results is that, in order to handle large datasets, high performance computing seems to be required (see the \emph{AP} dataset in Table 1 in the Supplementary File). For instance, even if one may use conventional servers with large random access memory, the time to compute $S$ via {\bf ESSCompress} becomes prohibitive for large datasets.  We are not aware of any effort to cast the compression of $k$-mer Dictionaries on high performance computing settings. In view of the comments already given in the previous section regarding the \textbf{Succinct on Disk Scenario}, we limit the remaining part of this section only to the {\bf Compressed on Disk Scenario}.
        
\begin{figure*}[!ht]
            \centering
          \includegraphics[width=\linewidth]{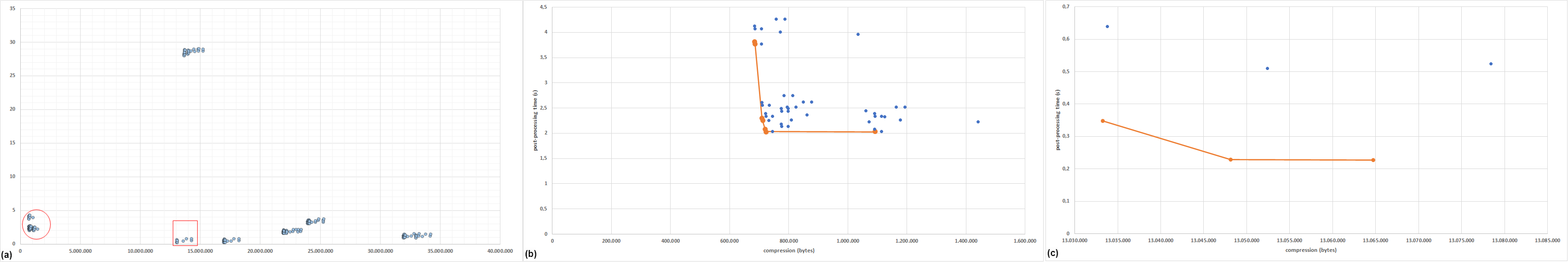}
            \caption{{\bf CD-NRAM Scenario: Optimal Pareto Points of the \textit{Staphylococcus Aureus} Dataset}. In each panel, the configurations, in terms of compression choices, of the various Cases are represented as points. The abscissa provides the number of output bytes stored on disk and the ordinate the post-processing time, expressed in seconds. Panel (a) represents the points corresponding to the 200 possible configurations that the tested methods can give raise to. Panel (b) provides the Pareto optimal frontier (red dots connected by line) regarding the points within the circle in Panel (a). Panel (c) provides the Pareto optimal frontier (red dots connected by line) regarding the points within the square in Panel (a).}
            \label{fig:Pareto_staph}
        \end{figure*}

\begin{figure*}[!ht]
            \centering
          \includegraphics[width=\linewidth]{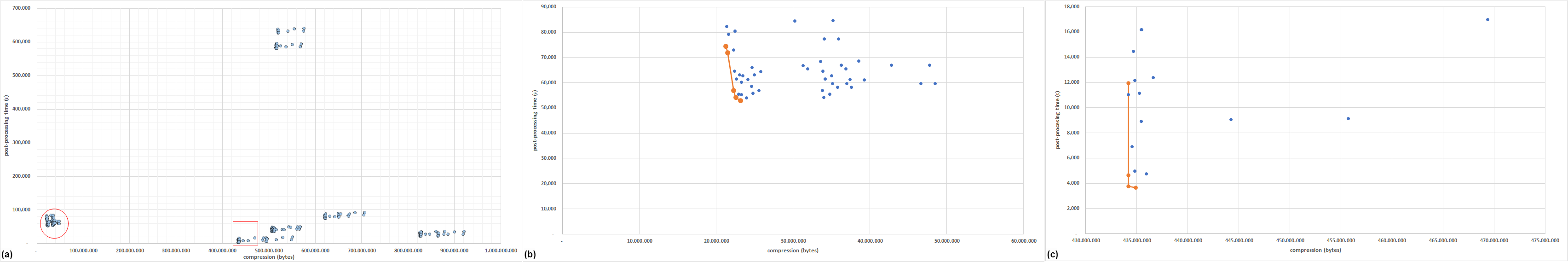}
            \caption{{\bf CD-NRAM Scenario: Optimal Pareto Points of the \textit{Human Chromosome 14} Dataset}. The Figure legend is as in Figure \ref{fig:Pareto_staph}. }
            \label{fig:Pareto_hc14}
        \end{figure*}

\begin{figure*}[!ht]
            \centering
          \includegraphics[width=\linewidth]{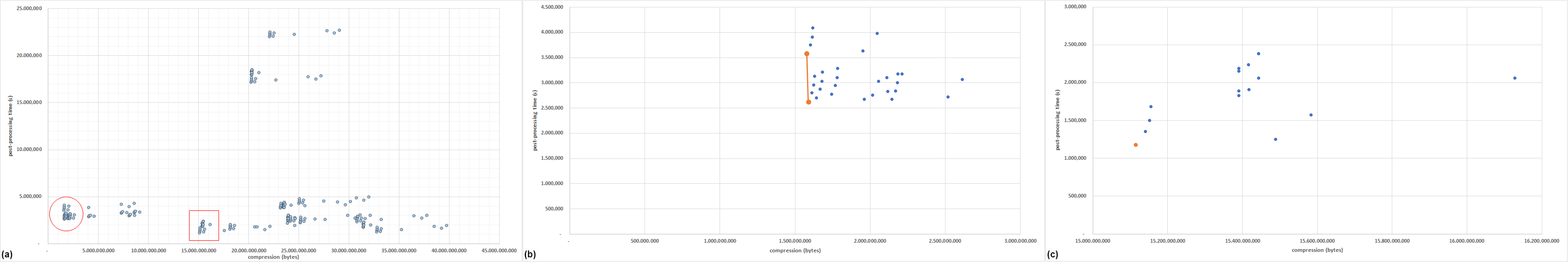}
            \caption{{\bf CD-NRAM Scenario: Optimal Pareto Points of the \textit{Assembled Plants} Dataset}. The Figure legend is as in Figure \ref{fig:Pareto_staph}. }
            \label{fig:Pareto_assp}
        \end{figure*}

        To this end, our experiments clearly show that, for Application Area 1, i.e., when the Dictionary cannot be rebuilt from the genomic sequences, in order to obtain a good solution balancing disk space occupancy and post-processing time, one needs to experiment with the full set of options that our software provides and pick the most suitable one for the specific dataset. This brings to light the usefulness of the software that we provide here. Although, admittedly, such a search is time consuming, it may be worthwhile when a good use of resources is critical, as we now exemplify. Consider Application Area 1 and fix $k=32$, Figures \ref{fig:Pareto_staph}-\ref{fig:Pareto_assp} report, for each dataset used in this research, the 200 possible configurations that the tested methods can give raise to. Each of them is a point, where number of output bytes is the abscissa and post-processing time the ordinate. In each of panels (b) and (c), we provide the magnified version of the same graph in (a) with the Pareto optimal points. That is, no other point is better than those ones both in terms of output bytes and post-processing time. The corresponding configurations are reported in Tables 18-20 in the Supplementary File. As evident from those Tables, there is a spectrum of choices that one can make to obtain suitable compression/post-processing time trade-off.    
        
        On the other hand, when the use of resources is not so critical, one may be willing to settle for \vir{a once and for all compromise solution}, i.e., a specific Case and compressor choices to be used on any dataset and value of $k$. Our experiments provide useful indication in this respect. Indeed, when favouring compression, the Case of choice is {\bf DP3} with {\bf MFC} to compress $S$ and {\bf bzip2} to compress $F_k$. An estimate of the potential losses of such a choice with respect to the use of the best Case in terms of compression is given in Table 19 in the Supplementary File. Likewise, when favouring post-processing time, the Case of choice is {\bf DP0}, with {\bf zstd} as a compressor both for $D_k$ and $F_k$. An estimate of the potential losses of such a choice with respect to the best one in terms of post-processing time is given in Table 20 in the Supplementary File.
	
	    As for Application Area 2, i.e., when the Dictionary can be rebuilt from the compressed genomic sequences, it is novel and somewhat surprising that the Cases we have considered turn out to provide an advantage with respect to rebuilding the Dictionary only in terms of post-processing time. Moreover, our extensive set of experiments,  reported in Tables 5-14 in the Supplementary File, provide a very consistent picture in this respect. The reason for such a finding seems to be in the fact that a compressed Dictionary needs not use $k$-mer counting software to extract $D_k$ and $F_k$.

        %% Figure Pareto_staph
    
        %% Figure Pareto_hc14
    
        %% Figure Pareto_assp

	\section{Conclusions}
	We have provided an extensive evaluation of methods to store in compressed form genomic $k$-mer Dictionaries. Our first main contribution is place this important topic on solid ground. Indeed, many our findings are either novel or substantial contributions to the pros/cons of the vast variety of choices one has available. Our second main contribution is a platform that collects State of the Art software and that can be readily used  to pick the most suitable configuration  by people interested in storing in compressed form their Dictionaries. In terms of future directions, apart from the need to design more efficient compressor methods, it emerges the acute need to have a system that automatically produces the Pareto optimal solutions in a reasonable amount, so that a user can identify the most useful configuration for the application at hand. Such a task is likely to require MapReduce/Spark approaches and it is a natural continuation of this work, along the lines of \cite{FerraroPetrillo2021}.

	\section{Acknowledgements}
	 Funded in part by the MUR-PRIN research project “Multicriteria Data Structures and Algorithms: from compressed to learned indexes, and beyond”, grant n. 2017WR7SHH. Additional support to RG provided by Indam (GNCS 2021) “Algorithms, methods and tools for knowledge discovery in the context of Precision Medicine” project.

	\section{Bibliographic Sketch}
	\textbf{Raffaele Giancarlo} is a Full Professor in Computer Science at the University of Palermo. He works on the design and analysis of algorithms and data structures for the solution of problems in the big data domain, with the main focus on biological data analysis and data compression.

    \textbf{Gennaro Grimaudo} is a PhD student in Information and Communication Technologies at the University of Palermo, on leave from the Technical Staff of the Sicily Regional Government. His research interests are focused on digital transformation and data ecosystems in the public sector.

%\bibliographystyle{unsrtnat}
%\bibliographystyle{plainnat}
%\bibliographystyle{plain}
%\bibliography{main}  

%%%%%%%%%% Merge with Supplementary Material %%%%%%%%%%
\pagebreak

\begin{center}
\textbf{\large Pareto Optimal Compression of Genomic Dictionaries, \\ with or without Random Access in Main Memory \\ \vspace{10pt}(Supplementary Material) }
\end{center}
%%%%%%%%%% Merge with Supplementary Material %%%%%%%%%%
%%%%%%%%%% Prefix a "S" to all equations, figures, tables and reset the counter 

%%%%%%%%%%
\setcounter{equation}{0}
\setcounter{section}{0}
\setcounter{figure}{0}
\setcounter{table}{0}
\setcounter{page}{1}
\makeatletter
\renewcommand{\theequation}{S\arabic{equation}}
\renewcommand{\thefigure}{S\arabic{figure}}
\renewcommand{\thetable}{S\arabic{table}}

%\renewcommand{\bibnumfmt}[1]{[S#1]}
%\renewcommand{\citenumfont}[1]{S#1}
%\renewcommand{\bibnumfmt}[1]{[#1]}
%\renewcommand{\citenumfont}[1]{#1}
%%%%%%%%%% Prefix a "S" to all equations, figures, tables and reset the counter %%%%%%%%%%

    \section{Comparison with the Base Case Scenario.} \label{Comparison_w_BaseCase}
    
    \begin{itemize}
        \item { {\bf{Succinct}} vs. {\bf{B-Succinct on Disk} Scenarios}.}
        We discuss explicitly the case in which $D_k$ is given Implicitly and $F_k$ is given Explicitly, since the case in which $D_k$ and $F_k$ are given both Explicitly is analogous. Assume that the compression step associated to this case, with the choice of the best compressor for {\bf BCSF}, e.g. {\bf zstd} to fix ideas, produces files for a total of \emph{m} bytes, and let \emph{g} be the number of bytes taken by the {\bf {FM-index}} of the genomic sequences. When the ratio $m/g$ is less than one, it is more convenient to keep the Dictionary compressed on disk via the Implicit Case rather than rebuilding it from scratch via the {\bf {FM-index}} stored on disk. 
     
        We take into account also another measure of performance, i.e., post-processing time, as motivated in the Introduction. Let $t_1$ be the post-processing time, taken by the Implicit Case, and let $t_2$ be the post-processing time of the {\bf{B-Succinct on Disk Scenario}} discussed earlier. When the ratio $t_1/t_2$ is less than one, it is more convenient in terms of post-processing time to keep the Dictionary compressed on disk via the Implicit Case rather than rebuilding it from scratch via the {\bf {FM-index}} stored on disk.

        \item { {\bf{Compressed}} vs. {\bf{B-Compressed on Disk} Scenarios}.}
        We discuss only the case {\bf{DP0}}, since the other cases are analogous. Assume that the compression step with {\bf{DP0}} in its best compression setting, e.g., {\bf bzip2} for $D_k$ and {\bf lz4} for $F_k$, for the second discussion, produces a sequence of \emph{m} bytes. Recalling the {\bf{B-Compressed on Disk Scenario}} and to fix ideas, assume that {\bf MFC} is the best compressor, among the textual compressors used in this research, to compress the genomic sequences, producing  \emph{g}  bytes of output.  When the ratio $m/g$ is less than one, it is more convenient to keep the Dictionary compressed on disk via the output of {\bf{DP0}} rather than rebuilding it from scratch, by decompressing the genomic sequences via {\bf MFC} followed by an application of {\bf DSK}. 
     
        Let $t_1$ be the post-processing time, taken by {\bf{DP0}} in its best post-processing time performance, and let $t_2$ be the post-processing time of the corresponding {\bf{B-Compressed on Disk Scenario}} again with {\bf MFC}. When the ratio $t_1/t_2$ is less than one, it is more convenient in terms of post-processing time to keep the Dictionary compressed on disk via {\bf{DP0}}, rather than rebuilding it from the compressed genomic sequences.
        
    \end{itemize}

    \section{Hardware}\label{Hardware}
      We use commodity hardware. That is, we do not consider high performance computing or Distributed Computing solutions, although both $k$-mer statistics collection \cite{FADE, KCH, FerraroPetrillo2019, Fastkmer, Xiao18} and compression have been considered in those settings \cite{FerraroPetrillo2021}. All our experiments have been performed,  single-threaded,  on a PC with Ubuntu Linux 64 bits Operating System, equipped with an Intel \textregistered  Xeon \textregistered  CPU W-2125 @ 4.00 GHz processor, 64GB of DDR4 ECC memory, 512GB SSD PCIe NVMe and 2TB HDD SATA III.

    \section{Ability to Compress the Input}
    
        \subsection{\textbf{Succinct on Disk Scenario}}
\begin{figure*}[ht]
            \centering
    	    \includegraphics[width=\linewidth]{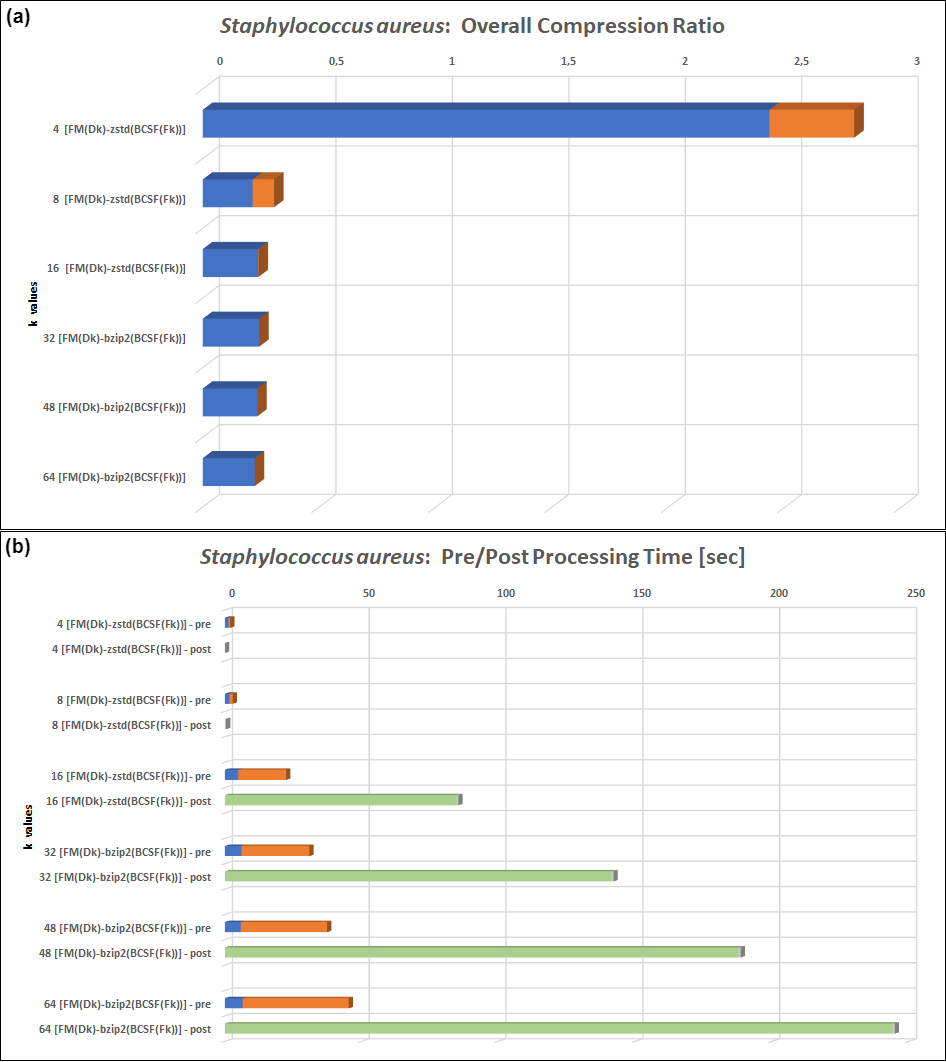}
    		\caption{{\bf SD-RAM Scenario, Case $D_k$ and $F_k$ Explicit - \emph{Staphylococcus Aureus}  Dataset}. Following the definitions and notation in the Main  text, for each value of $k$ included in this study, we report (a) overall compression ratio, where the blu bar refers to the contribution given by the compression of $D_k$ (via the {\bf FM-index})  and the orange bar to the contribution given the compression of the frequencies (via {\bf BCSF}, followed by a textual compressor - the one that provides the best compression ratio of {\bf BCSF}); (b) pre and post processing time, where the blu bar indicates the contribution of {\bf DSK}, the orange bar everything else (see the Main text),  the green bar indicates the contribution given by the {\bf FM-index} decompression, the yellow bar (hardly visible) indicates the contribution given by the {\bf BCSF} decompression, and the grey bar indicates the contribution given by the recovery of $F_k$ (see the Main text).}
    		\label{fig:staph_E}
        \end{figure*}

\begin{figure*}[ht]
            \centering
    	    \includegraphics[width=\linewidth]{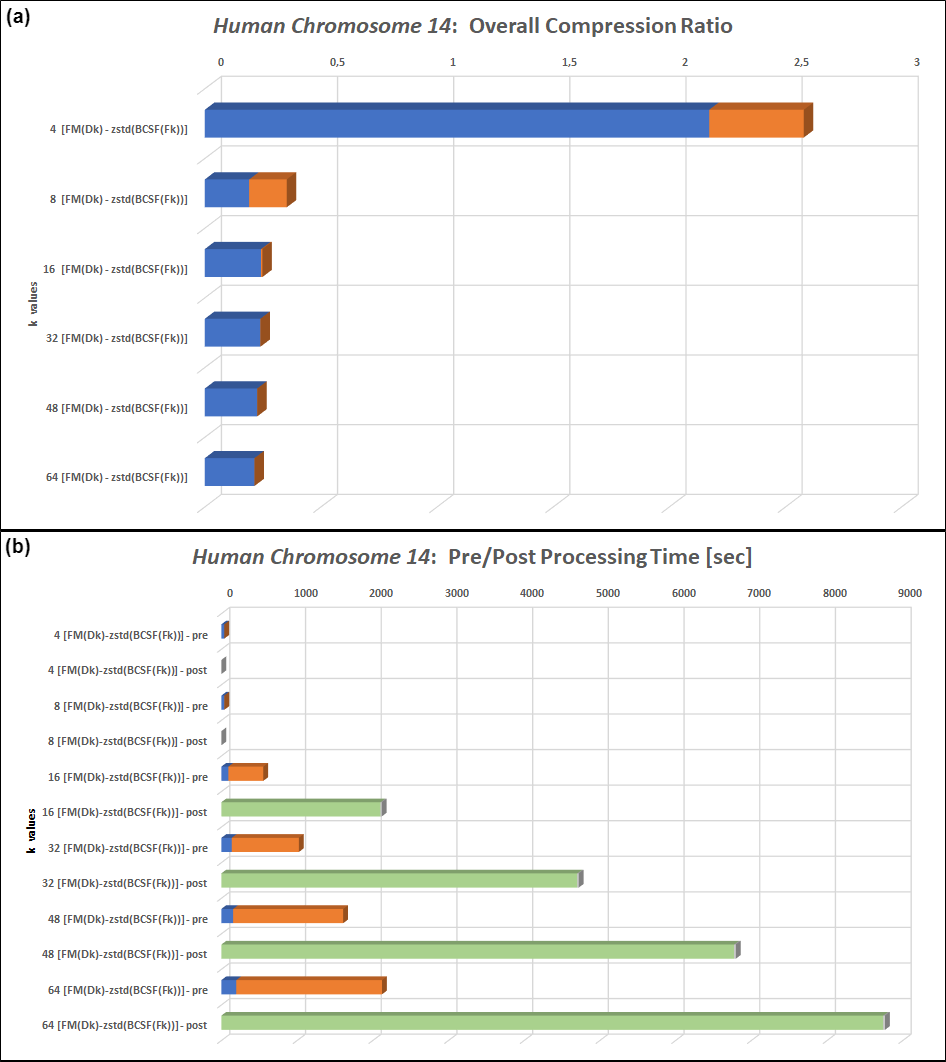}
    			\caption{{\bf SD-RAM Scenario, Case $D_k$ and $F_k$ Explicit - \emph{ Human Chromosome 14} Dataset}. The legend is as in Figure \ref{fig:staph_E}.}
    			\label{fig:hum_E}
        \end{figure*}

\begin{figure*}[ht]
            \centering
    	  \includegraphics[width=\linewidth]{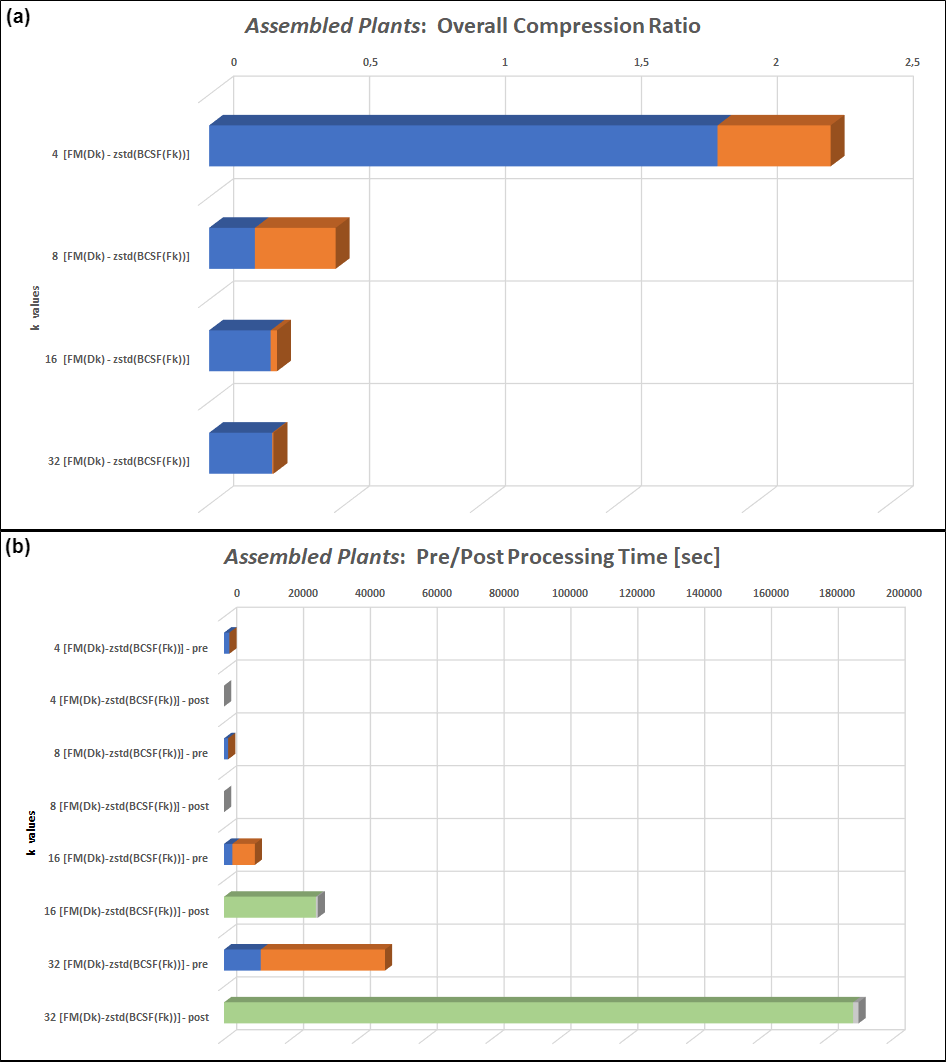}
            \caption{{\bf SD-RAM Scenario, Case $D_k$ and $F_k$ Explicit - \emph{Assembled Plants} Dataset}. The legend is as in Figure \ref{fig:staph_E}. The values of $k$ not reported correspond to experiments that were stopped after four days of computing time.}
            \label{fig:assplants_E}
        \end{figure*}

\begin{figure*}[ht]
            \centering
    	  \includegraphics[width=\linewidth]{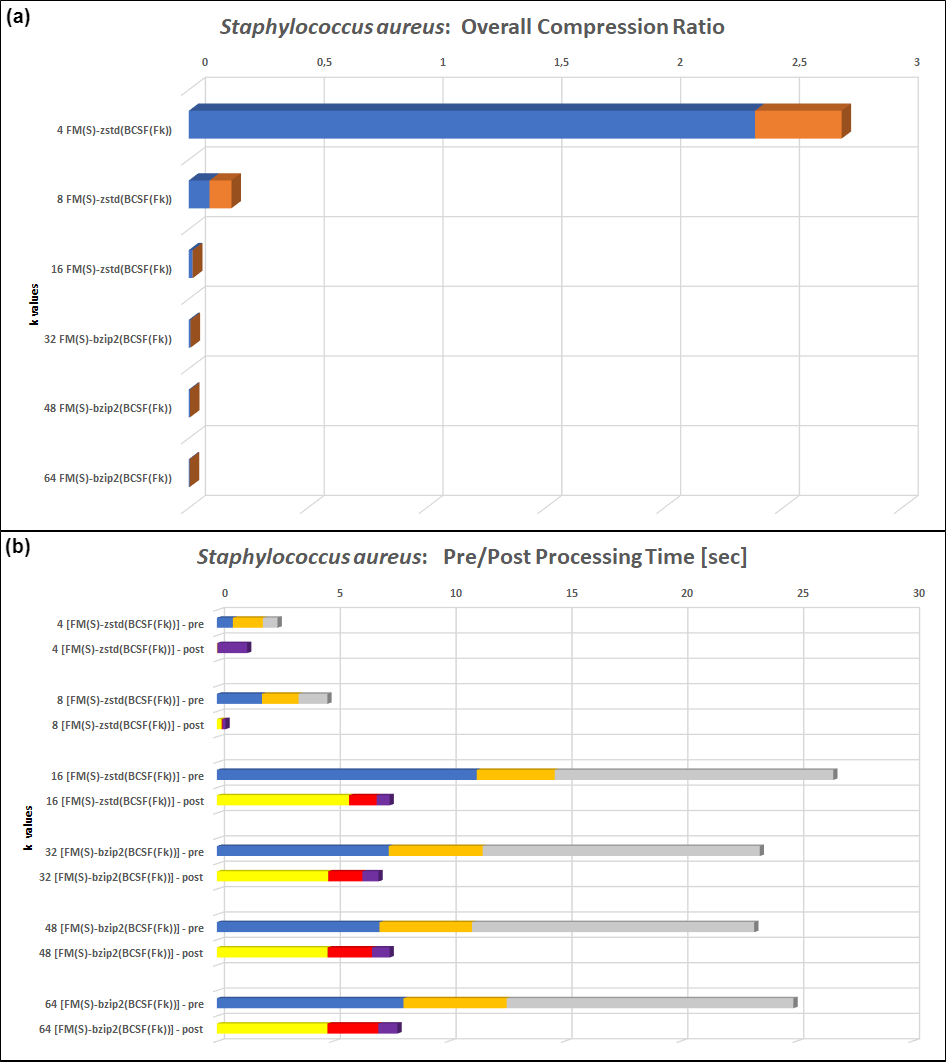}
    	\caption{{\bf SD-RAM Scenario, Case $D_k$ Implicit and $F_k$ Explicit - \emph{Staphylococcus Aureus} Dataset}. Following the definitions and notation in the Main text, for each value of $k$ included in this study, we report (a) overall compression ratio, where the blu bar refers to the contribution given by the compression of $S$ (via the {\bf FM-index})  and the orange bar to the contribution given the compression of the frequencies (via {\bf BCSF}, followed by a textual compressor - the one that provides the best compression ratio of {\bf BCSF})); (b) pre and post processing time, where the blu bar indicates the contribution of {\bf ESSCompress}, the orange bar indicates the contribution of {\bf DSK}, the grey bar indicates everything else (see the Main text), the yellow bar indicates the contribution given by the {\bf FM-index} decompression, the brown bar (hardly visible) indicates the contribution given by the {\bf BCSF} decompression, the red and violet bars indicate the contribution given by the recovery of $D_k$ and $F_k$, respectively (see the Main text).}	
    	\label{fig:staph_I}
        \end{figure*}

\begin{figure*}[ht]
            \centering
    	    \includegraphics[width=\linewidth]{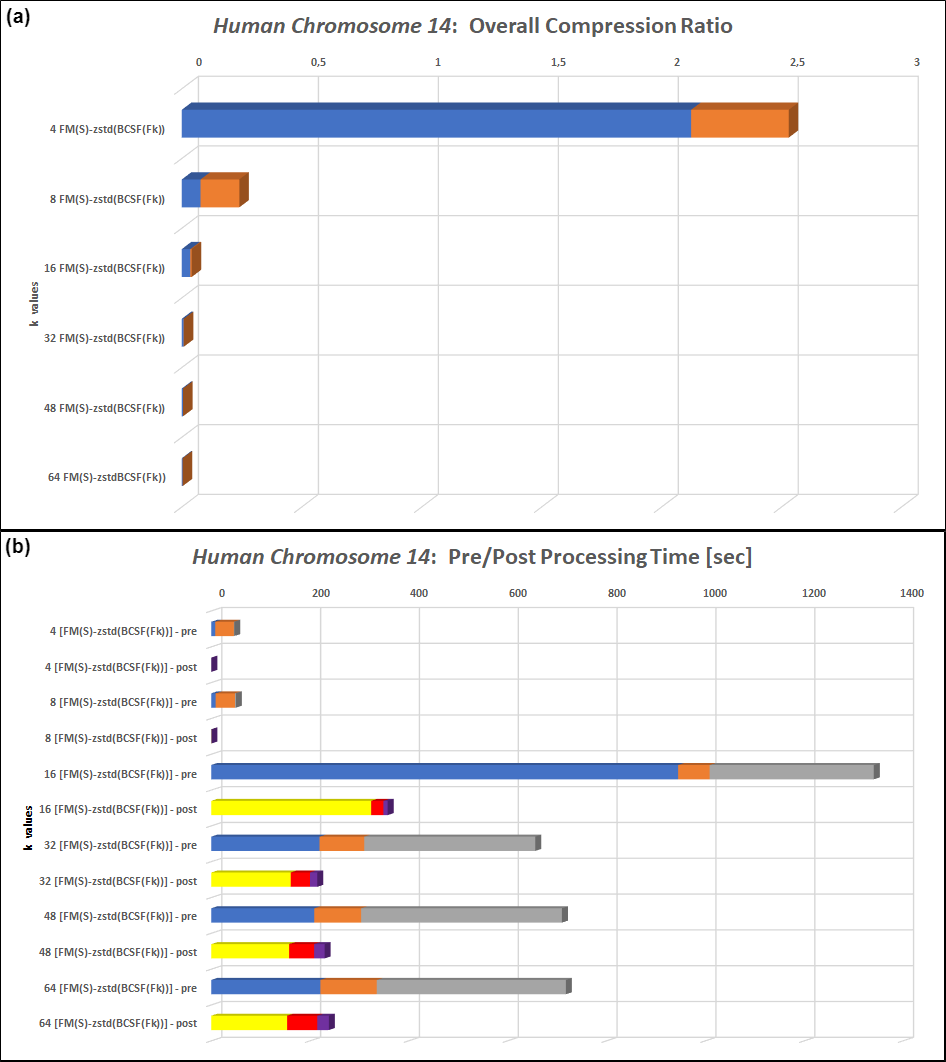}
    		\caption{{\bf SD-RAM Scenario, Case $D_k$ Implicit and $F_k$ Explicit - \emph{Human Chromosome 14} Dataset}. The legend is as in Figure \ref{fig:staph_I}.}
    		\label{fig:hum_I}
        \end{figure*}

\begin{figure*}[ht]
            \centering
    	    \includegraphics[width=\linewidth]{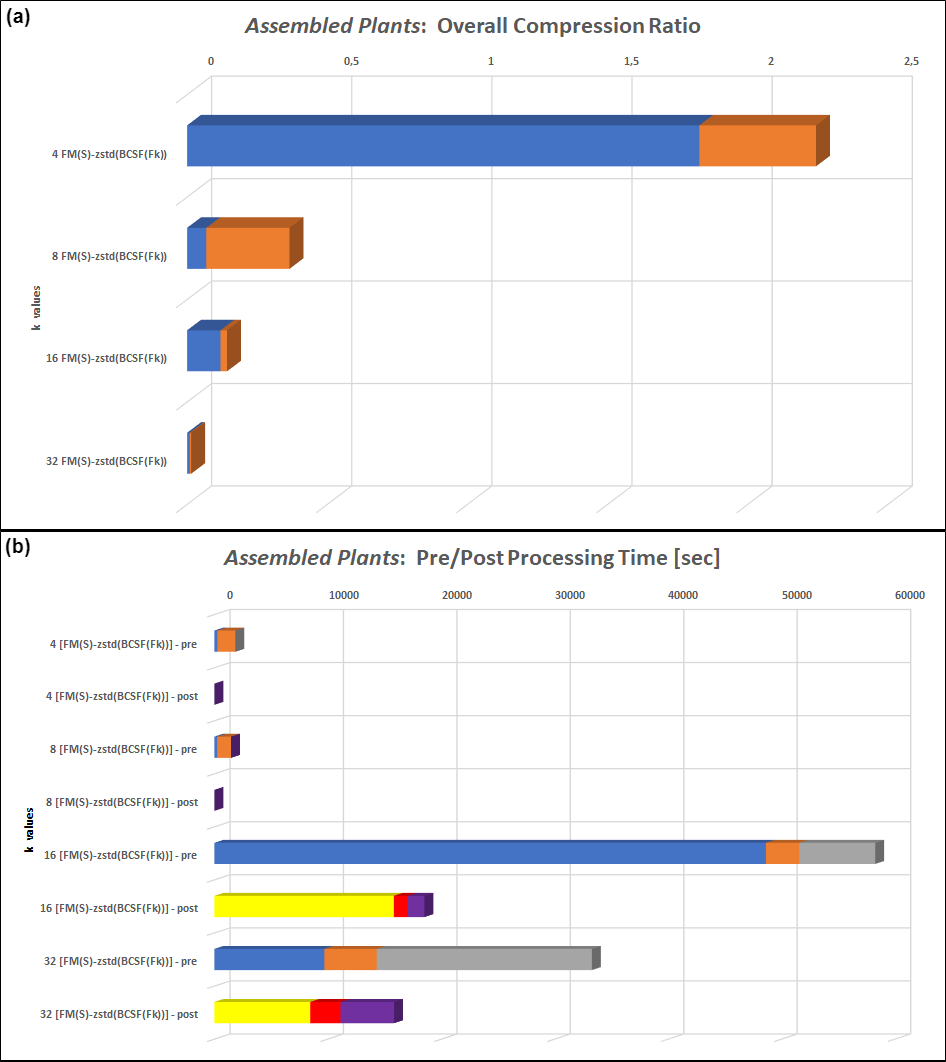}
            \caption{{\bf SD-RAM Scenario, Case $D_k$ Implicit and $F_k$ Explicit - \emph{Assembled Plants} Dataset}. The legend is as in Figure \ref{fig:staph_I}. The values of $k$ not reported correspond to experiments that were stopped after four days of computing time.}
            \label{fig:assplants_I}
        \end{figure*}

        Figures \ref{fig:staph_E}-\ref{fig:assplants_I} report the experiments with respect to standard Data Compression performance measures, as we now explain and recalling that the reported measures are incomparable among the various case. Consider the panel (a) in Figure \ref{fig:staph_E}. It provides the compression ratio, as defined in the {\bf Materials and Methods} Section of the Main text, for the case in which $D_k$ and $F_k$ are given Explicitly. The panel (b) in Figure \ref{fig:staph_E} provides the pre/post-processing times for the same case. In summary, those two panels quantify the compression effectiveness of the selected case with respect to the input dataset and how long it takes to compress/decompress the data, as in standard Data Compression studies.  From those Figures, it is easy to conclude that in both Cases, there is compression of the input, except for $k=4$. Apparently and intuitively, for this value of $k$, the space overhead associated to the succinct data structures does not allow for effective compression of the input datasets. Another point worthy of notice is that, except for the small dataset and for $k=32,48,64$, {\bf zstd} is the method of choice to compress {\bf BCSF}.

        \subsection{\textbf{Compressed on Disk Scenario}}
\begin{figure*}[ht]
            \centering
		    \includegraphics[width=\linewidth]{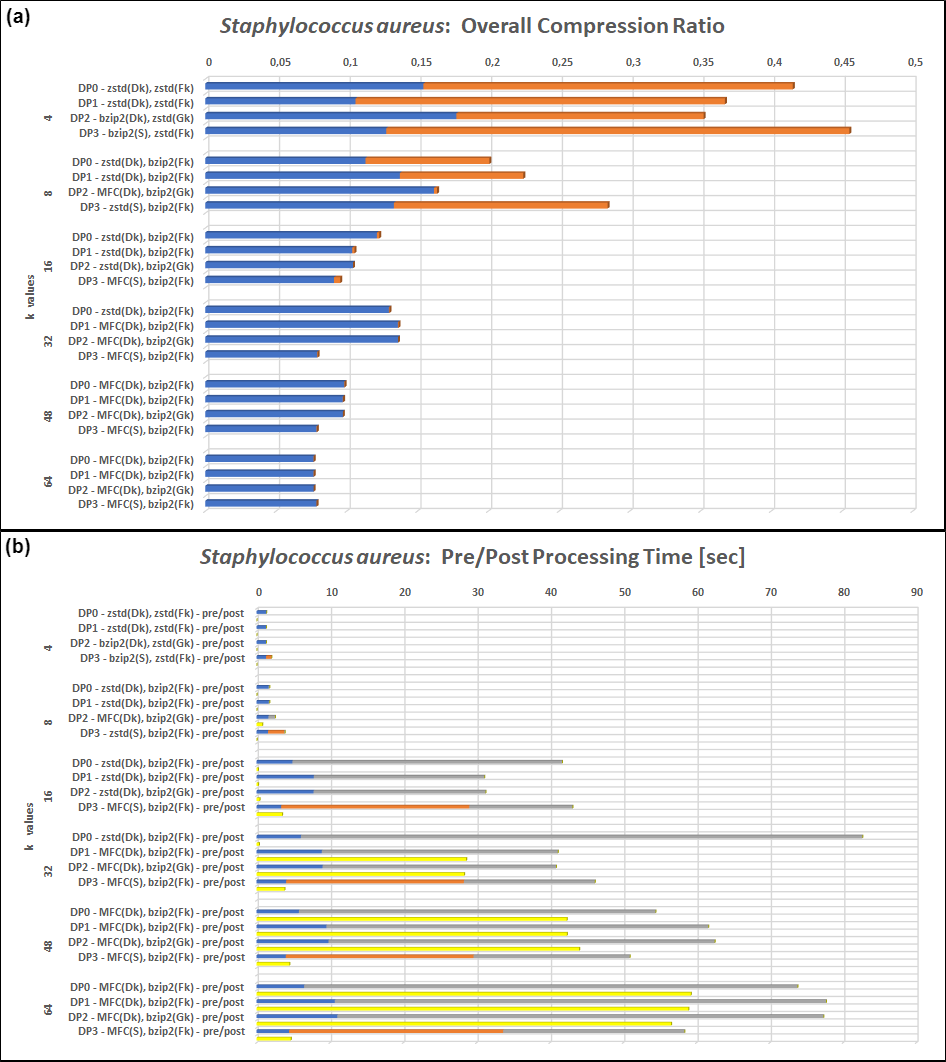}
			\caption{{\bf CD-NRAM  Scenario - \emph{Staphylococcus Aureus} Dataset}. Following the definitions and notation in the Main  text, we report  (a) overall compression ratio, where the blu bar refers to the contribution given by the compression of $D_k$ by a textual or specialized compressor and the orange bar to the contribution given the compression of the frequencies by a textual or specialized compressor; (b) pre and post processing time, where the blu bar indicates the contribution of {\bf DSK}, the orange bar indicates the contribution of {\bf ESSCompress}, the grey bar indicates the contribution of everything else for pre-processing (see the Main text), and the yellow bar indicates the post-processing time (see the Main text). For each $k$ and each of the sub-cases we report the method achieving  the best compression. That is, for $k=4$ and sub-case {\bf {DP0}}, $D_k$ is best compressed with {\bf{zstd}} and $F_k$ are best compressed with gap encoding followed by {\bf{bzip2}}. }
			\label{fig:staph_CD}
        \end{figure*}

\begin{figure*}[ht]
            \centering
		    \includegraphics[width=\linewidth]{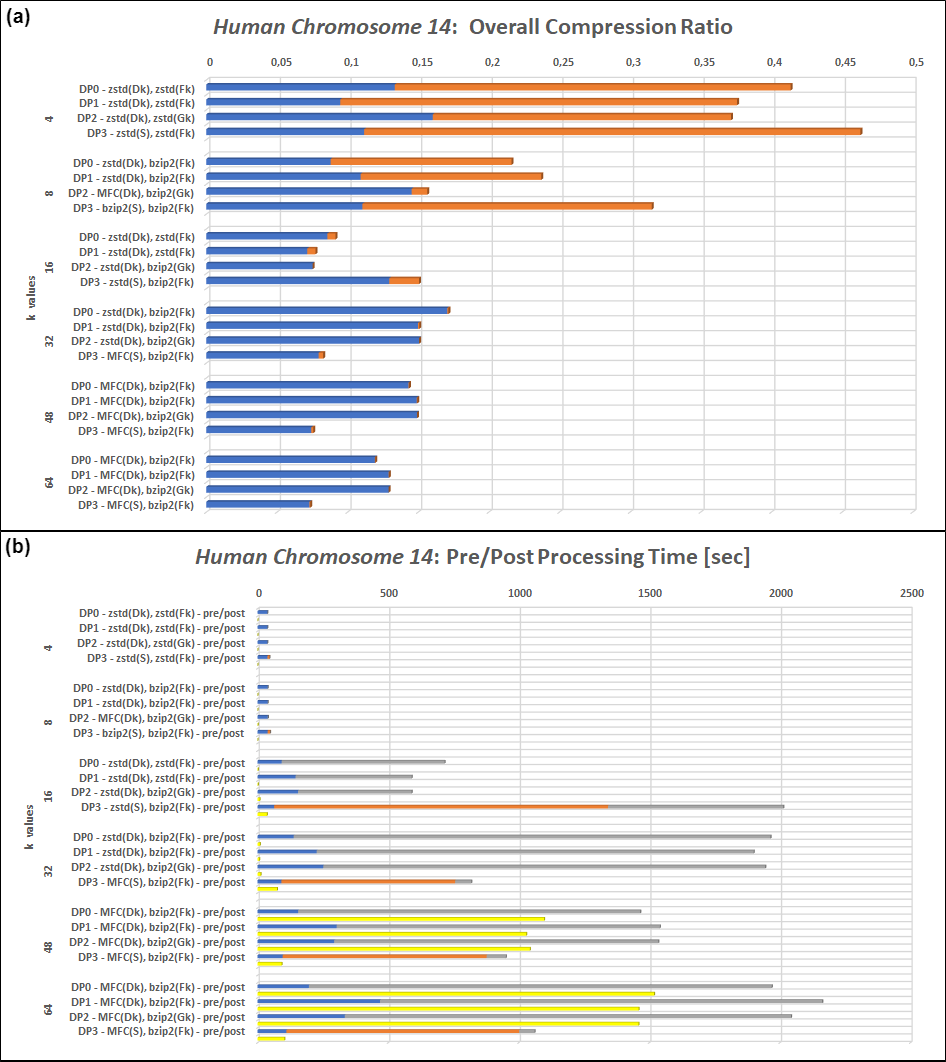}
		\caption{{\bf CD-NRAM Scenario - \emph{Human Chromosome 14} Dataset}. The legend is as in Figure \ref{fig:staph_CD}.}
		\label{fig:hum_CD}
        \end{figure*}

\begin{figure*}[ht]
            \centering
		    \includegraphics[width=\linewidth]{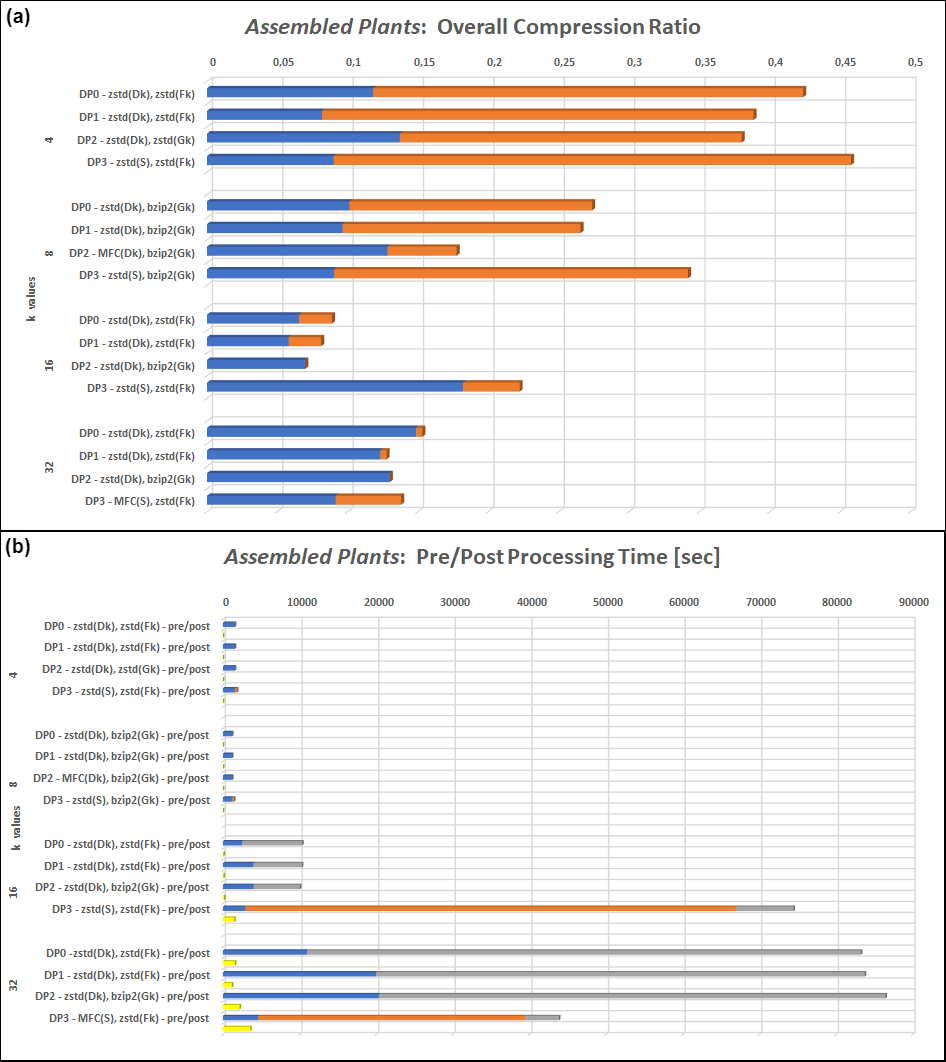}
        	\caption{{\bf CD-NRAM Scenario - \emph{Assembled Plants} Dataset}. The legend is as in Figure \ref{fig:staph_CD}.} 
        	\label{fig:assp_CD}
        \end{figure*}

        Figures \ref{fig:staph_CD}-\ref{fig:assp_CD} report the relevant experiments regarding the selected cases,  with attention to the compression ratio. That is, for each case, we select the best configuration in terms of compressors, based on the compression ratio.  Again, those Figures are divided into two panels, in analogy with the {\bf{Succinct on Disk Scenario}} and with the same role. We use Figure \ref{fig:staph_CD}, as a guiding example to illustrate the content of those panels. For each value of $k$ considered in this study, panel (a) provides the compression ratio, as defined in the Material and Methods section of the Main text, for each of the cases relevant for this scenario.  Moreover, for each case, the compressors yielding the best performance are also indicated. That is, consider $k=4$ and the case {\bf{DP0}}. In that panel, it is reported the best compression ratio obtained over the various choices of compressors we have available to process $D_k$ and $F_k$. Specifically, both $D_k$ and $F_k$ are  best compressed with {\bf{zstd}}. Panel (b) in Figure \ref{fig:staph_CD} provides pre/post-processing times and it is analogous to panel (a). From those Figures, it is evident that all of the considered cases are able to effectively compress their input.

    \section{\bf On the Choice on the $k$-mer Counter for Post-Processing Speed in the Compressed on Disk Scenario } 
    
        The comparison of the {\bf Compressed on Disk Scenario} with respect to the {\bf B-Compressed on Disk Scenario} shows  that it may be advantageous in terms of post-processing time. However, the {\bf B-Compressed on Disk Scenario} uses {\bf DSK} to recover the Dictionary. But this latter   is not the fastest $k$-mer counter available. Therefore, for the cases in which the {\bf B-Compressed on Disk Scenario} \vir{looses} in terms of post-processing time, it is natural to investigate what would happen if one uses a fast $k$-mer counter. For this part of our study, we have chosen {\bf KMC3} \cite{Kokot17},  since it is a \emph{de facto} standard in this area. Since such a counter returns canonical $k$-mers according to the standard lexicographic order relation and {\bf{DSK}} and {\bf{ESSCompress}} use a different order relation (see the Materials and Methods Section), it is necessary to transform the canonical output of {\bf{KMC3}} into the canonical output of {\bf{DSK}}, for consistency. Therefore we have added software that performs that conversion, and whose execution contributes  for the post-processing time of the {\bf{B-Compressed on Disk Scenario}}. We now notice that, from the results in Tables \ref{tab:CD-NRAM_Vs_BaseCase_MFC_Compr}-\ref{tab:CD-NRAM_Vs_BaseCase_SPRING_PostProc}, the use of {\bf KMC3} would only possibly change the \vir{Y entries} in Column \textbf{T}, since for the \vir{N entries}, the {\bf B-Compressed on Disk Scenario} is already better than the methods indicated in those entries and it would get only better with the use of {\bf KMC3}.  Accordingly, we have performed the new post-processing experiments of the {\bf B-Compressed on Disk Scenario} with {\bf{KMC3}} replacing {\bf{DSK}} on the \vir{Y entries} only. We do not provide the results of all of them for conciseness and we limit ourselves to report in Table \ref{tab:CD-NRAM_Vs_BaseCase_KMC3_ComprMFC_PostProcZSTD} the results for favouring compression with {\bf MFC} and favouring post-processing time with {\bf zstd}. 
        They are the analogues of Tables \ref{tab:CD-NRAM_Vs_BaseCase_MFC_Compr} and \ref{tab:CD-NRAM_Vs_BaseCase_zstd_PostProc}, respectively. Interestingly, across all the new experiments, the use of {\bf KMC3} rather than {\bf DSK} does not change the state of our findings. Therefore, the conclusions obtained regarding the advantage of the various scenarios over the {\bf{Base Case Scenario}} in terms of post-processing are robust with respect to the choice of $k$-mers counter.
                   
\ignore{
    \section{The Cost of Random Access for Compressed Dictionaries}

            \subsection{\bf Application Area 1}
            
            \begin{itemize}

                \item {{\bf Favouring  Compression}. For each value of $k$ and each dataset considered in this research, we take the ratio of the number of output bytes of each method reported in  Table \ref{tab:SD-RAM_Favour_Compr_PostProc}, with a reference to the Implicit Case of the {\bf Succinct on Disk Scenario} (Column \textbf{B}) with the corresponding best performing case in terms of output size in bytes reported in Table \ref{tab:CD-NRAM_favour_compression} for the {\bf Compressed on Disk Scenario}. The results are reported in Table \ref{tab:Cost_Random_Access_Appl-Area1_Compr_PostProc}, panel \textbf{Compression}. For the small and medium size datasets and except for small values of $k$, the cost of random access in terms of space occupancy is quite tolerable, i.e., below $20\%$ space in addition to the non-random access scenario. For the large dataset, an analogous but weaker indication may be drawn from the experiments. }

                \item {{\bf Favouring  Post-Processing Time}. For each value of $k$ and each dataset considered in this research, we take the ratio of the post-processing time of the  method reported in the corresponding entry in  Table \ref{tab:SD-RAM_Favour_Compr_PostProc}, with a reference to the Implicit Case of the {\bf Succinct on Disk Scenario} (Column \textbf{B}) with the method reported in the corresponding entry reported  in Table \ref{tab:CD-NRAM_favour_post_processing_time} for the {\bf Compressed on Disk Scenario}.  The results are reported in Table \ref{tab:Cost_Random_Access_Appl-Area1_Compr_PostProc}, panel \textbf{Post-Processing}.  It is evident that, as far as post-processing time is concerned, the cost of random access can be rather steep.

            \end{itemize}

            \subsection{\bf Application Area 2}
            
            \begin{itemize}

                \item {{\bf Favouring  Compression}. We proceed in analogy with the corresponding case in Application Area 1. We can take as reference the results in Table \ref{tab:Cost_Random_Access_Appl-Area1_Compr_PostProc}, except for the case in which the winner is the {\bf Base Case Scenario}. When this happens, we must consider the {\bf Base Case Scenario}. For instance, for $k=16$ and the small dataset, for both Scenarios, the corresponding {\bf Base Case} is the winner. Therefore, we proceed incrementally with respect to  Table \ref{tab:Cost_Random_Access_Appl-Area1_Compr_PostProc}. The new results are reported in Table \ref{tab:Cost_Random_Access_Appl-Area2_Compr_PostProc}, panel \textbf{Compression}. }

                \item {{\bf Favouring  Post-Processing Time}. Here we proceed in analogy with what just illustrated for compression. The resulting incremental table is reported in Table \ref{tab:Cost_Random_Access_Appl-Area2_Compr_PostProc}, panel \textbf{Post-Processing}. }        
            \end{itemize}
}}
           
    \section{Discussion: Additional Tables and Figures}
    
        The Tables relating this part of the manuscript and regarding the {\bf Compressed on Disk Scenario} are \ref{tab:CD-NRAM_favour_compression_wrt}-\ref{tab:CD-NRAM_favour_post_processing_time_wrt}.

        %% Figure 1
    
        %% Figure 2 

        %% Figure 3

        %% Figure 4

        %% Figure 5        

        %% Figure 6

        %% Figure 7 ESSCompress-CR     
        \begin{figure*}[ht]
            \centering
    	    \includegraphics[width=\linewidth]{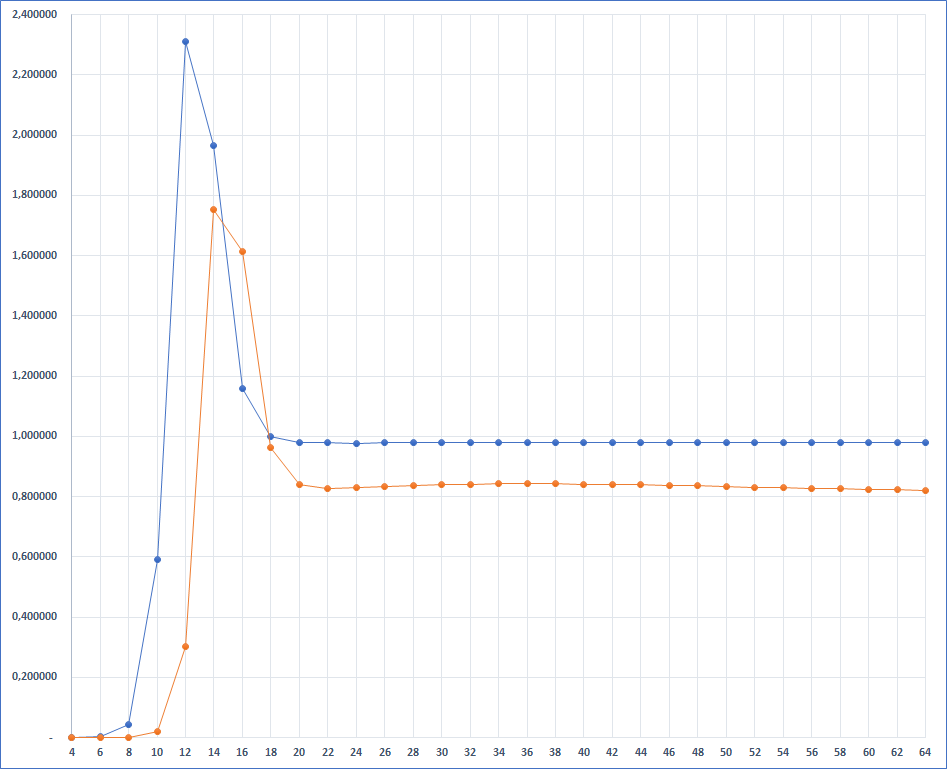}
            \caption{{\bf The Ratio of the Lengths of $S$ with Respect to the Total Lengths of the Genomic Set $G$, as a Function of $k$}. The abscissa reports the value of $k$, while the ordinate reports the corresponding ratio $|S|/|G|$. The blu curve refers to the \emph{Staphylococcus Aureus} dataset and the orange one to the \emph{Human Chromosome 14} dataset.}
            \label{fig:ESS-CR_SA_HC14}
        \end{figure*}    
    
        %% Figure 8         

        %% Figure 9	

        %% Figure 10

        %% Tabelle Sinossi SD
        %TAB.1new - SD-RAM: Favour Compression and Post-Processing Time
        \begin{table*}[]
            \centering
            \resizebox{.8\textwidth}{!}{%
            \begin{tabular}{|c|ccc|ccc|ccc|}
            \hline
            \multicolumn{1}{|l|}{} & \multicolumn{3}{c|}{\textit{SA}} & \multicolumn{3}{c|}{\textit{HC14}} & \multicolumn{3}{c|}{\textit{AP}} 
            \\[.75em]
            \hline

            \textbf{k} & \multicolumn{1}{c|}{\textbf{B}} & \multicolumn{1}{c|}{\textbf{C}} & \textbf{T} & \multicolumn{1}{c|}{\textbf{B}} & \multicolumn{1}{c|}{\textbf{C}} & \textbf{T} & \multicolumn{1}{c|}{\textbf{B}} & \multicolumn{1}{c|}{\textbf{C}} & \textbf{T} 
            \\[.75em] 
            \hline

            \textbf{4} & \multicolumn{1}{c|}{FM(Dk)-zstd(BCSF(Fk))  /    FM(S)-zstd(BCSF(Fk))} & \multicolumn{1}{c|}{1.018E+00} & 1.305E-02 & \multicolumn{1}{c|}{FM(Dk)-zstd(BCSF(Fk))  /    FM(S)-zstd(BCSF(Fk))} & \multicolumn{1}{c|}{1.019E+00} & 1.887E-01 & \multicolumn{1}{c|}{FM(Dk)-zstd(BCSF(Fk))  /    FM(S)-zstd(BCSF(Fk))} & \multicolumn{1}{c|}{1.019E+00} & 6.957E-02
            \\[1em] 
            \hline

            \textbf{8} & \multicolumn{1}{c|}{FM(Dk)-zstd(BCSF(Fk))  /    FM(S)-zstd(BCSF(Fk))} & \multicolumn{1}{c|}{1.704E+00} & 1.139E+00 & \multicolumn{1}{c|}{FM(Dk)-zstd(BCSF(Fk))  /    FM(S)-zstd(BCSF(Fk))} & \multicolumn{1}{c|}{1.471E+00} & 1.097E+00 & \multicolumn{1}{c|}{FM(Dk)-zstd(BCSF(Fk))  /    FM(S)-zstd(BCSF(Fk))} & \multicolumn{1}{c|}{1.273E+00} & 1.390E+00 
            \\[1em] 
            \hline

            \textbf{16} & \multicolumn{1}{c|}{FM(Dk)-zstd(BCSF(Fk))  /    FM(S)-zstd(BCSF(Fk))} & \multicolumn{1}{c|}{1.350E+01} & 1.145E+01 & \multicolumn{1}{c|}{FM(Dk)-zstd(BCSF(Fk))  /    FM(S)-zstd(BCSF(Fk))} & \multicolumn{1}{c|}{6.015E+00} & 5.935E+00 & \multicolumn{1}{c|}{FM(Dk)-zstd(BCSF(Fk))  /    FM(S)-zstd(BCSF(Fk))} & \multicolumn{1}{c|}{1.752E+00} & 1.513E+00 
            \\[1em] 
            \hline

            \textbf{32} & \multicolumn{1}{c|}{FM(Dk)-bzip2(BCSF(Fk))  /    FM(S)-bzip2(BCSF(Fk))} & \multicolumn{1}{c|}{3.118E+01} & 2.036E+01 & \multicolumn{1}{c|}{FM(Dk)-zstd(BCSF(Fk))  /    FM(S)-zstd(BCSF(Fk))} & \multicolumn{1}{c|}{2.784E+01} & 2.200E+01 & \multicolumn{1}{c|}{FM(Dk)-zstd(BCSF(Fk))  /    FM(S)-zstd(BCSF(Fk))} & \multicolumn{1}{c|}{1.632E+01} & 1.198E+01 
            \\[1em] 
            \hline

            \textbf{48} & \multicolumn{1}{c|}{FM(Dk)-bzip2(BCSF(Fk))  /    FM(S)-bzip2(BCSF(Fk))} & \multicolumn{1}{c|}{4.442E+01} & 2.529E+01 & \multicolumn{1}{c|}{FM(Dk)-zstd(BCSF(Fk))  /    FM(S)-zstd(BCSF(Fk))} & \multicolumn{1}{c|}{4.214E+01} & 2.961E+01 & \multicolumn{1}{c|}{\textbf{-}} & \multicolumn{1}{c|}{\textbf{-}} & \textbf{-} 
            \\[1em] 
            \hline

            \textbf{64} & \multicolumn{1}{c|}{FM(Dk)-bzip2(BCSF(Fk))  /    FM(S)-bzip2(BCSF(Fk))} & \multicolumn{1}{c|}{5.606E+01} & 3.142E+01 & \multicolumn{1}{c|}{FM(Dk)-zstd(BCSF(Fk))  /    FM(S)-zstd(BCSF(Fk))} & \multicolumn{1}{c|}{5.497E+01} & 3.683E+01 & \multicolumn{1}{c|}{\textbf{-}} & \multicolumn{1}{c|}{\textbf{-}} & \textbf{-} 
            \\[1em]
            \hline

            \end{tabular}
            }
            \vspace{5pt}
            \caption{ {\bf {SD-RAM Scenario: Favour Compression and Post-Processing Time}.} The values of $k$ we have considered are reported in the first column of the table, which is divided into three panels, one for each of the datasets used in this research. Those latter  are abbreviated as follows: \emph{SA, HC14} and \emph{AP} stand for \emph{Staphylococcus Aureus}, \emph{Human Chromosome 14}, and \emph{Assembled Plants}, respectively. For the first panel, the three columns within the dataset indicate the following. Column \textbf{B} indicates the ratio to be taken between the \textbf{Succinct on Disk Scenario} Explicit Case and the \textbf{Succinct on Disk Scenario} Implicit Case, both taken in the best compression configuration with respect to the \textbf{BCSF} succinct data structure. In Column \textbf{C} that ratio account for compression and, in Column \textbf{T}, it account for post-processing time. As for the second and third panels, their entries are analogous to the ones of the first one. An entry with a dash indicates that the experiment could not be completed, as explained in the Main text (see \textbf{Experimental Setup}). }             
            \label{tab:SD-RAM_Favour_Compr_PostProc}
        \end{table*}

        %TAB.2new (TAB.X1.old, TAB.X2.old)
        \begin{table*}[]
            \centering
            \resizebox{.8\textwidth}{!}{%
            \begin{tabular}{|c|cccccc|cccccc|}
                \hline
                \multicolumn{1}{|l|}{} &
                  \multicolumn{6}{c|}{\textbf{Logic Indication}} &
                  \multicolumn{6}{c|}{\textbf{Numeric Values}} \\ \hline
                \textbf{} &
                  \multicolumn{2}{c|}{\textit{SA}} &
                  \multicolumn{2}{c|}{\textit{HC14}} &
                  \multicolumn{2}{c|}{\textit{AP}} &
                  \multicolumn{2}{c|}{\textit{SA}} &
                  \multicolumn{2}{c|}{\textit{HC14}} &
                  \multicolumn{2}{c|}{\textit{AP}} \\ \hline
                \textbf{k} &
                  \multicolumn{1}{c|}{\textbf{C}} &
                  \multicolumn{1}{c|}{\textbf{T}} &
                  \multicolumn{1}{c|}{\textbf{C}} &
                  \multicolumn{1}{c|}{\textbf{T}} &
                  \multicolumn{1}{c|}{\textbf{C}} &
                  \textbf{T} &
                  \multicolumn{1}{c|}{\textbf{C}} &
                  \multicolumn{1}{c|}{\textbf{T}} &
                  \multicolumn{1}{c|}{\textbf{C}} &
                  \multicolumn{1}{c|}{\textbf{T}} &
                  \multicolumn{1}{c|}{\textbf{C}} &
                  \textbf{T} \\ \hline
                  
                \textbf{4} &
                  \multicolumn{1}{c|}{Y} &
                  \multicolumn{1}{c|}{Y} &
                  \multicolumn{1}{c|}{Y} &
                  \multicolumn{1}{c|}{Y} &
                  \multicolumn{1}{c|}{Y} &
                  \multicolumn{1}{c|}{Y} &
                  \multicolumn{1}{c|}{5.143E-03} &
                  \multicolumn{1}{c|}{1.984E-01} &
                  \multicolumn{1}{c|}{1.754E-04} &
                  \multicolumn{1}{c|}{7.778E-04} &
                  \multicolumn{1}{c|}{3.419E-06} &
                  \multicolumn{1}{c|}{1.031E-04} \\ \hline
                  
                \textbf{8} &
                  \multicolumn{1}{c|}{Y} &
                  \multicolumn{1}{c|}{Y} &
                  \multicolumn{1}{c|}{Y} &
                  \multicolumn{1}{c|}{Y} &
                  \multicolumn{1}{c|}{Y} &
                  \multicolumn{1}{c|}{Y} &
                  \multicolumn{1}{c|}{9.200E-02} &
                  \multicolumn{1}{c|}{5.351E-02} &
                  \multicolumn{1}{c|}{4.561E-03} &
                  \multicolumn{1}{c|}{1.800E-03} &
                  \multicolumn{1}{c|}{1.493E-04} &
                  \multicolumn{1}{c|}{2.779E-05} \\ \hline
                  
                \textbf{16} &
                  \multicolumn{1}{c|}{N} &
                  \multicolumn{1}{c|}{Y} &
                  \multicolumn{1}{c|}{N} &
                  \multicolumn{1}{c|}{N} &
                  \multicolumn{1}{c|}{N} &
                  \multicolumn{1}{c|}{N} &
                  \multicolumn{1}{c|}{1.219E+00} &
                  \multicolumn{1}{c|}{7.352E-01} &
                  \multicolumn{1}{c|}{2.225E+00} &
                  \multicolumn{1}{c|}{1.378E+00} &
                  \multicolumn{1}{c|}{2.115E+00} &
                  \multicolumn{1}{c|}{1.403E+00} \\ \hline
                  
                \textbf{32} &
                  \multicolumn{1}{c|}{Y} &
                  \multicolumn{1}{c|}{Y} &
                  \multicolumn{1}{c|}{N} &
                  \multicolumn{1}{c|}{Y} &
                  \multicolumn{1}{c|}{N} &
                  \multicolumn{1}{c|}{Y} &
                  \multicolumn{1}{c|}{9.974E-01} &
                  \multicolumn{1}{c|}{6.178E-01} &
                  \multicolumn{1}{c|}{1.053E+00} &
                  \multicolumn{1}{c|}{7.042E-01} &
                  \multicolumn{1}{c|}{1.379E+00} &
                  \multicolumn{1}{c|}{7.311E-01} \\ \hline
                  
                \textbf{48} &
                  \multicolumn{1}{c|}{Y} &
                  \multicolumn{1}{c|}{Y} &
                  \multicolumn{1}{c|}{Y} &
                  \multicolumn{1}{c|}{Y} &
                  \multicolumn{1}{c|}{-} &
                  \multicolumn{1}{c|}{-} &
                  \multicolumn{1}{c|}{9.874E-01} &
                  \multicolumn{1}{c|}{6.763E-01} &
                  \multicolumn{1}{c|}{9.938E-01} &
                  \multicolumn{1}{c|}{7.127E-01} &
                  \multicolumn{1}{c|}{-} &
                  \multicolumn{1}{c|}{-} \\ \hline
                  
                \textbf{64} &
                  \multicolumn{1}{c|}{Y} &
                  \multicolumn{1}{c|}{Y} &
                  \multicolumn{1}{c|}{Y} &
                  \multicolumn{1}{c|}{Y} &
                  \multicolumn{1}{c|}{-} &
                  \multicolumn{1}{c|}{-} &
                  \multicolumn{1}{c|}{9.824E-01} &
                  \multicolumn{1}{c|}{6.639E-01} &
                  \multicolumn{1}{c|}{9.607E-01} &
                  \multicolumn{1}{c|}{6.556E-01} &
                  \multicolumn{1}{c|}{-} &
                  \multicolumn{1}{c|}{-} \\ \hline
                  
                \end{tabular}
            }
            \vspace{5pt}
            \caption{\textbf{Synopsis of the Succinct Vs. B-Succinct on Disk Scenarios}. The values of $k$ we have considered are reported in the first column of the table, which  is divided into two panels: 
            {\bf Logic Indication}  and {\bf Numeric Values}. Dataset abbreviations are as in Table \ref{tab:SD-RAM_Favour_Compr_PostProc}. For the first panel, the first row indicates the dataset. For each of them, the two columns within the dataset indicate the compression ratio  (column \textbf{C}) and the post-processing time ratio (Column \textbf{T})  between the Implicit Case and the {\bf Base Case Scenario}, as specified in the Main text.  Y means that the Implicit Case is more convenient than the {\bf Base Case Scenario} and N its complement.  An entry with a dash indicates that the experiment could not be completed, as explained in the Main text (see {\bf{Experimental Setup}}). As for the second panel, its entries are analogous to the ones of the first one, except that the numeric values of the mentioned ratios are reported. } 
                   
            \label{tab:SD-RAM_Vs_BaseCase}
        \end{table*}

        % Tab.3new: CD-NRAM: Favour Compression (ex MAIN)
        \begin{table*}[]
            \centering
            \resizebox{.8\textwidth}{!}{    
                \begin{tabular}{|c|c|c|c|}
                \hline
                \multicolumn{1}{|c|}{\textbf{k}} & \multicolumn{1}{c|}{\textit{SA}} & \multicolumn{1}{c|}{\textit{HC14}} & \multicolumn{1}{c|}{\textit{AP}} 
                \\ & & & 
                \\ \hline           
                \textbf{4} & DP2(bzip2(Dk),   zstd(Gk)) - DP3(bzip2(S), zstd(Fk)) {[}1.27E+00{]} & DP2(zstd(Dk),   zstd(Gk)) - DP3(zstd(S), zstd(Fk)) {[}1.01E+00{]} & DP3(zstd(S),   zstd(Fk)) - DP2(zstd(Dk), zstd(Gk)) {[}1.01E+00{]} 
                \\ & & & 
                \\ \hline             
                \textbf{8} & DP2(MFC(Dk),   bzip2(Gk)) - DP3(zstd(S), bzip2(Fk)) {[}1.03E+00{]} & DP2(MFC(Dk),   bzip2(Gk)) - DP3(bzip2(S), bzip2(Fk)) {[}1.25E+00{]} & DP2(MFC(Dk),   bzip2(Gk)) - DP3(zstd(S), bzip2(Gk)) {[}1.28E+00{]} 
                \\ & & & 
                \\ \hline              
                \textbf{16} & DP3(MFC(S),   bzip2(Fk)) - DP2(zstd(Dk), bzip2(Gk)) {[}6.48E+00{]} & DP3(zstd(S),   bzip2(Fk)) - DP2(zstd(Dk), bzip2(Gk)) {[}2.05E+00{]} & DP2(zstd(Dk),   bzip2(Gk)) - DP3(zstd(S), zstd(Fk)) {[}1.83E+00{]} 
                \\ & & & 
                \\ \hline          
                \textbf{32} & DP3(MFC(S),   bzip2(Fk)) - DP0(zstd(Dk), bzip2(Fk)) {[}1.99E+01{]} & DP3(MFC(S),   bzip2(Fk)) - DP2(zstd(Dk), bzip2(Gk)) {[}2.05E+{01]} & DP3(MFC(S), zstd(Fk))   - DP1(zstd(Dk), zstd(Fk)) {[}9.58E+00{]} 
                \\ & & & 
                \\ \hline  
                \textbf{48} & DP3(MFC(S),   bzip2(Fk)) - DP2(MFC(Dk), bzip2(Gk)) {[}2.09E+01{]} & DP3(MFC(S),   bzip2(Fk)) - DP0(MFC(Dk), bzip2(Fk)) {[}3.16E+01{]} & \textbf{-} 
                \\ & & & 
                \\ \hline   
                \textbf{64} & DP3(MFC(S),   bzip2(Fk)) - DP2(MFC(Dk), bzip2(Gk)) {[}2.17E+01{]} & DP3(MFC(S),   bzip2(Fk)) - DP0(MFC(Dk), bzip2(Fk)) {[}3.59E+01{]} & \textbf{-} 
                \\ & & & 
                \\ \hline   
                \end{tabular}
            }
            \vspace{5pt}
            \caption{{\bf {Synopsis of the CD-NRAM Scenario: Favour Compression}}. The values of $k$ we have considered are reported in the first column of the table, which is divided into three panels, one for each of the datasets used in this research. Dataset abbreviations are as in Table \ref{tab:SD-RAM_Favour_Compr_PostProc}. For the first panel, we report the best performing case in terms of output size in bytes, together with the compressors used to achieve it. In case one of the Explicit Cases wins, we also report the best (minimum output size in bytes) compression obtained in the Implicit Case, with an indication of the loss in terms of output size (the numeric quantity in square brackets). When the Implicit Case wins, we proceed symmetrically. As for the second and third panels, their entries are analogous to the ones of the first one. An entry with a dash indicates that the experiment could not be completed, as explained in the Main text (see \textbf{Experimental Setup}). } 
            \label{tab:CD-NRAM_favour_compression}
        \end{table*}

        % Tab.4new: CD-NRAM: Favour Post-Processing Time (ex MAIN)
        \begin{table*}[]
            \centering
            \resizebox{.8\textwidth}{!}{%
                \begin{tabular}{|c|c|c|c|}
                \hline
                \multicolumn{1}{|c|}{\textbf{k}} & \multicolumn{1}{c|}{\textit{SA}} & \multicolumn{1}{c|}{\textit{HC14}} & \multicolumn{1}{c|}{\textit{AP}} 
                \\ & & & 
                \\ \hline
                
                \textbf{4} & DP1(zstd(Dk),   Opt-PFOR(Fk)) {[}1.80E+00{]},{[}1.50E+00{]} & DP1(zstd(Dk)) ,   Opt-PFOR(Fk)) {[}1.88E+00{]}, {[}2.50E+00{]} & DP2(zstd(Dk)) ,   zstd(Fk)) {[}1.12E+00{]}, {[}5.50E+00{]} 
                \\ & & & 
                \\ \hline
                
                \textbf{8} & DP0(lz4(Dk),   zstd(Fk)) {[}2.56E+00{]},{[}2.62E+02{]} & DP1(zstd(Dk)) ,   lz4(Fk))  {[}1.98E+00{]}, {[}2.87E+02{]} & DP1(zstd(Dk),   lz4(Fk))  {[}2.03E+00{]}, {[}2.90E+02{]} 
                \\ & & & 
                \\ \hline
                
                \textbf{16} & DP0(zstd(Dk),   zstd(Fk)) {[}7.65E+00{]},{[}4.03E+01{]} & DP2(zstd(Dk)) ,   zstd(Fk))  {[}2.05E+00{]}, {[}2.28E+01{]} & DP2(zstd(Dk)) ,   zstd(Fk))  {[}1.00E+00{]}, {[}2.56E+00{]}
                \\ & & & 
                \\ \hline
                
                \textbf{32} & DP0(zstd(Dk),   lz4(Fk)) {[}1.91E+01{]},{[}1.68E+01{]} & DP1(zstd(Dk)) ,   zstd(Fk))  {[}2.05E+01{]}, {[}2.04E+01{]} & DP1(zstd(Dk)) ,   zstd(Fk)) {[}9.58E+00{]}, {[}3.04E+00{]} 
                \\ & & & 
                \\ \hline
                
                \textbf{48} & DP0(zstd(Dk),   zstd(Fk))  {[}2.19E+01{]},{[}1.65E+01{]} & DP0(zstd(Dk),   zstd(Fk))  {[}4.01E+01{]}, {[}1.08E+01{]} & \textit{\textbf{-}} 
                \\ & & & 
                \\ \hline
                
                \textbf{64} & DP0(zstd(Dk),   zstd(Fk))  {[}2.44E+01{]},{[}1.32E+01{]} & DP2(zstd(Dk),   zstd(Fk))  {[}5.77E+01{]}, {[}1.06E+01{]} & \textit{\textbf{-}} 
                \\ & & & 
                \\ \hline
                
                \end{tabular}
            }
            \vspace{5pt}
            \caption{{\bf {Synopsis of the CD-NRAM Scenario: Favour Post-Processing Time}}. This table is analogous to Table \ref{tab:CD-NRAM_favour_compression}, except that the role of compression and post-processing are exchanged. For the first panel, we report the best performing case in terms of post-processing time, together with the compressors used to achieve it. We report how much the best case in terms of post-processing loses in terms of output size in bytes compared to the winner in compression. These two values provide an estimate of the trade-off between compression and post-processing. As for the second and third panels, their entries are analogous to the ones of the first one. An entry with a dash indicates that the experiment could not be completed, as explained in the Main text (see \textbf{Experimental Setup}). }
            \label{tab:CD-NRAM_favour_post_processing_time}
        \end{table*}

        %TAB.5new (TAB.X5.old, TAB.X6.old)
        \begin{table*}[]
            \centering
            \resizebox{.8\textwidth}{!}{%             
                \begin{tabular}{|c|ccccccccc|ccccccccc|}
                \hline
                 &
                  \multicolumn{3}{c|}{\textit{SA}} &
                  \multicolumn{3}{c|}{\textit{HC14}} &
                  \multicolumn{3}{c|}{\textit{AP}} &
                  \multicolumn{3}{c|}{\textit{SA}} &
                  \multicolumn{3}{c|}{\textit{HC14}} &
                  \multicolumn{3}{c|}{\textit{AP}} \\ \hline
                  
                \textbf{k} &
                  \multicolumn{1}{c|}{\textbf{C}} &
                  \multicolumn{1}{c|}{\textbf{T}} &
                  \multicolumn{1}{c|}{\textbf{B}} &
                  \multicolumn{1}{c|}{\textbf{C}} &
                  \multicolumn{1}{c|}{\textbf{T}} &
                  \multicolumn{1}{c|}{\textbf{B}} &
                  \multicolumn{1}{c|}{\textbf{C}} &
                  \multicolumn{1}{c|}{\textbf{T}} &
                  \multicolumn{1}{c|}{\textbf{B}} &
                  \multicolumn{1}{c|}{\textbf{C}} &
                  \multicolumn{1}{c|}{\textbf{T}} &
                  \multicolumn{1}{c|}{\textbf{B}} &
                  \multicolumn{1}{c|}{\textbf{C}} &
                  \multicolumn{1}{c|}{\textbf{T}} &
                  \multicolumn{1}{c|}{\textbf{B}} &
                  \multicolumn{1}{c|}{\textbf{C}} &
                  \multicolumn{1}{c|}{\textbf{T}} &
                  \multicolumn{1}{c|}{\textbf{B}} \\ \hline
                  
                \textbf{4} &
                  \multicolumn{1}{c|}{Y} &
                  \multicolumn{1}{c|}{Y} &
                  \multicolumn{1}{c|}{DP2} &
                  \multicolumn{1}{c|}{Y} &
                  \multicolumn{1}{c|}{Y} &
                  \multicolumn{1}{c|}{DP2} &
                  \multicolumn{1}{c|}{Y} &
                  \multicolumn{1}{c|}{Y} &
                  \multicolumn{1}{c|}{DP3} &
                  \multicolumn{1}{c|}{7.843E-04} &
                  \multicolumn{1}{c|}{6.416E-04} &
                  \multicolumn{1}{c|}{DP2} &
                  \multicolumn{1}{c|}{3.207E-05} &
                  \multicolumn{1}{c|}{4.206E-05} &
                  \multicolumn{1}{c|}{DP2} &
                  \multicolumn{1}{c|}{7.645E-07} &
                  \multicolumn{1}{c|}{4.739E-06} & 
                  \multicolumn{1}{c|}{DP3} \\ \hline
                  
                \textbf{8} &
                  \multicolumn{1}{c|}{Y} &
                  \multicolumn{1}{c|}{Y} &
                  \multicolumn{1}{c|}{DP2} &
                  \multicolumn{1}{c|}{Y} &
                  \multicolumn{1}{c|}{Y} &
                  \multicolumn{1}{c|}{DP2} &
                  \multicolumn{1}{c|}{Y} &
                  \multicolumn{1}{c|}{Y} &
                  \multicolumn{1}{c|}{DP2} &
                  \multicolumn{1}{c|}{9.976E-02} &
                  \multicolumn{1}{c|}{2.293E-01} &
                  \multicolumn{1}{c|}{DP2} &
                  \multicolumn{1}{c|}{3.704E-03} &
                  \multicolumn{1}{c|}{1.422E-02} &
                  \multicolumn{1}{c|}{DP2} &
                  \multicolumn{1}{c|}{9.624E-05} &
                  \multicolumn{1}{c|}{4.443E-04} & 
                  \multicolumn{1}{c|}{DP2} \\ \hline
                  
                \textbf{16} &
                  \multicolumn{1}{c|}{N} &
                  \multicolumn{1}{c|}{Y} &
                  \multicolumn{1}{c|}{DP3} &
                  \multicolumn{1}{c|}{N} &
                  \multicolumn{1}{c|}{Y} &
                  \multicolumn{1}{c|}{DP3} &
                  \multicolumn{1}{c|}{N} &
                  \multicolumn{1}{c|}{Y} &
                  \multicolumn{1}{c|}{DP2} &
                  \multicolumn{1}{c|}{1.316E+00} &
                  \multicolumn{1}{c|}{5.239E-01} &
                  \multicolumn{1}{c|}{DP3} &
                  \multicolumn{1}{c|}{2.464E+00} &
                  \multicolumn{1}{c|}{3.417E-01} &
                  \multicolumn{1}{c|}{DP3} &
                  \multicolumn{1}{c|}{1.376E+00} &
                  \multicolumn{1}{c|}{4.690E-02} &
                  \multicolumn{1}{c|}{DP2} \\ \hline
                \textbf{32} &
                  \multicolumn{1}{c|}{N} &
                  \multicolumn{1}{c|}{Y} &
                  \multicolumn{1}{c|}{DP3} &
                  \multicolumn{1}{c|}{N} &
                  \multicolumn{1}{c|}{Y} &
                  \multicolumn{1}{c|}{DP3} &
                  \multicolumn{1}{c|}{N} &
                  \multicolumn{1}{c|}{Y} &
                  \multicolumn{1}{c|}{DP3} &
                  \multicolumn{1}{c|}{1.043E+00} &
                  \multicolumn{1}{c|}{4.872E-01} &
                  \multicolumn{1}{c|}{DP3} &
                  \multicolumn{1}{c|}{1.121E+00} &
                  \multicolumn{1}{c|}{4.667E-01} &
                  \multicolumn{1}{c|}{DP3} &
                  \multicolumn{1}{c|}{1.722E+00} &
                  \multicolumn{1}{c|}{3.056E-01} & 
                  \multicolumn{1}{c|}{DP3} \\ \hline
                  
                \textbf{48} &
                  \multicolumn{1}{c|}{N} &
                  \multicolumn{1}{c|}{Y} &
                  \multicolumn{1}{c|}{DP3} &
                  \multicolumn{1}{c|}{N} &
                  \multicolumn{1}{c|}{Y} &
                  \multicolumn{1}{c|}{DP3} &
                  \multicolumn{1}{c|}{-} &
                  \multicolumn{1}{c|}{-} &
                  \multicolumn{1}{c|}{-} &
                  \multicolumn{1}{c|}{1.038E+00} &
                  \multicolumn{1}{c|}{5.931E-01} &
                  \multicolumn{1}{c|}{DP3} &
                  \multicolumn{1}{c|}{1.049E+00} &
                  \multicolumn{1}{c|}{5.215E-01} &
                  \multicolumn{1}{c|}{DP3} &
                  \multicolumn{1}{c|}{-} &
                  \multicolumn{1}{c|}{-} &
                  \multicolumn{1}{c|}{-} \\ \hline
                   
                \textbf{64} &
                  \multicolumn{1}{c|}{N} &
                  \multicolumn{1}{c|}{Y} &
                  \multicolumn{1}{c|}{DP3} &
                  \multicolumn{1}{c|}{N} &
                  \multicolumn{1}{c|}{Y} &
                  \multicolumn{1}{c|}{DP3} &
                  \multicolumn{1}{c|}{-} &
                  \multicolumn{1}{c|}{-} &
                  \multicolumn{1}{c|}{-} &
                  \multicolumn{1}{c|}{1.036E+00} &
                  \multicolumn{1}{c|}{5.650E-01} &
                  \multicolumn{1}{c|}{DP3} &
                  \multicolumn{1}{c|}{1.024E+00} &
                  \multicolumn{1}{c|}{4.724E-01} &
                  \multicolumn{1}{c|}{DP3} &
                  \multicolumn{1}{c|}{-} &
                  \multicolumn{1}{c|}{-} &
                  \multicolumn{1}{c|}{-} \\ \hline 
                \end{tabular}
            }
        \vspace{5pt}    
        \caption{\textbf{Synopsis of the CD-NRAM Scenarios Vs. The Base Case Scenario with MFC - Favour Compression.}
        The values of $k$ we have considered are reported in the first column of the table, which is divided into two panels: {\bf Logic Indication}  and {\bf Numeric Values}. For the first panel, the first row indicates the dataset. Dataset abbreviations are as in Table \ref{tab:SD-RAM_Favour_Compr_PostProc}. For each of them, the three columns within the dataset indicate the following. Column \textbf{B} indicates the best \textbf{Compressed on Disk Scenario} for the Case favouring compression. Columns \textbf{C} and \textbf{T} indicate, respectively, the compression ratio and the post-processing time ratio between the \textbf{Compressed on Disk Scenario}, indicated in the column \textbf{B}, and the corresponding \textbf{Base Case Scenario} with \textbf{MFC} compressor, as specified in the Main text. Y means that the \textbf{Compressed on Disk Scenario} is more convenient than the {\bf Base Case Scenario} and N its complement.  An entry with a dash indicates that the experiment could not be completed, as explained in the Main text (see {\bf{Experimental Setup}}). As for the second panel, its entries are analogous to the ones of the first one, except that the numeric values of the mentioned ratios are reported. }
        \label{tab:CD-NRAM_Vs_BaseCase_MFC_Compr}
        \end{table*}

        %TAB.6new (TAB.X9.old, TAB.X10.old)
        \begin{table*}[]
            \centering
            \resizebox{.8\textwidth}{!}{%             
                \begin{tabular}{|c|ccccccccc|ccccccccc|}
                \hline
                 &
                  \multicolumn{3}{c|}{\textit{SA}} &
                  \multicolumn{3}{c|}{\textit{HC14}} &
                  \multicolumn{3}{c|}{\textit{AP}} &
                  \multicolumn{3}{c|}{\textit{SA}} &
                  \multicolumn{3}{c|}{\textit{HC14}} &
                  \multicolumn{3}{c|}{\textit{AP}} \\ \hline
                  
                \textbf{k} &
                  \multicolumn{1}{c|}{\textbf{C}} &
                  \multicolumn{1}{c|}{\textbf{T}} &
                  \multicolumn{1}{c|}{\textbf{B}} &
                  \multicolumn{1}{c|}{\textbf{C}} &
                  \multicolumn{1}{c|}{\textbf{T}} &
                  \multicolumn{1}{c|}{\textbf{B}} &
                  \multicolumn{1}{c|}{\textbf{C}} &
                  \multicolumn{1}{c|}{\textbf{T}} &
                  \multicolumn{1}{c|}{\textbf{B}} &
                  \multicolumn{1}{c|}{\textbf{C}} &
                  \multicolumn{1}{c|}{\textbf{T}} &
                  \multicolumn{1}{c|}{\textbf{B}} &
                  \multicolumn{1}{c|}{\textbf{C}} &
                  \multicolumn{1}{c|}{\textbf{T}} &
                  \multicolumn{1}{c|}{\textbf{B}} &
                  \multicolumn{1}{c|}{\textbf{C}} &
                  \multicolumn{1}{c|}{\textbf{T}} &
                  \multicolumn{1}{c|}{\textbf{B}} \\ \hline
                  
                \textbf{4} &
                  \multicolumn{1}{c|}{Y} &
                  \multicolumn{1}{c|}{Y} &
                  \multicolumn{1}{c|}{ DP2} &
                  \multicolumn{1}{c|}{Y} &
                  \multicolumn{1}{c|}{Y} &
                  \multicolumn{1}{c|}{ DP2} &
                  \multicolumn{1}{c|}{Y} &
                  \multicolumn{1}{c|}{Y} &
                  \multicolumn{1}{c|}{ DP3} &
                  \multicolumn{1}{c|}{6.582E-04} &
                  \multicolumn{1}{c|}{1.499E-03} &
                  \multicolumn{1}{c|}{DP2} &
                  \multicolumn{1}{c|}{2.687E-05} &
                  \multicolumn{1}{c|}{6.552E-05} &
                  \multicolumn{1}{c|}{DP2} &
                  \multicolumn{1}{c|}{6.301E-07} &
                  \multicolumn{1}{c|}{6.948E-06} & 
                  \multicolumn{1}{c|}{DP3} \\ \hline 
                  
                \textbf{8} &
                  \multicolumn{1}{c|}{Y} &
                  \multicolumn{1}{c|}{Y} &
                  \multicolumn{1}{c|}{ DP2} &
                  \multicolumn{1}{c|}{Y} &
                  \multicolumn{1}{c|}{Y} &
                  \multicolumn{1}{c|}{ DP2} &
                  \multicolumn{1}{c|}{Y} &
                  \multicolumn{1}{c|}{Y} &
                  \multicolumn{1}{c|}{ DP2} &
                  \multicolumn{1}{c|}{8.372E-02} &
                  \multicolumn{1}{c|}{4.778E-01} &
                  \multicolumn{1}{c|}{DP2} &
                  \multicolumn{1}{c|}{3.104E-03} &
                  \multicolumn{1}{c|}{2.192E-02} &
                  \multicolumn{1}{c|}{DP2} &
                  \multicolumn{1}{c|}{7.932E-05} &
                  \multicolumn{1}{c|}{7.119E-04} & 
                  \multicolumn{1}{c|}{DP2} \\ \hline 
                   
                \textbf{16} &
                  \multicolumn{1}{c|}{N} &
                  \multicolumn{1}{c|}{Y} &    %SIST. ERRORE
                  \multicolumn{1}{c|}{DP3} &
                  \multicolumn{1}{c|}{N} &
                  \multicolumn{1}{c|}{Y} &
                  \multicolumn{1}{c|}{DP3} &
                  \multicolumn{1}{c|}{N} &
                  \multicolumn{1}{c|}{Y} &
                  \multicolumn{1}{c|}{DP2} &
                  \multicolumn{1}{c|}{1.105E+00} &
                  \multicolumn{1}{c|}{7.145E-01} &
                  \multicolumn{1}{c|}{DP3} &
                  \multicolumn{1}{c|}{2.065E+00} &
                  \multicolumn{1}{c|}{3.868E-01} &
                  \multicolumn{1}{c|}{DP3} &
                  \multicolumn{1}{c|}{1.134E+00} &
                  \multicolumn{1}{c|}{6.067E-02} & 
                  \multicolumn{1}{c|}{DP2} \\ \hline 
                   
                \textbf{32} &
                  \multicolumn{1}{c|}{Y} &
                  \multicolumn{1}{c|}{Y} &      %SIST. ERRORE
                  \multicolumn{1}{c|}{DP3} &
                  \multicolumn{1}{c|}{Y} &
                  \multicolumn{1}{c|}{Y} &
                  \multicolumn{1}{c|}{DP3} &
                  \multicolumn{1}{c|}{N} &
                  \multicolumn{1}{c|}{Y} &
                  \multicolumn{1}{c|}{DP3} &
                  \multicolumn{1}{c|}{8.754E-01} &
                  \multicolumn{1}{c|}{6.307E-01} &
                  \multicolumn{1}{c|}{DP3} &
                  \multicolumn{1}{c|}{9.395E-01} &
                  \multicolumn{1}{c|}{5.384E-01} &
                  \multicolumn{1}{c|}{DP3} &
                  \multicolumn{1}{c|}{1.420E+00} &
                  \multicolumn{1}{c|}{3.262E-01} & 
                  \multicolumn{1}{c|}{DP3} \\ \hline 
                   
                \textbf{48} &
                  \multicolumn{1}{c|}{Y} &
                  \multicolumn{1}{c|}{Y} &      %SIST. ERRORE
                  \multicolumn{1}{c|}{DP3} &
                  \multicolumn{1}{c|}{Y} &
                  \multicolumn{1}{c|}{Y} &
                  \multicolumn{1}{c|}{DP3} &
                  \multicolumn{1}{c|}{-} &
                  \multicolumn{1}{c|}{-} &
                  \multicolumn{1}{c|}{-} &
                  \multicolumn{1}{c|}{8.709E-01} &
                  \multicolumn{1}{c|}{7.758E-01} &
                  \multicolumn{1}{c|}{DP3} &
                  \multicolumn{1}{c|}{8.793E-01} &
                  \multicolumn{1}{c|}{5.927E-01} &
                  \multicolumn{1}{c|}{DP3} &
                  \multicolumn{1}{c|}{-} &
                  \multicolumn{1}{c|}{-} & 
                  \multicolumn{1}{c|}{-} \\ \hline 
                   
                \textbf{64} &
                  \multicolumn{1}{c|}{Y} &
                  \multicolumn{1}{c|}{Y} &      %SIST. ERRORE
                  \multicolumn{1}{c|}{DP3} &
                  \multicolumn{1}{c|}{Y} &
                  \multicolumn{1}{c|}{Y} &
                  \multicolumn{1}{c|}{DP3} &
                  \multicolumn{1}{c|}{-} &
                  \multicolumn{1}{c|}{-} &
                  \multicolumn{1}{c|}{-} &
                  \multicolumn{1}{c|}{8.697E-01} &
                  \multicolumn{1}{c|}{7.199E-01} &
                  \multicolumn{1}{c|}{DP3} &
                  \multicolumn{1}{c|}{8.577E-01} &
                  \multicolumn{1}{c|}{5.235E-01} &
                  \multicolumn{1}{c|}{DP3} &
                  \multicolumn{1}{c|}{-} &
                  \multicolumn{1}{c|}{-} & 
                  \multicolumn{1}{c|}{-} \\ \hline 
                   
                \end{tabular}
            }
        \vspace{5pt}    
        \caption{{{\bf {Synopsis of the CD-NRAM Scenarios Vs. The Base Case Scenario with zstd - Favour Compression.}} The table legend is as in Table \ref{tab:CD-NRAM_Vs_BaseCase_MFC_Compr}.}}
        \label{tab:CD-NRAM_Vs_BaseCase_zstd_Compr}
        \end{table*}

        %TAB.7new (TAB.X13.old, TAB.X14.old)
        \begin{table*}[]
            \centering
            \resizebox{.8\textwidth}{!}{%             
                \begin{tabular}{|c|ccccccccc|ccccccccc|}
                \hline
                 &
                  \multicolumn{3}{c|}{\textit{SA}} &
                  \multicolumn{3}{c|}{\textit{HC14}} &
                  \multicolumn{3}{c|}{\textit{AP}} &
                  \multicolumn{3}{c|}{\textit{SA}} &
                  \multicolumn{3}{c|}{\textit{HC14}} &
                  \multicolumn{3}{c|}{\textit{AP}} \\ \hline
                  
                \textbf{k} &
                  \multicolumn{1}{c|}{\textbf{C}} &
                  \multicolumn{1}{c|}{\textbf{T}} &
                  \multicolumn{1}{c|}{\textbf{B}} &
                  \multicolumn{1}{c|}{\textbf{C}} &
                  \multicolumn{1}{c|}{\textbf{T}} &
                  \multicolumn{1}{c|}{\textbf{B}} &
                  \multicolumn{1}{c|}{\textbf{C}} &
                  \multicolumn{1}{c|}{\textbf{T}} &
                  \multicolumn{1}{c|}{\textbf{B}} &
                  \multicolumn{1}{c|}{\textbf{C}} &
                  \multicolumn{1}{c|}{\textbf{T}} &
                  \multicolumn{1}{c|}{\textbf{B}} &
                  \multicolumn{1}{c|}{\textbf{C}} &
                  \multicolumn{1}{c|}{\textbf{T}} &
                  \multicolumn{1}{c|}{\textbf{B}} &
                  \multicolumn{1}{c|}{\textbf{C}} &
                  \multicolumn{1}{c|}{\textbf{T}} &
                  \multicolumn{1}{c|}{\textbf{B}} \\ \hline
                  
                \textbf{4} &
                  \multicolumn{1}{c|}{Y} &
                  \multicolumn{1}{c|}{Y} &
                  \multicolumn{1}{c|}{DP2} &
                  \multicolumn{1}{c|}{Y} &
                  \multicolumn{1}{c|}{Y} &
                  \multicolumn{1}{c|}{DP2} &
                  \multicolumn{1}{c|}{Y} &
                  \multicolumn{1}{c|}{Y} &
                  \multicolumn{1}{c|}{DP3} &
                  \multicolumn{1}{c|}{6.226E-04} &
                  \multicolumn{1}{c|}{1.367E-03} &
                  \multicolumn{1}{c|}{DP2} &
                  \multicolumn{1}{c|}{2.477E-05} &
                  \multicolumn{1}{c|}{5.902E-05} &
                  \multicolumn{1}{c|}{DP2} &
                  \multicolumn{1}{c|}{5.425E-07} &
                  \multicolumn{1}{c|}{6.268E-06} & 
                  \multicolumn{1}{c|}{DP3} \\ \hline 
                  
                \textbf{8} &
                  \multicolumn{1}{c|}{Y} &
                  \multicolumn{1}{c|}{Y} &
                  \multicolumn{1}{c|}{DP2} &
                  \multicolumn{1}{c|}{Y} &
                  \multicolumn{1}{c|}{Y} &
                  \multicolumn{1}{c|}{DP2} &
                  \multicolumn{1}{c|}{Y} &
                  \multicolumn{1}{c|}{Y} &
                  \multicolumn{1}{c|}{DP2} &
                  \multicolumn{1}{c|}{7.919E-02} &
                  \multicolumn{1}{c|}{4.431E-01} &
                  \multicolumn{1}{c|}{DP2} &
                  \multicolumn{1}{c|}{2.861E-03} &
                  \multicolumn{1}{c|}{1.981E-02} &
                  \multicolumn{1}{c|}{DP2} &
                  \multicolumn{1}{c|}{6.829E-05} &
                  \multicolumn{1}{c|}{6.243E-04} & 
                  \multicolumn{1}{c|}{DP2} \\ \hline 
                   
                \textbf{16} &
                  \multicolumn{1}{c|}{N} &
                  \multicolumn{1}{c|}{Y} &
                  \multicolumn{1}{c|}{DP3} &
                  \multicolumn{1}{c|}{N} &
                  \multicolumn{1}{c|}{Y} &
                  \multicolumn{1}{c|}{DP3} &
                  \multicolumn{1}{c|}{Y} &
                  \multicolumn{1}{c|}{Y} &
                  \multicolumn{1}{c|}{DP2} &
                  \multicolumn{1}{c|}{1.045E+00} &
                  \multicolumn{1}{c|}{6.961E-01} &
                  \multicolumn{1}{c|}{DP3} &
                  \multicolumn{1}{c|}{1.903E+00} &
                  \multicolumn{1}{c|}{3.700E-01} &
                  \multicolumn{1}{c|}{DP3} &
                  \multicolumn{1}{c|}{9.761E-01} &
                  \multicolumn{1}{c|}{5.679E-02} & 
                  \multicolumn{1}{c|}{DP2} \\ \hline 
                   
                \textbf{32} &
                  \multicolumn{1}{c|}{Y} &
                  \multicolumn{1}{c|}{Y} &
                  \multicolumn{1}{c|}{DP3} &
                  \multicolumn{1}{c|}{Y} &
                  \multicolumn{1}{c|}{Y} &
                  \multicolumn{1}{c|}{DP3} &
                  \multicolumn{1}{c|}{N} &
                  \multicolumn{1}{c|}{Y} &
                  \multicolumn{1}{c|}{DP3} &
                  \multicolumn{1}{c|}{8.280E-01} &
                  \multicolumn{1}{c|}{6.175E-01} &
                  \multicolumn{1}{c|}{DP3} &
                  \multicolumn{1}{c|}{8.660E-01} &
                  \multicolumn{1}{c|}{5.225E-01} &
                  \multicolumn{1}{c|}{DP3} &
                  \multicolumn{1}{c|}{1.222E+00} &
                  \multicolumn{1}{c|}{3.211E-01} & 
                  \multicolumn{1}{c|}{DP3} \\ \hline 
                   
                \textbf{48} &
                  \multicolumn{1}{c|}{Y} &
                  \multicolumn{1}{c|}{Y} &
                  \multicolumn{1}{c|}{DP3} &
                  \multicolumn{1}{c|}{Y} &
                  \multicolumn{1}{c|}{Y} &
                  \multicolumn{1}{c|}{DP3} &
                  \multicolumn{1}{c|}{-} &
                  \multicolumn{1}{c|}{-} &
                  \multicolumn{1}{c|}{-} &
                  \multicolumn{1}{c|}{8.238E-01} &
                  \multicolumn{1}{c|}{7.589E-01} &
                  \multicolumn{1}{c|}{DP3} &
                  \multicolumn{1}{c|}{8.105E-01} &
                  \multicolumn{1}{c|}{5.771E-01} &
                  \multicolumn{1}{c|}{DP3} &
                  \multicolumn{1}{c|}{-} &
                  \multicolumn{1}{c|}{-} & 
                  \multicolumn{1}{c|}{-} \\ \hline 
                   
                \textbf{64} &
                  \multicolumn{1}{c|}{Y} &
                  \multicolumn{1}{c|}{Y} &
                  \multicolumn{1}{c|}{DP3} &
                  \multicolumn{1}{c|}{Y} &
                  \multicolumn{1}{c|}{Y} &
                  \multicolumn{1}{c|}{DP3} &
                  \multicolumn{1}{c|}{-} &
                  \multicolumn{1}{c|}{-} &
                  \multicolumn{1}{c|}{-} &
                  \multicolumn{1}{c|}{8.227E-01} &
                  \multicolumn{1}{c|}{7.059E-01} &
                  \multicolumn{1}{c|}{DP3} &
                  \multicolumn{1}{c|}{7.906E-01} &
                  \multicolumn{1}{c|}{5.125E-01} &
                  \multicolumn{1}{c|}{DP3} &
                  \multicolumn{1}{c|}{-} &
                  \multicolumn{1}{c|}{-} & 
                  \multicolumn{1}{c|}{-} \\ \hline 
                   
                \end{tabular}
            }
        \vspace{5pt}    
        \caption{\textbf{Synopsis of the CD-NRAM Scenarios Vs. The Base Case Scenario with bzip2 - Favour Compression.} The table legend is as in Table \ref{tab:CD-NRAM_Vs_BaseCase_MFC_Compr}.}
        \label{tab:CD-NRAM_Vs_BaseCase_bzip2_Compr}
        \end{table*}

        %TAB.8new (TAB.X17.old, TAB.X18.old)
        \begin{table*}[]
            \centering
            \resizebox{.8\textwidth}{!}{%             
                \begin{tabular}{|c|ccccccccc|ccccccccc|}
                \hline
                 &
                  \multicolumn{3}{c|}{\textit{SA}} &
                  \multicolumn{3}{c|}{\textit{HC14}} &
                  \multicolumn{3}{c|}{\textit{AP}} &
                  \multicolumn{3}{c|}{\textit{SA}} &
                  \multicolumn{3}{c|}{\textit{HC14}} &
                  \multicolumn{3}{c|}{\textit{AP}} \\ \hline
                  
                \textbf{k} &
                  \multicolumn{1}{c|}{\textbf{C}} &
                  \multicolumn{1}{c|}{\textbf{T}} &
                  \multicolumn{1}{c|}{\textbf{B}} &
                  \multicolumn{1}{c|}{\textbf{C}} &
                  \multicolumn{1}{c|}{\textbf{T}} &
                  \multicolumn{1}{c|}{\textbf{B}} &
                  \multicolumn{1}{c|}{\textbf{C}} &
                  \multicolumn{1}{c|}{\textbf{T}} &
                  \multicolumn{1}{c|}{\textbf{B}} &
                  \multicolumn{1}{c|}{\textbf{C}} &
                  \multicolumn{1}{c|}{\textbf{T}} &
                  \multicolumn{1}{c|}{\textbf{B}} &
                  \multicolumn{1}{c|}{\textbf{C}} &
                  \multicolumn{1}{c|}{\textbf{T}} &
                  \multicolumn{1}{c|}{\textbf{B}} &
                  \multicolumn{1}{c|}{\textbf{C}} &
                  \multicolumn{1}{c|}{\textbf{T}} &
                  \multicolumn{1}{c|}{\textbf{B}} \\ \hline
                  
                \textbf{4} &
                  \multicolumn{1}{c|}{Y} &
                  \multicolumn{1}{c|}{Y} &
                  \multicolumn{1}{c|}{DP2} &
                  \multicolumn{1}{c|}{Y} &
                  \multicolumn{1}{c|}{Y} &
                  \multicolumn{1}{c|}{DP2} &
                  \multicolumn{1}{c|}{Y} &
                  \multicolumn{1}{c|}{Y} &
                  \multicolumn{1}{c|}{DP3} &
                  \multicolumn{1}{c|}{4.486E-04} &
                  \multicolumn{1}{c|}{1.488E-03} &
                  \multicolumn{1}{c|}{DP2} &
                  \multicolumn{1}{c|}{1.782E-05} &
                  \multicolumn{1}{c|}{6.323E-05} &
                  \multicolumn{1}{c|}{DP2} &
                  \multicolumn{1}{c|}{3.971E-07} &
                  \multicolumn{1}{c|}{6.687E-06} & 
                  \multicolumn{1}{c|}{DP3} \\ \hline 
                  
                \textbf{8} &
                  \multicolumn{1}{c|}{Y} &
                  \multicolumn{1}{c|}{Y} &
                  \multicolumn{1}{c|}{DP2} &
                  \multicolumn{1}{c|}{Y} &
                  \multicolumn{1}{c|}{Y} &
                  \multicolumn{1}{c|}{DP2} &
                  \multicolumn{1}{c|}{Y} &
                  \multicolumn{1}{c|}{Y} &
                  \multicolumn{1}{c|}{DP2} &
                  \multicolumn{1}{c|}{5.706E-02} &
                  \multicolumn{1}{c|}{4.749E-01} &
                  \multicolumn{1}{c|}{DP2} &
                  \multicolumn{1}{c|}{2.059E-03} &
                  \multicolumn{1}{c|}{2.118E-02} &
                  \multicolumn{1}{c|}{DP2} &
                  \multicolumn{1}{c|}{4.999E-05} &
                  \multicolumn{1}{c|}{6.777E-04} & 
                  \multicolumn{1}{c|}{DP2} \\ \hline 
                   
                \textbf{16} &
                  \multicolumn{1}{c|}{Y} &
                  \multicolumn{1}{c|}{Y} &
                  \multicolumn{1}{c|}{DP3} &
                  \multicolumn{1}{c|}{N} &
                  \multicolumn{1}{c|}{Y} &
                  \multicolumn{1}{c|}{DP3} &
                  \multicolumn{1}{c|}{Y} &
                  \multicolumn{1}{c|}{Y} &
                  \multicolumn{1}{c|}{DP2} &
                  \multicolumn{1}{c|}{7.529E-01} &
                  \multicolumn{1}{c|}{7.130E-01} &
                  \multicolumn{1}{c|}{DP3} &
                  \multicolumn{1}{c|}{1.370E+00} &
                  \multicolumn{1}{c|}{3.811E-01} &
                  \multicolumn{1}{c|}{DP3} &
                  \multicolumn{1}{c|}{7.145E-01} &
                  \multicolumn{1}{c|}{5.921E-02} & 
                  \multicolumn{1}{c|}{DP2} \\ \hline 
                   
                \textbf{32} &
                  \multicolumn{1}{c|}{Y} &
                  \multicolumn{1}{c|}{Y} &
                  \multicolumn{1}{c|}{DP3} &
                  \multicolumn{1}{c|}{Y} &
                  \multicolumn{1}{c|}{Y} &
                  \multicolumn{1}{c|}{DP3} &
                  \multicolumn{1}{c|}{Y} &
                  \multicolumn{1}{c|}{Y} &
                  \multicolumn{1}{c|}{DP3} &
                  \multicolumn{1}{c|}{5.966E-01} &
                  \multicolumn{1}{c|}{6.296E-01} &
                  \multicolumn{1}{c|}{DP3} &
                  \multicolumn{1}{c|}{6.232E-01} &
                  \multicolumn{1}{c|}{5.331E-01} &
                  \multicolumn{1}{c|}{DP3} &
                  \multicolumn{1}{c|}{8.947E-01} &
                  \multicolumn{1}{c|}{3.243E-01} & 
                  \multicolumn{1}{c|}{DP3} \\ \hline 
                   
                \textbf{48} &
                  \multicolumn{1}{c|}{Y} &
                  \multicolumn{1}{c|}{Y} &
                  \multicolumn{1}{c|}{DP3} &
                  \multicolumn{1}{c|}{Y} &
                  \multicolumn{1}{c|}{Y} &
                  \multicolumn{1}{c|}{DP3} &
                  \multicolumn{1}{c|}{-} &
                  \multicolumn{1}{c|}{-} &
                  \multicolumn{1}{c|}{-} &
                  \multicolumn{1}{c|}{5.935E-01} &
                  \multicolumn{1}{c|}{7.745E-01} &
                  \multicolumn{1}{c|}{DP3} &
                  \multicolumn{1}{c|}{5.833E-01} &
                  \multicolumn{1}{c|}{5.875E-01} &
                  \multicolumn{1}{c|}{DP3} &
                  \multicolumn{1}{c|}{-} &
                  \multicolumn{1}{c|}{-} & 
                  \multicolumn{1}{c|}{-} \\ \hline 
                   
                \textbf{64} &
                  \multicolumn{1}{c|}{Y} &
                  \multicolumn{1}{c|}{Y} &
                  \multicolumn{1}{c|}{DP3} &
                  \multicolumn{1}{c|}{Y} &
                  \multicolumn{1}{c|}{Y} &
                  \multicolumn{1}{c|}{DP3} &
                  \multicolumn{1}{c|}{-} &
                  \multicolumn{1}{c|}{-} &
                  \multicolumn{1}{c|}{-} &
                  \multicolumn{1}{c|}{5.927E-01} &
                  \multicolumn{1}{c|}{7.188E-01} &
                  \multicolumn{1}{c|}{DP3} &
                  \multicolumn{1}{c|}{5.689E-01} &
                  \multicolumn{1}{c|}{5.198E-01} &
                  \multicolumn{1}{c|}{DP3} &
                  \multicolumn{1}{c|}{-} &
                  \multicolumn{1}{c|}{-} & 
                  \multicolumn{1}{c|}{-} \\ \hline 
                   
                \end{tabular}
            }
            \vspace{5pt}
            \caption{\textbf{Synopsis of the CD-NRAM Scenarios Vs. The Base Case Scenario with lz4 - Favour Compression.} The table legend is as in Table \ref{tab:CD-NRAM_Vs_BaseCase_MFC_Compr}.}
            \label{tab:CD-NRAM_Vs_BaseCase_lz4_Compr}
        \end{table*}

        %TAB.9new (TAB.X21.old, TAB.X22.old)
        \begin{table*}[]
            \centering
            \resizebox{.8\textwidth}{!}{%             
                \begin{tabular}{|c|ccccccccc|ccccccccc|}
                \hline
                 &
                  \multicolumn{3}{c|}{\textit{SA}} &
                  \multicolumn{3}{c|}{\textit{HC14}} &
                  \multicolumn{3}{c|}{\textit{AP}} &
                  \multicolumn{3}{c|}{\textit{SA}} &
                  \multicolumn{3}{c|}{\textit{HC14}} &
                  \multicolumn{3}{c|}{\textit{AP}} \\ \hline
                  
                \textbf{k} &
                  \multicolumn{1}{c|}{\textbf{C}} &
                  \multicolumn{1}{c|}{\textbf{T}} &
                  \multicolumn{1}{c|}{\textbf{B}} &
                  \multicolumn{1}{c|}{\textbf{C}} &
                  \multicolumn{1}{c|}{\textbf{T}} &
                  \multicolumn{1}{c|}{\textbf{B}} &
                  \multicolumn{1}{c|}{\textbf{C}} &
                  \multicolumn{1}{c|}{\textbf{T}} &
                  \multicolumn{1}{c|}{\textbf{B}} &
                  \multicolumn{1}{c|}{\textbf{C}} &
                  \multicolumn{1}{c|}{\textbf{T}} &
                  \multicolumn{1}{c|}{\textbf{B}} &
                  \multicolumn{1}{c|}{\textbf{C}} &
                  \multicolumn{1}{c|}{\textbf{T}} &
                  \multicolumn{1}{c|}{\textbf{B}} &
                  \multicolumn{1}{c|}{\textbf{C}} &
                  \multicolumn{1}{c|}{\textbf{T}} &
                  \multicolumn{1}{c|}{\textbf{B}} \\ \hline
                  
                \textbf{4} &
                  \multicolumn{1}{c|}{Y} &
                  \multicolumn{1}{c|}{Y} &
                  \multicolumn{1}{c|}{DP2} &
                  \multicolumn{1}{c|}{Y} &
                  \multicolumn{1}{c|}{Y} &
                  \multicolumn{1}{c|}{DP2} &
                  \multicolumn{1}{c|}{Y} &
                  \multicolumn{1}{c|}{Y} &
                  \multicolumn{1}{c|}{DP3} &
                  \multicolumn{1}{c|}{7.171E-04} &
                  \multicolumn{1}{c|}{1.356E-03} &
                  \multicolumn{1}{c|}{DP2} &
                  \multicolumn{1}{c|}{2.815E-05} &
                  \multicolumn{1}{c|}{6.306E-05} &
                  \multicolumn{1}{c|}{DP2} &
                  \multicolumn{1}{c|}{6.523E-07} &
                  \multicolumn{1}{c|}{6.676E-06} & 
                  \multicolumn{1}{c|}{DP3} \\ \hline 
                  
                \textbf{8} &
                  \multicolumn{1}{c|}{Y} &
                  \multicolumn{1}{c|}{Y} &
                  \multicolumn{1}{c|}{DP2} &
                  \multicolumn{1}{c|}{Y} &
                  \multicolumn{1}{c|}{Y} &
                  \multicolumn{1}{c|}{DP2} &
                  \multicolumn{1}{c|}{Y} &
                  \multicolumn{1}{c|}{Y} &
                  \multicolumn{1}{c|}{DP2} &
                  \multicolumn{1}{c|}{9.121E-02} &
                  \multicolumn{1}{c|}{4.401E-01} &
                  \multicolumn{1}{c|}{DP2} &
                  \multicolumn{1}{c|}{3.251E-03} &
                  \multicolumn{1}{c|}{2.112E-02} &
                  \multicolumn{1}{c|}{DP2} &
                  \multicolumn{1}{c|}{8.212E-05} &
                  \multicolumn{1}{c|}{6.762E-04} & 
                  \multicolumn{1}{c|}{DP2} \\ \hline 
                   
                \textbf{16} &
                  \multicolumn{1}{c|}{N} &
                  \multicolumn{1}{c|}{Y} &
                  \multicolumn{1}{c|}{DP3} &
                  \multicolumn{1}{c|}{N} &
                  \multicolumn{1}{c|}{Y} &
                  \multicolumn{1}{c|}{DP3} &
                  \multicolumn{1}{c|}{N} &
                  \multicolumn{1}{c|}{Y} &
                  \multicolumn{1}{c|}{DP2} &
                  \multicolumn{1}{c|}{1.204E+00} &
                  \multicolumn{1}{c|}{6.945E-01} &
                  \multicolumn{1}{c|}{DP3} &
                  \multicolumn{1}{c|}{2.163E+00} &
                  \multicolumn{1}{c|}{3.807E-01} &
                  \multicolumn{1}{c|}{DP3} &
                  \multicolumn{1}{c|}{1.174E+00} &
                  \multicolumn{1}{c|}{5.915E-02} & 
                  \multicolumn{1}{c|}{DP2} \\ \hline 
                   
                \textbf{32} &
                  \multicolumn{1}{c|}{Y} &
                  \multicolumn{1}{c|}{Y} &
                  \multicolumn{1}{c|}{DP3} &
                  \multicolumn{1}{c|}{Y} &
                  \multicolumn{1}{c|}{Y} &
                  \multicolumn{1}{c|}{DP3} &
                  \multicolumn{1}{c|}{N} &
                  \multicolumn{1}{c|}{Y} &
                  \multicolumn{1}{c|}{DP3} &
                  \multicolumn{1}{c|}{9.537E-01} &
                  \multicolumn{1}{c|}{6.163E-01} &
                  \multicolumn{1}{c|}{DP3} &
                  \multicolumn{1}{c|}{9.842E-01} &
                  \multicolumn{1}{c|}{5.327E-01} &
                  \multicolumn{1}{c|}{DP3} &
                  \multicolumn{1}{c|}{1.470E+00} &
                  \multicolumn{1}{c|}{3.243E-01} & 
                  \multicolumn{1}{c|}{DP3} \\ \hline 
                   
                \textbf{48} &
                  \multicolumn{1}{c|}{Y} &
                  \multicolumn{1}{c|}{Y} &
                  \multicolumn{1}{c|}{DP3} &
                  \multicolumn{1}{c|}{Y} &
                  \multicolumn{1}{c|}{Y} &
                  \multicolumn{1}{c|}{DP3} &
                  \multicolumn{1}{c|}{-} &
                  \multicolumn{1}{c|}{-} &
                  \multicolumn{1}{c|}{-} &
                  \multicolumn{1}{c|}{9.488E-01} &
                  \multicolumn{1}{c|}{7.574E-01} &
                  \multicolumn{1}{c|}{DP3} &
                  \multicolumn{1}{c|}{9.212E-01} &
                  \multicolumn{1}{c|}{5.871E-01} &
                  \multicolumn{1}{c|}{DP3} &
                  \multicolumn{1}{c|}{-} &
                  \multicolumn{1}{c|}{-} & 
                  \multicolumn{1}{c|}{-} \\ \hline 
                   
                \textbf{64} &
                  \multicolumn{1}{c|}{Y} &
                  \multicolumn{1}{c|}{Y} &
                  \multicolumn{1}{c|}{DP3} &
                  \multicolumn{1}{c|}{Y} &
                  \multicolumn{1}{c|}{Y} &
                  \multicolumn{1}{c|}{DP3} &
                  \multicolumn{1}{c|}{-} &
                  \multicolumn{1}{c|}{-} &
                  \multicolumn{1}{c|}{-} &
                  \multicolumn{1}{c|}{9.475E-01} &
                  \multicolumn{1}{c|}{7.047E-01} &
                  \multicolumn{1}{c|}{DP3} &
                  \multicolumn{1}{c|}{8.985E-01} &
                  \multicolumn{1}{c|}{5.195E-01} &
                  \multicolumn{1}{c|}{DP3} &
                  \multicolumn{1}{c|}{-} &
                  \multicolumn{1}{c|}{-} & 
                  \multicolumn{1}{c|}{-} \\ \hline 
                   
                \end{tabular}
            }
        \vspace{5pt}    
        \caption{\textbf{Synopsis of the CD-NRAM Scenarios Vs. The Base Case Scenario with SPRING - Favour Compression.} The table legend is as in Table \ref{tab:CD-NRAM_Vs_BaseCase_MFC_Compr}.}
        \label{tab:CD-NRAM_Vs_BaseCase_SPRING_Compr}
        \end{table*}

\begin{table*}[]
            \centering
            \resizebox{.8\textwidth}{!}{%             
                \begin{tabular}{|c|ccccccccc|ccccccccc|}
                \hline
                 &
                  \multicolumn{3}{c|}{\textit{SA}} &
                  \multicolumn{3}{c|}{\textit{HC14}} &
                  \multicolumn{3}{c|}{\textit{AP}} &
                  \multicolumn{3}{c|}{\textit{SA}} &
                  \multicolumn{3}{c|}{\textit{HC14}} &
                  \multicolumn{3}{c|}{\textit{AP}} \\ \hline
                  
                \textbf{k} &
                  \multicolumn{1}{c|}{\textbf{C}} &
                  \multicolumn{1}{c|}{\textbf{T}} &
                  \multicolumn{1}{c|}{\textbf{B}} &
                  \multicolumn{1}{c|}{\textbf{C}} &
                  \multicolumn{1}{c|}{\textbf{T}} &
                  \multicolumn{1}{c|}{\textbf{B}} &
                  \multicolumn{1}{c|}{\textbf{C}} &
                  \multicolumn{1}{c|}{\textbf{T}} &
                  \multicolumn{1}{c|}{\textbf{B}} &
                  \multicolumn{1}{c|}{\textbf{C}} &
                  \multicolumn{1}{c|}{\textbf{T}} &
                  \multicolumn{1}{c|}{\textbf{B}} &
                  \multicolumn{1}{c|}{\textbf{C}} &
                  \multicolumn{1}{c|}{\textbf{T}} &
                  \multicolumn{1}{c|}{\textbf{B}} &
                  \multicolumn{1}{c|}{\textbf{C}} &
                  \multicolumn{1}{c|}{\textbf{T}} &
                  \multicolumn{1}{c|}{\textbf{B}} \\ \hline
                  
                \textbf{4} &
                  \multicolumn{1}{c|}{Y} &
                  \multicolumn{1}{c|}{Y} &
                  \multicolumn{1}{c|}{DP1} &
                  \multicolumn{1}{c|}{Y} &
                  \multicolumn{1}{c|}{Y} &
                  \multicolumn{1}{c|}{DP1} &
                  \multicolumn{1}{c|}{Y} &
                  \multicolumn{1}{c|}{Y} &
                  \multicolumn{1}{c|}{DP2} &
                  \multicolumn{1}{c|}{1.184E-03} &
                  \multicolumn{1}{c|}{1.124E-03} &
                  \multicolumn{1}{c|}{DP1} &
                  \multicolumn{1}{c|}{5.064E-05} &
                  \multicolumn{1}{c|}{2.621E-05} &
                  \multicolumn{1}{c|}{DP1} &
                  \multicolumn{1}{c|}{7.065E-07} &
                  \multicolumn{1}{c|}{1.263E-06} & 
                  \multicolumn{1}{c|}{DP2} \\ \hline 
                  
                \textbf{8} &
                  \multicolumn{1}{c|}{Y} &
                  \multicolumn{1}{c|}{Y} &
                  \multicolumn{1}{c|}{DP0} &
                  \multicolumn{1}{c|}{Y} &
                  \multicolumn{1}{c|}{Y} &
                  \multicolumn{1}{c|}{DP1} &
                  \multicolumn{1}{c|}{Y} &
                  \multicolumn{1}{c|}{Y} &
                  \multicolumn{1}{c|}{DP1} &
                  \multicolumn{1}{c|}{2.145E-01} &
                  \multicolumn{1}{c|}{1.824E-03} &
                  \multicolumn{1}{c|}{DP0} &
                  \multicolumn{1}{c|}{6.137E-03} &
                  \multicolumn{1}{c|}{7.630E-05} &
                  \multicolumn{1}{c|}{DP1} &
                  \multicolumn{1}{c|}{1.613E-04} &
                  \multicolumn{1}{c|}{2.449E-06} & 
                  \multicolumn{1}{c|}{DP1} \\ \hline 
                   
                \textbf{16} &
                  \multicolumn{1}{c|}{N} &
                  \multicolumn{1}{c|}{Y} &
                  \multicolumn{1}{c|}{DP0} &
                  \multicolumn{1}{c|}{N} &
                  \multicolumn{1}{c|}{Y} &
                  \multicolumn{1}{c|}{DP2} &
                  \multicolumn{1}{c|}{N} &
                  \multicolumn{1}{c|}{Y} &
                  \multicolumn{1}{c|}{DP2} &
                  \multicolumn{1}{c|}{8.446E+00} &
                  \multicolumn{1}{c|}{1.774E-02} &
                  \multicolumn{1}{c|}{DP0} &
                  \multicolumn{1}{c|}{4.238E+00} &
                  \multicolumn{1}{c|}{1.694E-02} &
                  \multicolumn{1}{c|}{DP2} &
                  \multicolumn{1}{c|}{1.134E+00} &
                  \multicolumn{1}{c|}{2.363E-02} & 
                  \multicolumn{1}{c|}{DP2} \\ \hline 
                   
                \textbf{32} &
                  \multicolumn{1}{c|}{N} &
                  \multicolumn{1}{c|}{Y} &
                  \multicolumn{1}{c|}{DP0} &
                  \multicolumn{1}{c|}{N} &
                  \multicolumn{1}{c|}{Y} &
                  \multicolumn{1}{c|}{DP1} &
                  \multicolumn{1}{c|}{N} &
                  \multicolumn{1}{c|}{Y} &
                  \multicolumn{1}{c|}{DP1} &
                  \multicolumn{1}{c|}{1.673E+01} &
                  \multicolumn{1}{c|}{3.766E-02} &
                  \multicolumn{1}{c|}{DP0} &
                  \multicolumn{1}{c|}{1.925E+01} &
                  \multicolumn{1}{c|}{2.641E-02} &
                  \multicolumn{1}{c|}{DP1} &
                  \multicolumn{1}{c|}{1.328E+01} &
                  \multicolumn{1}{c|}{1.071E-01} & 
                  \multicolumn{1}{c|}{DP1} \\ \hline 
                   
                \textbf{48} &
                  \multicolumn{1}{c|}{N} &
                  \multicolumn{1}{c|}{Y} &
                  \multicolumn{1}{c|}{DP0} &
                  \multicolumn{1}{c|}{N} &
                  \multicolumn{1}{c|}{Y} &
                  \multicolumn{1}{c|}{DP0} &
                  \multicolumn{1}{c|}{-} &
                  \multicolumn{1}{c|}{-} &
                  \multicolumn{1}{c|}{-} &
                  \multicolumn{1}{c|}{1.912E+01} &
                  \multicolumn{1}{c|}{4.701E-02} &
                  \multicolumn{1}{c|}{DP0} &
                  \multicolumn{1}{c|}{3.518E+01} &
                  \multicolumn{1}{c|}{5.492E-02} &
                  \multicolumn{1}{c|}{DP0} &
                  \multicolumn{1}{c|}{-} &
                  \multicolumn{1}{c|}{-} & 
                  \multicolumn{1}{c|}{-} \\ \hline 
                   
                \textbf{64} &
                  \multicolumn{1}{c|}{N} &
                  \multicolumn{1}{c|}{Y} &
                  \multicolumn{1}{c|}{DP0} &
                  \multicolumn{1}{c|}{N} &
                  \multicolumn{1}{c|}{Y} &
                  \multicolumn{1}{c|}{DP2} &
                  \multicolumn{1}{c|}{-} &
                  \multicolumn{1}{c|}{-} &
                  \multicolumn{1}{c|}{-} &
                  \multicolumn{1}{c|}{2.126E+01} &
                  \multicolumn{1}{c|}{5.475E-02} &
                  \multicolumn{1}{c|}{DP0} &
                  \multicolumn{1}{c|}{4.948E+01} &
                  \multicolumn{1}{c|}{4.921E-02} &
                  \multicolumn{1}{c|}{DP2} &
                  \multicolumn{1}{c|}{-} &
                  \multicolumn{1}{c|}{-} & 
                  \multicolumn{1}{c|}{-} \\ \hline 
                   
                \end{tabular}
            }
        \vspace{5pt}    
        \caption{\textbf{Synopsis of the CD-NRAM Scenarios Vs. The Base Case Scenario with zstd - Favour Post-Processing. } The values of $k$ we have considered are reported in the first column of the table, which is divided into two panels: {\bf Logic Indication}  and {\bf Numeric Values}. For the first panel, the first row indicates the dataset. Dataset abbreviations are as in Table \ref{tab:SD-RAM_Favour_Compr_PostProc}. For each of them, the three columns within the dataset indicate the following. Column \textbf{B} indicates the best \textbf{Compressed on Disk Scenario} for the Case favouring post-processing time. Columns \textbf{C} and \textbf{T} indicate, respectively, the compression ratio and the post-processing time ratio between the \textbf{Compressed on Disk Scenario}, indicated in the column \textbf{B}, and the corresponding \textbf{Base Case Scenario} with \textbf{zstd} compressor, as specified in the Main text. Y means that the \textbf{Compressed on Disk Scenario} is more convenient than the {\bf Base Case Scenario} and N its complement.  An entry with a dash indicates that the experiment could not be completed, as explained in the Main text (see {\bf{Experimental Setup}}). As for the second panel, its entries are analogous to the ones of the first one, except that the numeric values of the mentioned ratios are reported.}
        \label{tab:CD-NRAM_Vs_BaseCase_zstd_PostProc}
        \end{table*}

        %TAB.10new (TAB.X7.old, TAB.X8.old)
        \begin{table*}[]
            \centering
            \resizebox{.8\textwidth}{!}{%             
                \begin{tabular}{|c|ccccccccc|ccccccccc|}
                \hline
                 &
                  \multicolumn{3}{c|}{\textit{SA}} &
                  \multicolumn{3}{c|}{\textit{HC14}} &
                  \multicolumn{3}{c|}{\textit{AP}} &
                  \multicolumn{3}{c|}{\textit{SA}} &
                  \multicolumn{3}{c|}{\textit{HC14}} &
                  \multicolumn{3}{c|}{\textit{AP}} \\ \hline
                  
                \textbf{k} &
                  \multicolumn{1}{c|}{\textbf{C}} &
                  \multicolumn{1}{c|}{\textbf{T}} &
                  \multicolumn{1}{c|}{\textbf{B}} &
                  \multicolumn{1}{c|}{\textbf{C}} &
                  \multicolumn{1}{c|}{\textbf{T}} &
                  \multicolumn{1}{c|}{\textbf{B}} &
                  \multicolumn{1}{c|}{\textbf{C}} &
                  \multicolumn{1}{c|}{\textbf{T}} &
                  \multicolumn{1}{c|}{\textbf{B}} &
                  \multicolumn{1}{c|}{\textbf{C}} &
                  \multicolumn{1}{c|}{\textbf{T}} &
                  \multicolumn{1}{c|}{\textbf{B}} &
                  \multicolumn{1}{c|}{\textbf{C}} &
                  \multicolumn{1}{c|}{\textbf{T}} &
                  \multicolumn{1}{c|}{\textbf{B}} &
                  \multicolumn{1}{c|}{\textbf{C}} &
                  \multicolumn{1}{c|}{\textbf{T}} &
                  \multicolumn{1}{c|}{\textbf{B}} \\ \hline
                  
                \textbf{4} &
                  \multicolumn{1}{c|}{Y} &
                  \multicolumn{1}{c|}{Y} &
                  \multicolumn{1}{c|}{DP1} &
                  \multicolumn{1}{c|}{Y} &
                  \multicolumn{1}{c|}{Y} &
                  \multicolumn{1}{c|}{DP1} &
                  \multicolumn{1}{c|}{Y} &
                  \multicolumn{1}{c|}{Y} &
                  \multicolumn{1}{c|}{DP2} &
                  \multicolumn{1}{c|}{1.411E-03} &
                  \multicolumn{1}{c|}{4.812E-04} &
                  \multicolumn{1}{c|}{DP1} &
                  \multicolumn{1}{c|}{6.044E-05} &
                  \multicolumn{1}{c|}{1.682E-05} &
                  \multicolumn{1}{c|}{DP1} &
                  \multicolumn{1}{c|}{8.573E-07} &
                  \multicolumn{1}{c|}{8.617E-07} & 
                  \multicolumn{1}{c|}{DP2} \\ \hline 
                  
                \textbf{8} &
                  \multicolumn{1}{c|}{Y} &
                  \multicolumn{1}{c|}{Y} &
                  \multicolumn{1}{c|}{DP0} &
                  \multicolumn{1}{c|}{Y} &
                  \multicolumn{1}{c|}{Y} &
                  \multicolumn{1}{c|}{DP1} &
                  \multicolumn{1}{c|}{Y} &
                  \multicolumn{1}{c|}{Y} &
                  \multicolumn{1}{c|}{DP1} &
                  \multicolumn{1}{c|}{2.556E-01} &
                  \multicolumn{1}{c|}{8.751E-04} &
                  \multicolumn{1}{c|}{DP0} &
                  \multicolumn{1}{c|}{7.325E-03} &
                  \multicolumn{1}{c|}{4.950E-05} &
                  \multicolumn{1}{c|}{DP1} &
                  \multicolumn{1}{c|}{1.957E-04} &
                  \multicolumn{1}{c|}{1.528E-06} & 
                  \multicolumn{1}{c|}{DP1} \\ \hline 
                   
                \textbf{16} &
                  \multicolumn{1}{c|}{N} &
                  \multicolumn{1}{c|}{Y} &
                  \multicolumn{1}{c|}{DP0} &
                  \multicolumn{1}{c|}{N} &
                  \multicolumn{1}{c|}{Y} &
                  \multicolumn{1}{c|}{DP2} &
                  \multicolumn{1}{c|}{N} &
                  \multicolumn{1}{c|}{Y} &
                  \multicolumn{1}{c|}{DP2} &
                  \multicolumn{1}{c|}{1.006E+01} &
                  \multicolumn{1}{c|}{1.301E-02} &
                  \multicolumn{1}{c|}{DP0} &
                  \multicolumn{1}{c|}{5.058E+00} &
                  \multicolumn{1}{c|}{1.378E-02} &
                  \multicolumn{1}{c|}{DP2} &
                  \multicolumn{1}{c|}{1.376E+00} &
                  \multicolumn{1}{c|}{1.827E-02} & 
                  \multicolumn{1}{c|}{DP2} \\ \hline 
                   
                \textbf{32} &
                  \multicolumn{1}{c|}{N} &
                  \multicolumn{1}{c|}{Y} &
                  \multicolumn{1}{c|}{DP0} &
                  \multicolumn{1}{c|}{N} &
                  \multicolumn{1}{c|}{Y} &
                  \multicolumn{1}{c|}{DP1} &
                  \multicolumn{1}{c|}{N} &
                  \multicolumn{1}{c|}{Y} &
                  \multicolumn{1}{c|}{DP1} &
                  \multicolumn{1}{c|}{1.994E+01} &
                  \multicolumn{1}{c|}{2.909E-02} &
                  \multicolumn{1}{c|}{DP0} &
                  \multicolumn{1}{c|}{2.298E+01} &
                  \multicolumn{1}{c|}{2.289E-02} &
                  \multicolumn{1}{c|}{DP1} &
                  \multicolumn{1}{c|}{1.612E+01} &
                  \multicolumn{1}{c|}{1.004E-01} & 
                  \multicolumn{1}{c|}{DP1} \\ \hline 
                   
                \textbf{48} &
                  \multicolumn{1}{c|}{N} &
                  \multicolumn{1}{c|}{Y} &
                  \multicolumn{1}{c|}{DP0} &
                  \multicolumn{1}{c|}{N} &
                  \multicolumn{1}{c|}{Y} &
                  \multicolumn{1}{c|}{DP0} &
                  \multicolumn{1}{c|}{-} &
                  \multicolumn{1}{c|}{-} &
                  \multicolumn{1}{c|}{-} &
                  \multicolumn{1}{c|}{2.278E+01} &
                  \multicolumn{1}{c|}{3.594E-02} &
                  \multicolumn{1}{c|}{DP0} &
                  \multicolumn{1}{c|}{4.199E+01} &
                  \multicolumn{1}{c|}{4.832E-02} &
                  \multicolumn{1}{c|}{DP0} &
                  \multicolumn{1}{c|}{-} &
                  \multicolumn{1}{c|}{-} & 
                  \multicolumn{1}{c|}{-} \\ \hline 
                   
                \textbf{64} &
                  \multicolumn{1}{c|}{N} &
                  \multicolumn{1}{c|}{Y} &
                  \multicolumn{1}{c|}{DP0} &
                  \multicolumn{1}{c|}{N} &
                  \multicolumn{1}{c|}{Y} &
                  \multicolumn{1}{c|}{DP2} &
                  \multicolumn{1}{c|}{-} &
                  \multicolumn{1}{c|}{-} &
                  \multicolumn{1}{c|}{-} &
                  \multicolumn{1}{c|}{2.534E+01} &
                  \multicolumn{1}{c|}{4.297E-02} &
                  \multicolumn{1}{c|}{DP0} &
                  \multicolumn{1}{c|}{5.905E+01} &
                  \multicolumn{1}{c|}{4.441E-02} &
                  \multicolumn{1}{c|}{DP2} &
                  \multicolumn{1}{c|}{-} &
                  \multicolumn{1}{c|}{-} & 
                  \multicolumn{1}{c|}{-} \\ \hline 
                   
                \end{tabular}
            }
        \vspace{5pt}    
        \caption{\textbf{Synopsis of the CD-NRAM Scenarios Vs. The Base Case Scenario with MFC - Favour Post-Processing.}  The table legend is as in Table \ref{tab:CD-NRAM_Vs_BaseCase_zstd_PostProc}.}
        \label{tab:CD-NRAM_Vs_BaseCase_MFC_PostProc}
        \end{table*}

        %TAB.11new (TAB.X11.old, TAB.X12.old)

        %TAB.12new (TAB.X15.old, TAB.X16.old)
        \begin{table*}[]
            \centering
            \resizebox{.8\textwidth}{!}{%             
                \begin{tabular}{|c|ccccccccc|ccccccccc|}
                \hline
                 &
                  \multicolumn{3}{c|}{\textit{SA}} &
                  \multicolumn{3}{c|}{\textit{HC14}} &
                  \multicolumn{3}{c|}{\textit{AP}} &
                  \multicolumn{3}{c|}{\textit{SA}} &
                  \multicolumn{3}{c|}{\textit{HC14}} &
                  \multicolumn{3}{c|}{\textit{AP}} \\ \hline
                  
                \textbf{k} &
                  \multicolumn{1}{c|}{\textbf{C}} &
                  \multicolumn{1}{c|}{\textbf{T}} &
                  \multicolumn{1}{c|}{\textbf{B}} &
                  \multicolumn{1}{c|}{\textbf{C}} &
                  \multicolumn{1}{c|}{\textbf{T}} &
                  \multicolumn{1}{c|}{\textbf{B}} &
                  \multicolumn{1}{c|}{\textbf{C}} &
                  \multicolumn{1}{c|}{\textbf{T}} &
                  \multicolumn{1}{c|}{\textbf{B}} &
                  \multicolumn{1}{c|}{\textbf{C}} &
                  \multicolumn{1}{c|}{\textbf{T}} &
                  \multicolumn{1}{c|}{\textbf{B}} &
                  \multicolumn{1}{c|}{\textbf{C}} &
                  \multicolumn{1}{c|}{\textbf{T}} &
                  \multicolumn{1}{c|}{\textbf{B}} &
                  \multicolumn{1}{c|}{\textbf{C}} &
                  \multicolumn{1}{c|}{\textbf{T}} &
                  \multicolumn{1}{c|}{\textbf{B}} \\ \hline
                  
                \textbf{4} &
                  \multicolumn{1}{c|}{Y} &
                  \multicolumn{1}{c|}{Y} &
                  \multicolumn{1}{c|}{DP1} &
                  \multicolumn{1}{c|}{Y} &
                  \multicolumn{1}{c|}{Y} &
                  \multicolumn{1}{c|}{DP1} &
                  \multicolumn{1}{c|}{Y} &
                  \multicolumn{1}{c|}{Y} &
                  \multicolumn{1}{c|}{DP2} &
                  \multicolumn{1}{c|}{1.120E-03} &
                  \multicolumn{1}{c|}{1.025E-03} &
                  \multicolumn{1}{c|}{DP1} &
                  \multicolumn{1}{c|}{4.667E-05} &
                  \multicolumn{1}{c|}{2.361E-05} &
                  \multicolumn{1}{c|}{DP1} &
                  \multicolumn{1}{c|}{6.083E-07} &
                  \multicolumn{1}{c|}{1.140E-06} & 
                  \multicolumn{1}{c|}{DP2} \\ \hline 
                  
                \textbf{8} &
                  \multicolumn{1}{c|}{Y} &
                  \multicolumn{1}{c|}{Y} &
                  \multicolumn{1}{c|}{DP0} &
                  \multicolumn{1}{c|}{Y} &
                  \multicolumn{1}{c|}{Y} &
                  \multicolumn{1}{c|}{DP1} &
                  \multicolumn{1}{c|}{Y} &
                  \multicolumn{1}{c|}{Y} &
                  \multicolumn{1}{c|}{DP1} &
                  \multicolumn{1}{c|}{2.029E-01} &
                  \multicolumn{1}{c|}{1.691E-03} &
                  \multicolumn{1}{c|}{DP0} &
                  \multicolumn{1}{c|}{5.657E-03} &
                  \multicolumn{1}{c|}{6.894E-05} &
                  \multicolumn{1}{c|}{DP1} &
                  \multicolumn{1}{c|}{1.388E-04} &
                  \multicolumn{1}{c|}{2.148E-06} & 
                  \multicolumn{1}{c|}{DP1} \\ \hline 
                   
                \textbf{16} &
                  \multicolumn{1}{c|}{N} &
                  \multicolumn{1}{c|}{Y} &
                  \multicolumn{1}{c|}{DP0} &
                  \multicolumn{1}{c|}{N} &
                  \multicolumn{1}{c|}{Y} &
                  \multicolumn{1}{c|}{DP2} &
                  \multicolumn{1}{c|}{Y} &
                  \multicolumn{1}{c|}{Y} &
                  \multicolumn{1}{c|}{DP2} &
                  \multicolumn{1}{c|}{7.989E+00} &
                  \multicolumn{1}{c|}{1.729E-02} &
                  \multicolumn{1}{c|}{DPO} &
                  \multicolumn{1}{c|}{3.906E+00} &
                  \multicolumn{1}{c|}{1.620E-02} &
                  \multicolumn{1}{c|}{DP2} &
                  \multicolumn{1}{c|}{9.762E-01} &
                  \multicolumn{1}{c|}{2.212E-02} & 
                  \multicolumn{1}{c|}{DP2} \\ \hline 
                   
                \textbf{32} &
                  \multicolumn{1}{c|}{N} &
                  \multicolumn{1}{c|}{Y} &
                  \multicolumn{1}{c|}{DP0} &
                  \multicolumn{1}{c|}{N} &
                  \multicolumn{1}{c|}{Y} &
                  \multicolumn{1}{c|}{DP1} &
                  \multicolumn{1}{c|}{N} &
                  \multicolumn{1}{c|}{Y} &
                  \multicolumn{1}{c|}{DP1} &
                  \multicolumn{1}{c|}{1.582E+01} &
                  \multicolumn{1}{c|}{3.588E-02} &
                  \multicolumn{1}{c|}{DP0} &
                  \multicolumn{1}{c|}{1.774E+01} &
                  \multicolumn{1}{c|}{2.563E-02} &
                  \multicolumn{1}{c|}{DP1} &
                  \multicolumn{1}{c|}{1.144E+01} &
                  \multicolumn{1}{c|}{1.055E-01} & 
                  \multicolumn{1}{c|}{DP1} \\ \hline 
                   
                \textbf{48} &
                  \multicolumn{1}{c|}{N} &
                  \multicolumn{1}{c|}{Y} &
                  \multicolumn{1}{c|}{DP0} &
                  \multicolumn{1}{c|}{N} &
                  \multicolumn{1}{c|}{Y} &
                  \multicolumn{1}{c|}{DP0} &
                  \multicolumn{1}{c|}{-} &
                  \multicolumn{1}{c|}{-} &
                  \multicolumn{1}{c|}{-} &
                  \multicolumn{1}{c|}{1.808E+01} &
                  \multicolumn{1}{c|}{4.598E-02} &
                  \multicolumn{1}{c|}{DP0} &
                  \multicolumn{1}{c|}{3.242E+01} &
                  \multicolumn{1}{c|}{5.348E-02} &
                  \multicolumn{1}{c|}{DP0} &
                  \multicolumn{1}{c|}{-} &
                  \multicolumn{1}{c|}{-} & 
                  \multicolumn{1}{c|}{-} \\ \hline 
                   
                \textbf{64} &
                  \multicolumn{1}{c|}{N} &
                  \multicolumn{1}{c|}{Y} &
                  \multicolumn{1}{c|}{DP0} &
                  \multicolumn{1}{c|}{N} &
                  \multicolumn{1}{c|}{Y} &
                  \multicolumn{1}{c|}{DP2} &
                  \multicolumn{1}{c|}{-} &
                  \multicolumn{1}{c|}{-} &
                  \multicolumn{1}{c|}{-} &
                  \multicolumn{1}{c|}{2.011E+01} &
                  \multicolumn{1}{c|}{5.369E-02} &
                  \multicolumn{1}{c|}{DP0} &
                  \multicolumn{1}{c|}{4.560E+01} &
                  \multicolumn{1}{c|}{4.819E-02} &
                  \multicolumn{1}{c|}{DP2} &
                  \multicolumn{1}{c|}{-} &
                  \multicolumn{1}{c|}{-} &
                  \multicolumn{1}{c|}{-} \\ \hline 
                   
                \end{tabular}
            }
            \vspace{5pt}
            \caption{\textbf{Synopsis of the CD-NRAM Scenarios Vs. The Base Case Scenario with bzip2 - Favour Post-Processing.} The table legend is as in Table \ref{tab:CD-NRAM_Vs_BaseCase_zstd_PostProc}.}
            \label{tab:CD-NRAM_Vs_BaseCase_bzip2_PostProc}
        \end{table*}

        %TAB.13new (TAB.X19.old, TAB.X20.old)
        \begin{table*}[]
            \centering
            \resizebox{.8\textwidth}{!}{%             
                \begin{tabular}{|c|ccccccccc|ccccccccc|}
                \hline
                 &
                  \multicolumn{3}{c|}{\textit{SA}} &
                  \multicolumn{3}{c|}{\textit{HC14}} &
                  \multicolumn{3}{c|}{\textit{AP}} &
                  \multicolumn{3}{c|}{\textit{SA}} &
                  \multicolumn{3}{c|}{\textit{HC14}} &
                  \multicolumn{3}{c|}{\textit{AP}} \\ \hline
                  
                \textbf{k} &
                  \multicolumn{1}{c|}{\textbf{C}} &
                  \multicolumn{1}{c|}{\textbf{T}} &
                  \multicolumn{1}{c|}{\textbf{B}} &
                  \multicolumn{1}{c|}{\textbf{C}} &
                  \multicolumn{1}{c|}{\textbf{T}} &
                  \multicolumn{1}{c|}{\textbf{B}} &
                  \multicolumn{1}{c|}{\textbf{C}} &
                  \multicolumn{1}{c|}{\textbf{T}} &
                  \multicolumn{1}{c|}{\textbf{B}} &
                  \multicolumn{1}{c|}{\textbf{C}} &
                  \multicolumn{1}{c|}{\textbf{T}} &
                  \multicolumn{1}{c|}{\textbf{B}} &
                  \multicolumn{1}{c|}{\textbf{C}} &
                  \multicolumn{1}{c|}{\textbf{T}} &
                  \multicolumn{1}{c|}{\textbf{B}} &
                  \multicolumn{1}{c|}{\textbf{C}} &
                  \multicolumn{1}{c|}{\textbf{T}} &
                  \multicolumn{1}{c|}{\textbf{B}} \\ \hline
                  
                \textbf{4} &
                  \multicolumn{1}{c|}{Y} &
                  \multicolumn{1}{c|}{Y} &
                  \multicolumn{1}{c|}{DP1} &
                  \multicolumn{1}{c|}{Y} &
                  \multicolumn{1}{c|}{Y} &
                  \multicolumn{1}{c|}{DP1} &
                  \multicolumn{1}{c|}{Y} &
                  \multicolumn{1}{c|}{Y} &
                  \multicolumn{1}{c|}{DP2} &
                  \multicolumn{1}{c|}{8.073E-04} &
                  \multicolumn{1}{c|}{1.116E-03} &
                  \multicolumn{1}{c|}{DP1} &
                  \multicolumn{1}{c|}{3.359E-05} &
                  \multicolumn{1}{c|}{2.529E-05} &
                  \multicolumn{1}{c|}{DP1} &
                  \multicolumn{1}{c|}{4.453E-07} &
                  \multicolumn{1}{c|}{1.216E-06} & 
                  \multicolumn{1}{c|}{DP2} \\ \hline 
                  
                \textbf{8} &
                  \multicolumn{1}{c|}{Y} &
                  \multicolumn{1}{c|}{Y} &
                  \multicolumn{1}{c|}{DP0} &
                  \multicolumn{1}{c|}{Y} &
                  \multicolumn{1}{c|}{Y} &
                  \multicolumn{1}{c|}{DP1} &
                  \multicolumn{1}{c|}{Y} &
                  \multicolumn{1}{c|}{Y} &
                  \multicolumn{1}{c|}{DP1} &
                  \multicolumn{1}{c|}{1.462E-01} &
                  \multicolumn{1}{c|}{1.813E-03} &
                  \multicolumn{1}{c|}{DP0} &
                  \multicolumn{1}{c|}{4.071E-03} &
                  \multicolumn{1}{c|}{7.371E-05} &
                  \multicolumn{1}{c|}{DP1} &
                  \multicolumn{1}{c|}{1.016E-04} &
                  \multicolumn{1}{c|}{2.331E-06} & 
                  \multicolumn{1}{c|}{DP1} \\ \hline 
                   
                \textbf{16} &
                  \multicolumn{1}{c|}{N} &
                  \multicolumn{1}{c|}{Y} &
                  \multicolumn{1}{c|}{DP0} &
                  \multicolumn{1}{c|}{N} &
                  \multicolumn{1}{c|}{Y} &
                  \multicolumn{1}{c|}{DP2} &
                  \multicolumn{1}{c|}{Y} &
                  \multicolumn{1}{c|}{Y} &
                  \multicolumn{1}{c|}{DP2} &
                  \multicolumn{1}{c|}{5.756E+00} &
                  \multicolumn{1}{c|}{1.771E-02} &
                  \multicolumn{1}{c|}{DP0} &
                  \multicolumn{1}{c|}{2.811E+00} &
                  \multicolumn{1}{c|}{1.669E-02} &
                  \multicolumn{1}{c|}{DP2} &
                  \multicolumn{1}{c|}{7.146E-01} &
                  \multicolumn{1}{c|}{2.306E-02} & 
                  \multicolumn{1}{c|}{DP2} \\ \hline 
                   
                \textbf{32} &
                  \multicolumn{1}{c|}{N} &
                  \multicolumn{1}{c|}{Y} &
                  \multicolumn{1}{c|}{DP0} &
                  \multicolumn{1}{c|}{N} &
                  \multicolumn{1}{c|}{Y} &
                  \multicolumn{1}{c|}{DP1} &
                  \multicolumn{1}{c|}{N} &
                  \multicolumn{1}{c|}{Y} &
                  \multicolumn{1}{c|}{DP1} &
                  \multicolumn{1}{c|}{1.140E+01} &
                  \multicolumn{1}{c|}{3.760E-02} &
                  \multicolumn{1}{c|}{DP0} &
                  \multicolumn{1}{c|}{1.277E+01} &
                  \multicolumn{1}{c|}{2.615E-02} &
                  \multicolumn{1}{c|}{DP1} &
                  \multicolumn{1}{c|}{8.371E+00} &
                  \multicolumn{1}{c|}{1.065E-01} & 
                  \multicolumn{1}{c|}{DP1} \\ \hline 
                   
                \textbf{48} &
                  \multicolumn{1}{c|}{N} &
                  \multicolumn{1}{c|}{Y} &
                  \multicolumn{1}{c|}{DP0} &
                  \multicolumn{1}{c|}{N} &
                  \multicolumn{1}{c|}{Y} &
                  \multicolumn{1}{c|}{DP0} &
                  \multicolumn{1}{c|}{-} &
                  \multicolumn{1}{c|}{-} &
                  \multicolumn{1}{c|}{-} &
                  \multicolumn{1}{c|}{1.303E+01} &
                  \multicolumn{1}{c|}{4.693E-02} &
                  \multicolumn{1}{c|}{DP0} &
                  \multicolumn{1}{c|}{2.333E+01} &
                  \multicolumn{1}{c|}{5.444E-02} &
                  \multicolumn{1}{c|}{DP0} &
                  \multicolumn{1}{c|}{-} &
                  \multicolumn{1}{c|}{-} & 
                  \multicolumn{1}{c|}{-} \\ \hline 
                   
                \textbf{64} &
                  \multicolumn{1}{c|}{N} &
                  \multicolumn{1}{c|}{Y} &
                  \multicolumn{1}{c|}{DP0} &
                  \multicolumn{1}{c|}{N} &
                  \multicolumn{1}{c|}{Y} &
                  \multicolumn{1}{c|}{DP2} &
                  \multicolumn{1}{c|}{-} &
                  \multicolumn{1}{c|}{-} &
                  \multicolumn{1}{c|}{-} &
                  \multicolumn{1}{c|}{1.449E+01} &
                  \multicolumn{1}{c|}{5.467E-02} &
                  \multicolumn{1}{c|}{DP0} &
                  \multicolumn{1}{c|}{3.282E+01} &
                  \multicolumn{1}{c|}{4.887E-02} &
                  \multicolumn{1}{c|}{DP2} &
                  \multicolumn{1}{c|}{-} &
                  \multicolumn{1}{c|}{-} & 
                  \multicolumn{1}{c|}{-} \\ \hline 
                   
                \end{tabular}
            }
        \vspace{5pt}    
        \caption{\textbf{Synopsis of the CD-NRAM Scenarios Vs. The Base Case Scenario with lz4 - Favour Post-Processing.} The table legend is as in Table \ref{tab:CD-NRAM_Vs_BaseCase_zstd_PostProc}.}
        \label{tab:CD-NRAM_Vs_BaseCase_lz4_PostProc}
        \end{table*}

        %TAB.14new (TAB.X23.old, TAB.X24.old)
        \begin{table*}[]
            \centering
            \resizebox{.8\textwidth}{!}{%             
                \begin{tabular}{|c|ccccccccc|ccccccccc|}
                \hline
                 &
                  \multicolumn{3}{c|}{\textit{SA}} &
                  \multicolumn{3}{c|}{\textit{HC14}} &
                  \multicolumn{3}{c|}{\textit{AP}} &
                  \multicolumn{3}{c|}{\textit{SA}} &
                  \multicolumn{3}{c|}{\textit{HC14}} &
                  \multicolumn{3}{c|}{\textit{AP}} \\ \hline
                  
                \textbf{k} &
                  \multicolumn{1}{c|}{\textbf{C}} &
                  \multicolumn{1}{c|}{\textbf{T}} &
                  \multicolumn{1}{c|}{\textbf{B}} &
                  \multicolumn{1}{c|}{\textbf{C}} &
                  \multicolumn{1}{c|}{\textbf{T}} &
                  \multicolumn{1}{c|}{\textbf{B}} &
                  \multicolumn{1}{c|}{\textbf{C}} &
                  \multicolumn{1}{c|}{\textbf{T}} &
                  \multicolumn{1}{c|}{\textbf{B}} &
                  \multicolumn{1}{c|}{\textbf{C}} &
                  \multicolumn{1}{c|}{\textbf{T}} &
                  \multicolumn{1}{c|}{\textbf{B}} &
                  \multicolumn{1}{c|}{\textbf{C}} &
                  \multicolumn{1}{c|}{\textbf{T}} &
                  \multicolumn{1}{c|}{\textbf{B}} &
                  \multicolumn{1}{c|}{\textbf{C}} &
                  \multicolumn{1}{c|}{\textbf{T}} &
                  \multicolumn{1}{c|}{\textbf{B}} \\ \hline
                  
                \textbf{4} &
                  \multicolumn{1}{c|}{Y} &
                  \multicolumn{1}{c|}{Y} &
                  \multicolumn{1}{c|}{DP1} &
                  \multicolumn{1}{c|}{Y} &
                  \multicolumn{1}{c|}{Y} &
                  \multicolumn{1}{c|}{DP1} &
                  \multicolumn{1}{c|}{Y} &
                  \multicolumn{1}{c|}{Y} &
                  \multicolumn{1}{c|}{DP2} &
                  \multicolumn{1}{c|}{1.290E-03} &
                  \multicolumn{1}{c|}{1.017E-03} &
                  \multicolumn{1}{c|}{DP1} &
                  \multicolumn{1}{c|}{5.305E-05} &
                  \multicolumn{1}{c|}{2.522E-05} &
                  \multicolumn{1}{c|}{DP1} &
                  \multicolumn{1}{c|}{7.315E-07} &
                  \multicolumn{1}{c|}{1.214E-06} & 
                  \multicolumn{1}{c|}{DP2} \\ \hline 
                  
                \textbf{8} &
                  \multicolumn{1}{c|}{Y} &
                  \multicolumn{1}{c|}{Y} &
                  \multicolumn{1}{c|}{DP0} &
                  \multicolumn{1}{c|}{Y} &
                  \multicolumn{1}{c|}{Y} &
                  \multicolumn{1}{c|}{DP1} &
                  \multicolumn{1}{c|}{Y} &
                  \multicolumn{1}{c|}{Y} &
                  \multicolumn{1}{c|}{DP1} &
                  \multicolumn{1}{c|}{2.337E-01} &
                  \multicolumn{1}{c|}{1.680E-03} &
                  \multicolumn{1}{c|}{DP0} &
                  \multicolumn{1}{c|}{6.429E-03} &
                  \multicolumn{1}{c|}{7.352E-05} &
                  \multicolumn{1}{c|}{DP1} &
                  \multicolumn{1}{c|}{1.670E-04} &
                  \multicolumn{1}{c|}{2.326E-06} & 
                  \multicolumn{1}{c|}{DP1} \\ \hline 
                   
                \textbf{16} &
                  \multicolumn{1}{c|}{N} &
                  \multicolumn{1}{c|}{Y} &
                  \multicolumn{1}{c|}{DP0} &
                  \multicolumn{1}{c|}{N} &
                  \multicolumn{1}{c|}{Y} &
                  \multicolumn{1}{c|}{DP2} &
                  \multicolumn{1}{c|}{N} &
                  \multicolumn{1}{c|}{Y} &
                  \multicolumn{1}{c|}{DP2} &
                  \multicolumn{1}{c|}{9.201E+00} &
                  \multicolumn{1}{c|}{1.725E-02} &
                  \multicolumn{1}{c|}{DP0} &
                  \multicolumn{1}{c|}{4.439E+00} &
                  \multicolumn{1}{c|}{1.667E-02} &
                  \multicolumn{1}{c|}{DP2} &
                  \multicolumn{1}{c|}{1.174E+00} &
                  \multicolumn{1}{c|}{2.304E-02} & 
                  \multicolumn{1}{c|}{DP2} \\ \hline 
                   
                \textbf{32} &
                  \multicolumn{1}{c|}{N} &
                  \multicolumn{1}{c|}{Y} &
                  \multicolumn{1}{c|}{DP0} &
                  \multicolumn{1}{c|}{N} &
                  \multicolumn{1}{c|}{Y} &
                  \multicolumn{1}{c|}{DP1} &
                  \multicolumn{1}{c|}{N} &
                  \multicolumn{1}{c|}{Y} &
                  \multicolumn{1}{c|}{DP1} &
                  \multicolumn{1}{c|}{1.823E+01} &
                  \multicolumn{1}{c|}{3.680E-02} &
                  \multicolumn{1}{c|}{DP0} &
                  \multicolumn{1}{c|}{2.017E+01} &
                  \multicolumn{1}{c|}{2.613E-02} &
                  \multicolumn{1}{c|}{DP1} &
                  \multicolumn{1}{c|}{1.375E+01} &
                  \multicolumn{1}{c|}{1.065E-01} & 
                  \multicolumn{1}{c|}{DP1} \\ \hline 
                   
                \textbf{48} &
                  \multicolumn{1}{c|}{N} &
                  \multicolumn{1}{c|}{Y} &
                  \multicolumn{1}{c|}{DP0} &
                  \multicolumn{1}{c|}{N} &
                  \multicolumn{1}{c|}{Y} &
                  \multicolumn{1}{c|}{DP0} &
                  \multicolumn{1}{c|}{-} &
                  \multicolumn{1}{c|}{-} &
                  \multicolumn{1}{c|}{-} &
                  \multicolumn{1}{c|}{2.083E+01} &
                  \multicolumn{1}{c|}{4.589E-02} &
                  \multicolumn{1}{c|}{DP0} &
                  \multicolumn{1}{c|}{3.685E+01} &
                  \multicolumn{1}{c|}{5.440E-02} &
                  \multicolumn{1}{c|}{DP0} &
                  \multicolumn{1}{c|}{-} &
                  \multicolumn{1}{c|}{-} & 
                  \multicolumn{1}{c|}{-} \\ \hline 
                   
                \textbf{64} &
                  \multicolumn{1}{c|}{N} &
                  \multicolumn{1}{c|}{Y} &
                  \multicolumn{1}{c|}{DP0} &
                  \multicolumn{1}{c|}{N} &
                  \multicolumn{1}{c|}{Y} &
                  \multicolumn{1}{c|}{DP2} &
                  \multicolumn{1}{c|}{-} &
                  \multicolumn{1}{c|}{-} &
                  \multicolumn{1}{c|}{-} &
                  \multicolumn{1}{c|}{2.317E+01} &
                  \multicolumn{1}{c|}{5.359E-02} &
                  \multicolumn{1}{c|}{DP0} &
                  \multicolumn{1}{c|}{5.183E+01} &
                  \multicolumn{1}{c|}{4.885E-02} &
                  \multicolumn{1}{c|}{DP2} &
                  \multicolumn{1}{c|}{-} &
                  \multicolumn{1}{c|}{-} & 
                  \multicolumn{1}{c|}{-} \\ \hline 
                   
                \end{tabular}
            }
        \vspace{5pt}      
        \caption{\textbf{Synopsis of the CD-NRAM Scenarios Vs. The Base Case Scenario with SPRING - Favour Post-Processing.} 
          The table legend is as in Table \ref{tab:CD-NRAM_Vs_BaseCase_zstd_PostProc}.}
        \label{tab:CD-NRAM_Vs_BaseCase_SPRING_PostProc}
        \end{table*}

        %TAB.15new (TAB.X25.old, TAB.X26.old)
        \begin{table*}[]
        \centering
        \resizebox{.8\textwidth}{!}{%            
            \begin{tabular}{|c|cccccc|cccccc|}
            \hline

            \multicolumn{1}{|c|}{\textbf{}} & \multicolumn{6}{c|}{\textbf{Compression}} & \multicolumn{6}{c|}{\textbf{Post-Processing}} \\ \cline{2-13} &
              \multicolumn{2}{c|}{\textit{SA}} &
              \multicolumn{2}{c|}{\textit{HC14}} &
              \multicolumn{2}{c|}{\textit{AP}} &
              \multicolumn{2}{c|}{\textit{SA}} &
              \multicolumn{2}{c|}{\textit{HC14}} &
              \multicolumn{2}{c|}{\textit{AP}} \\ \hline
              
            \textbf{k} &
              \multicolumn{1}{c|}{\textbf{T}} &
              \multicolumn{1}{c|}{\textbf{B}} &
              \multicolumn{1}{c|}{\textbf{T}} &
              \multicolumn{1}{c|}{\textbf{B}} &
              \multicolumn{1}{c|}{\textbf{T}} &
              \textbf{B} &
              \multicolumn{1}{c|}{\textbf{T}} &
              \multicolumn{1}{c|}{\textbf{B}} &
              \multicolumn{1}{c|}{\textbf{T}} &
              \multicolumn{1}{c|}{\textbf{B}} &
              \multicolumn{1}{c|}{\textbf{T}} &
              \multicolumn{1}{c|}{\textbf{B}} \\ \hline
              
            \textbf{4} &
              \multicolumn{1}{c|}{Y} &
              \multicolumn{1}{c|}{DP2} &
              \multicolumn{1}{c|}{Y} &
              \multicolumn{1}{c|}{DP2} &
              \multicolumn{1}{c|}{Y} &
              \multicolumn{1}{c|}{DP3} &
              \multicolumn{1}{c|}{Y} &
              \multicolumn{1}{c|}{DP1} &
              \multicolumn{1}{c|}{Y} &
              \multicolumn{1}{c|}{DP1} &
              \multicolumn{1}{c|}{Y} &
              \multicolumn{1}{c|}{DP2} \\ \hline
              
            \textbf{8} &
              \multicolumn{1}{c|}{Y} &
              \multicolumn{1}{c|}{DP2} &
              \multicolumn{1}{c|}{Y} &
              \multicolumn{1}{c|}{DP2} &
              \multicolumn{1}{c|}{Y} &
              \multicolumn{1}{c|}{DP2} &
              \multicolumn{1}{c|}{Y} &
              \multicolumn{1}{c|}{DP0} &
              \multicolumn{1}{c|}{Y} &
              \multicolumn{1}{c|}{DP1} &
              \multicolumn{1}{c|}{Y} &
              \multicolumn{1}{c|}{DP1} \\ \hline
              
            \textbf{16} &
              \multicolumn{1}{c|}{Y} &
              \multicolumn{1}{c|}{DP3} &
              \multicolumn{1}{c|}{Y} &
              \multicolumn{1}{c|}{DP3} &
              \multicolumn{1}{c|}{Y} &
              \multicolumn{1}{c|}{DP2} &
              \multicolumn{1}{c|}{Y} &
              \multicolumn{1}{c|}{DP0} &
              \multicolumn{1}{c|}{Y} &
              \multicolumn{1}{c|}{DP2} &
              \multicolumn{1}{c|}{Y} &
              \multicolumn{1}{c|}{DP2} \\ \hline
              
            \textbf{32} &
              \multicolumn{1}{c|}{Y} &
              \multicolumn{1}{c|}{DP3} &
              \multicolumn{1}{c|}{Y} &
              \multicolumn{1}{c|}{DP3} &
              \multicolumn{1}{c|}{Y} &
              \multicolumn{1}{c|}{DP3} &
              \multicolumn{1}{c|}{Y} &
              \multicolumn{1}{c|}{DP0} &
              \multicolumn{1}{c|}{Y} &
              \multicolumn{1}{c|}{DP1} &
              \multicolumn{1}{c|}{Y} &
              \multicolumn{1}{c|}{DP1} \\ \hline
              
            \textbf{48} &
              \multicolumn{1}{c|}{Y} &
              \multicolumn{1}{c|}{DP3} &
              \multicolumn{1}{c|}{Y} &
              \multicolumn{1}{c|}{DP3} &
              \multicolumn{1}{c|}{-} &
              \multicolumn{1}{c|}{-} &
              \multicolumn{1}{c|}{Y} &
              \multicolumn{1}{c|}{DP0} &
              \multicolumn{1}{c|}{Y} &
              \multicolumn{1}{c|}{DP0} &
              \multicolumn{1}{c|}{-} &
              \multicolumn{1}{c|}{-} \\ \hline
              
            \textbf{64} &
              \multicolumn{1}{c|}{Y} &
              \multicolumn{1}{c|}{DP3} &
              \multicolumn{1}{c|}{Y} &
              \multicolumn{1}{c|}{DP3} &
              \multicolumn{1}{c|}{-} &
              \multicolumn{1}{c|}{-} &
              \multicolumn{1}{c|}{Y} &
              \multicolumn{1}{c|}{DP0} &
              \multicolumn{1}{c|}{Y} &
              \multicolumn{1}{c|}{DP2} &
              \multicolumn{1}{c|}{-} &
              \multicolumn{1}{c|}{-} \\ \hline
              
        \end{tabular}
        }
        \vspace{5pt}
        \caption{\textbf{Synopsis of the CD-RAM Scenarios Vs. The Base Case Scenario with KMC3 as a $k$-mer counter - Favour Compression with MFC, Logic Indication; Favour Post-Processing with zstd, Logic Indication}. Dataset abbreviations are as in Table \ref{tab:SD-RAM_Favour_Compr_PostProc}. An entry with a dash indicates that the experiment could not be completed, as explained in the Main text (see \textbf{Experimental Setup}). } 
        \label{tab:CD-NRAM_Vs_BaseCase_KMC3_ComprMFC_PostProcZSTD}
        \end{table*}

%%% PARETO TABLES %%%

        %TAB.16new Staphylococcus Aureus Pareto Table 
        \begin{table*}[]
            \centering
            \resizebox{.9\textwidth}{!}{%
                \begin{tabular}{|l|c|c|}
                \hline
                \multicolumn{1}{|c|}{\textbf{Compression Setting}} & \textbf{Compression (bytes)} & \textbf{Post-Processing Time (s)} \\ \hline
                    DP3(MFC(S), bzip2(Fk))  & 683.623    & 3,818 \\ \hline
                    DP3(MFC(S), zstd(Fk)) & 685.163    & 3,771 \\ \hline
                    DP3(SPRING(S), bzip2(Fk)) & 709.530    & 2,303 \\ \hline
                    DP3(SPRING(S), zstd(Fk)) & 711.070    & 2,256 \\ \hline
                    DP3(zstd(S), bzip2(Fk)) & 720.870    & 2,082 \\ \hline
                    DP3(zstd(S), zstd(Fk)) & 722.410    & 2,035 \\ \hline
                    DP3(lz4(S), zstd(Fk)) & 1.091.374  & 2,034 \\ \hline
                    DP0(zstd(Dk), bzip2(Fk)) & 13.033.265 & 0,348 \\ \hline
                    DP0(zstd(Dk), zstd(Fk)) & 13.048.104 & 0,229 \\ \hline
                    DP0(zstd(Dk), lz4(Fk)) & 13.064.701 & 0,228 \\ \hline
                \end{tabular}
            }
            \vspace{5pt}
            \caption{\textbf{CD-NRAM Scenario: \textit{Staphylococcus Aureus} Dataset - Pareto Optimality}. The first column indicates the pareto optimal configurations, sorted by compression. The other two columns indicate the total output bytes stored on disk by the methods reported in the first column and the related post-processing time, respectively. }
            \label{tab:ParetoPoints_SA}
        \end{table*}

        %TAB.17new Human Chromosome 14 Pareto Table 
        \begin{table*}[]
            \centering
            \resizebox{.9\textwidth}{!}{%
                \begin{tabular}{|l|c|c|}
                \hline
                \multicolumn{1}{|c|}{\textbf{Compression Setting}} & \textbf{Compression (bytes)} & \textbf{Post-Processing Time (s)} \\ \hline
                    DP3(MFC(S), bzip2(Fk))    & 21.225.154  & 74,567 \\ \hline
                    DP3(MFC(S), zstd(Fk))     & 21.448.976  & 71,896 \\ \hline
                    DP3(SPRING(S), bzip2(Fk)) & 22.268.947  & 56,867 \\ \hline
                    DP3(SPRING(S), zstd(Fk))  & 22.492.769  & 54,196 \\ \hline
                    DP3(zstd(S), zstd(Fk))    & 23.129.127  & 52,817 \\ \hline
                    DP2(zstd(Dk), bzip2(Gk))  & 434.172.272 & 11,936 \\ \hline
                    DP2(zstd(Dk), bzip2(Fk))  & 434.174.212 & 4,656  \\ \hline
                    DP2(zstd(Dk), zstd(Fk))   & 434.180.711 & 3,774  \\ \hline
                    DP1(zstd(Dk), zstd(Fk))   & 434.908.387 & 3,658  \\ \hline
                \end{tabular}
            }
            \vspace{5pt}
            \caption{\textbf{CD-NRAM Scenario: \textit{Human Chromosome 14} Dataset - Pareto Optimality}. The table legend is as in Table \ref{tab:ParetoPoints_SA}.}
            \label{tab:ParetoPoints_HC14}
        \end{table*}

        %TAB.18new Assembled Plants Pareto Table 
        \begin{table*}[]
            \centering
            \resizebox{.9\textwidth}{!}{%
                \begin{tabular}{|l|c|c|}
                \hline
                \multicolumn{1}{|c|}{\textbf{Compression Setting}} & \textbf{Compression (bytes)} & \textbf{Post-Processing Time (s)} \\ \hline
                    DP3(MFC(S), zstd(Fk))   & 1.577.641.106  & 3.581,911 \\ \hline
                    DP3(zstd(S), zstd(Fk))  & 1.587.689.427  & 2.626,511 \\ \hline
                    DP1(zstd(Dk), zstd(Fk)) & 15.114.316.478 & 1.176,631 \\ \hline
                \end{tabular}
            }
            \vspace{5pt}
            \caption{\textbf{CD-NRAM Scenario: \textit{Assembled Plants} Dataset - Pareto Optimality}. The table legend is as in Table \ref{tab:ParetoPoints_SA}.}
            \label{tab:ParetoPoints_AP}
        \end{table*}

        % TAB.19new: Favour Compression using DP3(MFC(S), bzip2(Fk))
        \begin{table*}[]
            \centering
            \resizebox{.8\textwidth}{!}{%
                \begin{tabular}{|c|c|c|c|}
                \hline
                \multicolumn{1}{|c|}{\textbf{k}} & \multicolumn{1}{c|}{\textit{SA}} & \multicolumn{1}{c|}{\textit{HC14}} & \multicolumn{1}{c|}{\textit{AP}} 
                \\ & & & 
                \\ \hline
                
                \textbf{4} & DP2 (bzip2(Dk),   zstd(Gk))  {[}1.32E+00{]} & DP2(zstd(Dk),   zstd(Gk))  {[}1.24E+00{]} & DP3(zstd(S),   zstd(Fk))  {[}1.22E+00{]} 
                \\ & & & 
                \\ \hline
                
                \textbf{8} & DP2 (MFC(Dk),   bzip2(Gk))  {[}1.03E+00{]} & DP2(MFC(Dk),   bzip2(Gk))  {[}1.29E+00{]} & DP2(MFC(Dk),   bzip2(Gk))  {[}1.36E+00{]} 
                \\ & & & 
                \\ \hline
                
                \textbf{16} & *  {[}1.00E+00{]} & DP3(zstd(S),   bzip2(Fk))  {[}1.08E+00{]} & DP2(zstd(Dk),   bzip2(Gk))  {[}2.03E+00{]} 
                \\ & & & 
                \\ \hline
                
                \textbf{32} & *  {[}1.00E+00{]} & *  {[}1.00E+00{]} & DP3(MFC(S),   zstd(Fk))  {[}1.02E+00{]} 
                \\ & & & 
                \\ \hline
                
                \textbf{48} & *  {[}1.00E+00{]} & *  {[}1.00E+00{]} & \textit{\textbf{-}} 
                \\ & & & 
                \\ \hline
                
                \textbf{64} & *  {[}1.00E+00{]} & *  {[}1.00E+00{]} & \textit{\textbf{-}} 
                \\ & & & 
                \\ \hline
                
                \end{tabular}
            }
            \vspace{5pt}
            \caption{{\bf {Synopsis of the CD-NRAM Scenario: Favour Compression using DP3(MFC(S), bzip2(Fk)) with respect to the use of the best Case in terms of compression}}. The values of $k$ we have considered are reported in the first column of the table, which is divided into three panels, one for each of the datasets used in this research. Dataset abbreviations are as in Table \ref{tab:SD-RAM_Favour_Compr_PostProc}. For each panel, we report an estimate of the potential losses of choosing to use {\bf DP3} with {\bf MFC} to compress $S$ and {\bf bzip2} to compress $F_k$ with respect to the use of the best Case in terms of compression. An entry with a dash indicates that the experiment could not be completed, as explained in the Main text (see \textbf{Experimental Setup}). } 
            \label{tab:CD-NRAM_favour_compression_wrt}
        \end{table*}

        % TAB.20new: Favour Post-Processing using DP0(zstd(Dk), zstd(Fk))
        \begin{table*}[]
            \centering
            \resizebox{.8\textwidth}{!}{%
                    \begin{tabular}{|c|c|c|c|}
                    \hline
                \multicolumn{1}{|c|}{\textbf{k}} & \multicolumn{1}{c|}{\textit{SA}} & \multicolumn{1}{c|}{\textit{HC14}} & \multicolumn{1}{c|}{\textit{AP}} 
                    \\ & & & 
                    \\ \hline
                    
                    \textbf{4} & DP1 (zstd(Dk),   Opt-PFOR(Fk))  {[}4.66E+00{]} & DP1 (zstd(Dk),   Opt-PFOR(Fk))  {[}2.00E+00{]} & DP2 (zstd(Dk),   zstd(Fk))  {[}1.50E+00{]} 
                    \\ & & & 
                    \\ \hline
                    
                    \textbf{8} & DP0 (lz4(Dk),   zstd(Fk))  {[}1.33E+00{]} & DP1 (zstd(Dk),   lz4(Fk))  {[}1.33E+00{]} & DP1 (zstd(Dk),   lz4(Fk))  {[}5.33E+00{]} 
                    \\ & & & 
                    \\ \hline
                    
                    \textbf{16} & * {[}1.00E+00{]} & DP2 (zstd(Dk),   zstd(Fk))  {[}1.00E+00{]} & DP2 (zstd(Dk),   zstd(Fk))  {[}1.46E+00{]} 
                    \\ & & & 
                    \\ \hline
                    
                    \textbf{32} & DP0 (zstd(Dk),   lz4(Fk)) {[}1.01E+00{]} & DP1 (zstd(Dk),   zstd(Fk)) {[}1.61E+00{]} & DP1 (zstd(Dk),   zstd(Fk))  {[}1.34E+00{]}
                    \\ & & & 
                    \\ \hline
                    
                    \textbf{48} & *  {[}1.00E+00{]} & *  {[}1.00E+00{]} & \textit{\textbf{-}} 
                    \\ & & & 
                    \\ \hline
                    
                    \textbf{64} & *  {[}1.00E+00{]} & *  DP2 (zstd(Dk),   zstd(Fk)) {[}1.45E+00{]} & \textit{\textbf{-}} 
                    \\ & & & 
                    \\ \hline
                    
                    \end{tabular}
               }
            \vspace{5pt}
            \caption{{\bf {Synopsis of the CD-NRAM Scenario: Favour Post-Processing using DP0(zstd(Dk), zstd(Fk)) with respect to the use of the best Case in terms of post-processing time}}. The values of $k$ we have considered are reported in the first column of the table, which is divided into three panels, one for each of the datasets used in this research. Dataset abbreviations are as in Table \ref{tab:SD-RAM_Favour_Compr_PostProc}. For each panel, we report an estimate of the potential losses of choosing to use {\bf DP0} with {\bf zstd} to compress $D_k$ and {\bf zstd} to compress $F_k$ with respect to the use of the best Case in terms of post-processing time. An entry with a dash indicates that the experiment could not be completed, as explained in the Main text (see \textbf{Experimental Setup}). } 
            \label{tab:CD-NRAM_favour_post_processing_time_wrt}
        \end{table*}

\end{document}